\documentclass[12pt, letterpaper]{article}

\usepackage{amssymb, amsthm, amsmath, bm, color}
\usepackage{natbib}
\usepackage{graphicx}
\usepackage[figuresright]{rotating}
\usepackage{subfig}
\usepackage{soul}
\usepackage{multirow}
\usepackage{fullpage}
\usepackage{setspace}
\usepackage{pdflscape}
\usepackage[table]{xcolor}
\usepackage[hidelinks]{hyperref}

\usepackage{JASA_manu}

\newcommand{\blind}{1}

\usepackage{threeparttable}
\usepackage{color, colortbl}

\definecolor{red}{rgb}{1,0,0}
\definecolor{blue}{rgb}{0,0,1}
\definecolor{green}{rgb}{0,0.6,0.4}

\definecolor{Gray}{gray}{0.9}

\newtheorem{theorem}{Theorem}

\newcommand\indep{\protect\mathpalette{\protect\independenT}{\perp}}
\def\independenT#1#2{\mathrel{\rlap{$#1#2$}\mkern2mu{#1#2}}}
	
\begin{document}

\if1\blind
{
  \title{\bf Risk Projection for Time-to-event Outcome Leveraging Summary Statistics With Source Individual-level Data}
  \author{Jiayin Zheng, Yingye Zheng, and Li Hsu\thanks{Corresponding author (lih@fhcrc.org).
    The work is supported in part by the grants from the National Institutes of Health (R01 CA189532, R01 CA195789, R01 CA236558 and U01CA86368).  
    The authors are grateful to the generosity of WHI investigators for allowing them to use the WHI data to illustrate the method proposed in this paper.
    The WHI program is funded by the National Heart, Lung, and Blood Institute, National Institutes of Health, U.S. Department of Health and Human Services through contracts HHSN268201600018C, HHSN268201600001C, HHSN268201600002C, HHSN268201600003C, and HHSN268201600004C. The list of
    investigators is provided in Web Appendix F  of the Supplementary Materials. Part of this research has been conducted using the UK Biobank Resource under Application Number 8614.}\hspace{.2cm}\\
    Fred Hutchinson Cancer Research Center}
  \date{}
  \maketitle
} \fi

\if0\blind
{
  \bigskip
  \bigskip
  \bigskip
  \begin{center}
    {\LARGE\bf Risk Projection for Time-to-event Outcome Leveraging Summary Statistics With Source Individual-level Data}
\end{center}
  \medskip
} \fi
	
\bigskip
\begin{abstract}

	Predicting risks of chronic diseases has become increasingly important in clinical practice. When a prediction model is developed in a given source cohort, there is often a great interest to apply the model to other cohorts. However, due to potential discrepancy in baseline disease incidences between different cohorts and shifts in patient composition, the risk predicted by the original model often under- or over-estimates the risk in the new cohort. The remedy of such a poorly calibrated prediction is needed for proper medical decision-making. In this article, we assume the relative risks of predictors are the same between the two cohorts, and propose a novel weighted estimating equation approach to re-calibrating the projected risk for the targeted population through updating the baseline risk. The recalibration leverages the knowledge about the overall  survival probabilities for the disease of interest and competing events, and the summary information of risk factors from the targeted population. The proposed re-calibrated risk estimators gain efficiency if the risk factor distributions are the same for both the source and target cohorts, and are robust with little bias if they differ. We establish the consistency and asymptotic normality of the proposed estimators. Extensive simulation studies demonstrate that the proposed estimators perform very well in terms of robustness and efficiency in finite samples. A real data application to colorectal cancer risk prediction also illustrates that the proposed method can be used in practice for model recalibration.  
\end{abstract}

\noindent%

\section{INTRODUCTION}
\label{s:intro}

Accurate individual risk prediction for disease occurrence and progression is critical in clinical decision-making to identify high-risk patients that need medical intervention while preventing unnecessary treatment of low-risk patients \citep{alba2017discrimination}. 
An important aspect of predictive model assessment is calibration, i.e., the agreement between predicted risks and observed outcomes of the target population \citep{steyerberg2010assessing}. Ideally, the cohort used for model development reflects well the target population to which the model will be applied. It can then be expected that the predicted risk will be  calibrated well to the observed risk of the target population.  However, this is seldom the case. In practice, the data used for model building are often from a clinical trial or research cohort that is assembled for specific purposes. Due to self-selection and inclusion/exclusion criteria, the risk profiles of study participants may not represent that of the population from which they are sampled. In other situations, researchers may be interested in applying an established risk prediction model developed for the general population to their own cohort of patients with possibly different risk profiles from the general population. When not all the risk factors are included in the risk prediction model because they are unknown or not collected, the model can not be generalized across cohorts. Consider a hypothetical example. Suppose the Charlson comorbidity index, a summary measure of various comorbidities, is an important risk factor but has not been included in the risk prediction model for a clinical outcome. Further the model is built in a healthy cohort with on average a lower Charlson comorbidity index than the general population. As expected, the baseline risk obtained from the healthy cohort would be lower than the general population. Thus if one naively applies this model to the general population, it can lead to an under-estimation of risk. 

The poor calibration of prediction models across cohorts occurs frequently for many disease risk prediction models. For example, \citet{collins2009independent,collins2010independent,collins2012predicting} evaluated a series of models for predicting 10-year cardiovascular disease in several independent United Kingdom (UK) cohorts of patients and found that the Framingham cardiovascular model over-estimated the risk, especially in men; \citet{hippisley2013derivation} observed the Framingham stroke score under-predicted stroke risk in UK patients; \citet{abbasi2012prediction}  validated 12 prediction models for the risk of developing type 2 diabetes over 7.5 years and found most of them over-estimated the observed risk of diabetes, particularly at higher observed risks; 
recently, \citet{drost2019predicting} assessed two risk calculators for prostate cancer progression in five active surveillance cohorts, and found one calculator  over- or under-estimated risks in two cohorts, and the other calculator over-estimated risks in most of the cohorts. These diverse examples show that there is a substantial heterogeneity among cohorts and insufficient calibration is common.  Therefore, re-calibration of a risk prediction model before applying it to the target population is critically needed for accurate risk projection but is frequently under-appreciated \citep{van2015calibration}.

Estimating disease risk involves two components: the baseline risk and relative risks of risk factors. Suppose $T$ is the time-to-event and $\boldsymbol{Z}$ a $p\times 1$ vector of risk factors. Under the Cox proportional hazards model, the probability of developing the disease before $t_1$ given being disease-free at $t_0$ in the absence of competing risks can be expressed as
\begin{eqnarray}
\label{absrisk}
\mbox{Pr}(t_0\leq T <t_1|T\geq t_0, \boldsymbol{Z})=1-\exp[\{\Lambda_0(t_0)-\Lambda_0(t_1)\}\exp(\boldsymbol{\beta_0}^T \boldsymbol{Z})],
\end{eqnarray} where $\Lambda_0(t)$ is cumulative baseline hazard function and $\exp{(\boldsymbol{\beta_0})}$ are the hazard ratios of $\boldsymbol{Z}$. As the risk prediction is for the target cohort, ideally both parameters $\boldsymbol{\beta_0}$ and $\Lambda_0(t)$ should represent the target population. When the parameter estimates are obtained from a source cohort that differs from the target population, they may not represent well the target population and  need to be recalibrated. It is common to assume that $\boldsymbol{\beta_0}$ is generalizable  from the source cohort to the target population, as \citet{keiding2016perils} noted that conditional features or distributions are more likely to be ``transportable'' from one population to another than are marginal distributions. However, $\Lambda_0(t)$ can be quite different between the source cohort and the target population as demonstrated in the above examples. Since disease risk involves both $\boldsymbol{\beta_0}$ and $\Lambda_0(t)$, this poses a major challenge when applying a risk prediction model developed in one cohort to another.

To re-calibrate the baseline hazard function, the most direct approach is to refit the prediction model using data from the targeted cohort. However, individual-level cohort data linking predictors to the outcome are usually not readily available, particularly for rare diseases such as cancer. Assembling a cohort for rare diseases requires an extended follow-up of a large number of individuals with detailed risk factor information, which can be prohibitive in logistics and cost. In contrast, summary information such as age-specific disease incidence rates and summary statistics (e.g., mean and prevalence) of aspects of patient characteristics is often available from disease registries, historical patient data, census data, and population-based survey data. A more practical solution for re-calibration is to leverage this aggregated information  from the target population together with the individual level data of source cohort.

Leveraging summary information on disease incidence rates has been proposed for obtaining the baseline hazard function  \citep{gail1989projecting,liu2014estimating}. These approaches require the joint distribution of all risk factors $f(\boldsymbol{Z})$ to be common between the source cohort and the population from which the disease incidence rates are obtained. However, this assumption is strong and often violated in practice.  For example, in our real data application, many of the risk factors have very different frequencies across cohorts (Table \ref{table:example1} in Section \ref{s:application}). To relax this assumption, \citet{chen2006projecting} obtained $f(\boldsymbol{Z})=f(\boldsymbol{Z_1},\boldsymbol{Z_2})$ through combining the partial distribution $f(\boldsymbol{Z_1})$ from the target cohort and  $f(\boldsymbol{Z_2}|\boldsymbol{Z_1})$ from the source cohort, however still assuming the conditional distribution is same in the two populations. This approach requires individual-level data on $\boldsymbol{Z_1}$ from the target population. Therefore existing methods are limited in their assumptions on the covariate distributions and the requirement for individual level data from the target cohort when such  assumptions are unjustified.

In this article, we propose a novel method to re-calibrate the baseline hazard function for the target population, leveraging only the summary-level information without the need for a reference sample from the target population. Specifically, we propose an estimating equation for $\Lambda_0(t)$ such that the overall disease probability is constrained to be the same as that for the target cohort.  To  relax the assumption that the joint or conditional covariate distributions need to be the same between cohorts as in the existing methods \citep{gail1989projecting, chen2006projecting}, we propose to further re-weigh the contribution of each individual from the source cohort in the estimating equation. This effectively shifts the observed distribution of the source population  towards the target population and makes the sample more representative of the target population. Without using individual level data, we propose to identify the weights based on an empirical likelihood approach \citep{owen1988empirical,owen2001empirical, qin1994empirical}. We note that such a technique has been considered in the literature, for example, to improve the efficiency of parameter estimation (e.g., regression coefficients) by considering data sources external to the study  \citep{chatterjee2016constrained,huang2016efficient, han2017empirical}. We adapt it here in the setting of correcting for bias due to difference in covariate distributions. By solving the weighted estimating equation, we obtain an estimator that gains substantial efficiency if the risk factor distributions are same for the two populations, and reduces the bias considerably when they differ.  Finally, as the event of interest is often subject to competing risks, we extend our method allowing the individuals to fail from multiple events and provide a more accurate and practical solution to projecting absolute risk. To our knowledge, this is the first work to recalibrate the baseline hazard function and thus risk prediction in survival analysis for the target population with only the summary information. 

The rest of this article is organized as follows. We describe the proposed estimation procedure in Section \ref{s:methods} and the large sample properties in Section \ref{s:largesample}. In Section \ref{s:cmprsk} we extend the method for absolute risk when the competing risk is present. We evaluate the finite sample performance of the proposed estimator through extensive simulation studies, with the results presented in Section \ref{s:simulation}. We illustrate the proposed method by an application to re-calibrate a risk prediction model for colorectal cancer developed in the Women’s Health Initiative (WHI) study \citep{study1998design} as the source cohort to five different cohorts from  the Prostate, Lung, Colorectal and Ovarian Trial (PLCO) study \citep{prorok2000design} and the UK Biobank \citep{sudlow2015uk} in Section \ref{s:application}.  Some concluding remarks are given in Section \ref{s:discussion}.  Technical details and additional simulation results are provided in the Web-based Supplementary Materials.

\section{METHODS}
\label{s:methods}

\subsection{Notation and Models}
\label{subs:notation}

Consider a source study in which a risk prediction model is developed for the time to the event of interest, $T$, given a $p\times 1$ vector of risk factors $\boldsymbol{Z}$. We denote the underlying population for this study by $P^*$, and refer $\mbox{Pr}^*$ and $\mbox{E}^*$ generically as the probability and  expectation of random variables with respect to $P^*$. The interest is to generalize this risk prediction model to a target population $P$, for which the corresponding probability and expectation are denoted by $\mbox{Pr}$ and $\mbox{E}$, respectively. We assume that  individual-level data are only available for $P^*$, but for $P$ only some summary information such as overall disease incidence rates and/or mean values of some risk factors. 

For both populations, we postulate the Cox proportional hazards model \citep{cox1972regression} for the relationship between the failure time $T$ and the risk factors $\boldsymbol{Z}$. Specifically, for the source population $P^*$, 
\begin{eqnarray}
\label{def:coxin}
\lambda^*(t|\boldsymbol{Z})=\lambda^*_0(t)\exp(\boldsymbol{\beta_0}^T \boldsymbol{Z}),
\end{eqnarray}
where $\lambda^*(t|\boldsymbol{Z})=\lim_{dt\rightarrow 0}\mbox{Pr}^*(t<T<t+dt|T\geq t,\boldsymbol{Z})/dt$, $\lambda^*_0(t)$ is an unspecified baseline hazard function for $P^*$, and $\exp(\boldsymbol{\beta_0})$ is a $p\times 1$ vector of hazard ratio parameters. Following the convention, we assume the hazard ratios for $\boldsymbol{Z}$ are common between the two populations. Then for the target population $P$, we have
\begin{eqnarray}
\label{def:coxex}
\lambda(t|\boldsymbol{Z})=\lambda_0(t)\exp(\boldsymbol{\beta_0}^T \boldsymbol{Z}),
\end{eqnarray}
where $\lambda(t|\boldsymbol{Z})=\lim_{dt\rightarrow 0}\mbox{Pr}(t<T<t+dt|T\geq t,\boldsymbol{Z})/dt$ and  $\lambda_0(t)$ is an unspecified baseline hazard function for $P$.  Note that no assumption is made about $f(\boldsymbol{Z})$ being same for the two populations. 

For the source study from which the model is developed, let $T_i$, $L_i$, and $C_i$ be the failure time, the left truncation time (e.g., study entry), and the right censoring time for the $i$th ($i=1,...,n$) subject, respectively. Let $\boldsymbol{Z_i}$ be a $p\times 1$  vector of risk factors in the prediction model.  Define $X_i=\min(T_i,C_i)$ if $X_i\geq L_i$, and the disease indicator $\delta_i=I{(L_i<T_i\leq C_i)}$, where $I{(\cdot)}$ is an indicator function. Therefore $\delta_i=1$ if the failure time is observed and $\delta_i=0$ otherwise. We assume that $\{(X_i,\delta_i,\boldsymbol{Z_i}),i=1,...,n\}$ are independently and identically distributed (i.i.d.), and both $L_i$ and $C_i$ are independent of $T_i$ conditional on $\boldsymbol{Z_i}$. Define the counting process $N_i(t)=I{(L_i<X_i\leq t,\delta_i=1)}$ and the at-risk process $Y_i(t)=I{(L_i<t\leq X_i)}$. In addition, define $\overline{N}(t)=n^{-1}\sum^n_{i=1}N_i(t)$, $H(t)=n^{-1}\sum^n_{i=1}Y_i(t)$, and $\boldsymbol{H}_r(t;\boldsymbol{\beta})=n^{-1}\sum^n_{i=1} Y_i(t)\boldsymbol{Z}^{\otimes r}_i \exp(\boldsymbol{\beta}^T \boldsymbol{Z_i})$, where $r=0,1,2$, $\boldsymbol{a}^{\otimes 0}=1$, $\boldsymbol{a}^{\otimes 1}=\boldsymbol{a}$, and $\boldsymbol{a}^{\otimes 2}=\boldsymbol{a} \boldsymbol{a}^T$. The maximum partial likelihood estimator $\boldsymbol{\hat{\beta}}$ can be obtained by solving the following score equations
$\boldsymbol{U}(\boldsymbol{\beta})=\sum^n_{i=1} \int_{0}^{\tau} \{ \boldsymbol{Z_i}-\overline{\boldsymbol{Z}}(t;\boldsymbol{\beta})\}dN_i(t)$,
where $\overline{\boldsymbol{Z}}(t;\boldsymbol{\beta})=\boldsymbol{H}_1(t;\boldsymbol{\beta})/H_0(t;\boldsymbol{\beta})$ and $\tau$ is the end of the follow-up period such that $\Pr(H(\tau)>0)>0$. The cumulative baseline hazard function $\Lambda^*_0(t)=\int_{0}^{\tau}\lambda^*_0(s)ds$ can be estimated by the Breslow estimator \citep{breslow1972disussion}, denoted as $\widehat{\Lambda}^\textup{b}_0(t)=\int_{0}^{t} d\overline{N}(u)/H_0(u;\boldsymbol{\hat{\beta}})$. These estimators have been widely used for estimating $\boldsymbol{\beta}$ and $\Lambda^*_0(t)$ under the Cox model for cohort studies.

\subsection{Proposed Estimation Methods}

For the target population $P$, the  available information may include $S(t) = \Pr(T > t)$, the disease-free probability at time $t$, and/or $\boldsymbol{\mu_0} \equiv \mbox{E}\{\boldsymbol{h(Z)}\}$, where $\boldsymbol{h(Z)}\equiv (h_1(\boldsymbol{Z}),...,h_q(\boldsymbol{Z}))^T$ is a 
$q\times 1$ (known) mapping function. For example,  a common function $h(\cdot)$ is  $h(\boldsymbol{Z})=Z_j$, then the summary information is the mean of $j$th risk factor $Z_j$. Other functions include $h(\boldsymbol{Z}) = Z_j^2$ or $Z_jI(Z_k=1)$ if $k$th risk factor $Z_k$ is binary. The expectation of these functions is the second moment of $Z_j$ or the mean of $Z_j$ for those with $Z_k=1$.

Let $V(t)$ be a generic baseline cumulative hazard function and $\Phi(\boldsymbol{Z}; V(t),\boldsymbol{\beta},S(t))=\exp\{-V(t)\exp(\boldsymbol{\beta}^T \boldsymbol{Z})\}-S(t)$, for $t \in [0, \tau]$. 
It is easy to see that for target population 
\begin{eqnarray}
\label{def:ee}
\mbox{E}\{\Phi(\boldsymbol{Z}; \Lambda_0(t),\boldsymbol{\beta_0},S(t))\}=0,
\end{eqnarray}
where $\Lambda_0(t)=\int^t_0 \lambda_0(u)du$. 
If we assume that $f(\boldsymbol{Z})$ is identical between $P$ and $P^*$, we can estimate $\mbox{E}$ by using the empirical risk factor distribution estimator from $P^*$, placing 1/n density mass on each of the observed data points $\{\boldsymbol{Z_i}, i=1,...,n\}$. When $S(t)$ at time points $t=t_1,...,t_s$ are available from $P$, we can re-calibrate $\Lambda_0(t)$ at these time points by solving the following estimating equation for $V(t)$  :
\begin{eqnarray}
\label{def:eedata}
n^{-1}\sum^n_{i=1}\Phi(\boldsymbol{Z_i};  V(t),\boldsymbol{\hat{\beta}},S(t))
=n^{-1}\sum^n_{i=1} \left [ \exp\{-V(t)\exp(\boldsymbol{\hat{\beta}}^T \boldsymbol{Z_i})\}-S(t) \right]
=0. 
\end{eqnarray}
Since $\Phi(\boldsymbol{Z}; V(t),\boldsymbol{\hat{\beta}},S(t))$ is a continuous and strictly decreasing function of $V(t)$ for any fixed $(\boldsymbol{Z}, \boldsymbol{\hat{\beta}},S(t))$, a unique non-negative solution to equation (\ref{def:eedata}) exists when $S(t)>0$. We denote this solution as $\widehat{\Lambda}^\textup{u}_0(t)$, called an unweighted estimator. The Newton-Raphson algorithm can be used to solve equation (\ref{def:eedata}). 

The proposed estimating equation $\mbox{E}\{\Phi(\boldsymbol{Z}; \Lambda_0(t),\boldsymbol{\beta_0},S(t))\}=0$ has connection with the existing methods of estimating baseline hazard by combining case-control or cohort data with disease registry data \citep{pfeiffer2017absolute}.  They proposed to estimate $\lambda_0(t)$ from external composite incidence rates through $\lambda_0(t)=\{1-AR(t)\}\lambda(t)$, where $AR(t)$ is an attributable hazard function at time $t$, which can be estimated from the case-control or cohort data \citep{liu2014estimating,zhao2019adjusted},  and $\lambda(t)$ is the composite incidence rate at time $t$ for the target population. 
This equation is equivalent to the expectation of proposed estimating equation, $\mbox{E}\{\Phi( \boldsymbol{Z}; \Lambda_0(t),\boldsymbol{\beta_0},S(t))\}=0$. This can be seen by taking the derivative of the expected estimating equation with respect to $t$, yielding 
$\lambda_0(t) = \lambda(t) S(t) / \mbox{E}[\exp(\boldsymbol{\beta_0}^T \boldsymbol{Z}) \exp\{-\Lambda_0(t)\exp(\boldsymbol{\beta_0}^T \boldsymbol{Z})\}] $, 
where the right hand side is $\lambda(t)\{1-AR(t)\}$ under the Cox model. In fact, this equivalent relationship between our proposed estimating equation and  $AR(t)$ holds for a general hazard model and the proof is provided in Web Appendix E.  

As noted, in practice the assumption of having a same risk factor distribution for both the source and the target  populations is stringent and implausible \citep{keiding2016perils}, and it is difficult to verify. If it is violated, the risk factor data $\{\boldsymbol{Z_i}, i=1,...,n\}$ from $P^*$ will not represent $f(\boldsymbol{Z})$ in the target population $P$, and the equation (\ref{def:eedata}) will yield biased $\widehat\Lambda_0(t)$. To reduce the bias, a natural approach is to assign  weights to $\{\boldsymbol{Z_i}, i=1,...,n\}$, aiming at making the weighted data more representative of the distribution of $\boldsymbol{Z}$ in $P$. Using the information of $S(t)$ and $\boldsymbol{\mu_0} = \mbox{E}\{\boldsymbol{h(Z)}\}$ from the target population, we can employ empirical likelihood  \citep{owen1988empirical,owen2001empirical, qin1994empirical} to obtain the weights, and then assign them to $\{\boldsymbol{Z_i}, i=1,...,n\}$ in the estimating equation (\ref{def:eedata}). Specifically, the weights can be obtained by maximizing the log empirical likelihood $\sum^n_{i=1}\log (w_i)$ under the constraints of
\begin{eqnarray}
\label{def:elconstraint}
\sum^n_{i=1} w_i \boldsymbol{h(Z_i)} = \boldsymbol{\mu_0}, \quad
w_i>0, \quad
\sum^n_{i=1} w_i = 1.
\end{eqnarray}
Standard empirical likelihood programs can be used to derive the weights (see e.g., \url{http://statweb.stanford.edu/$\sim$owen/empirical/}). Note that as a common premise to use the empirical likelihood method, $\boldsymbol{\mu_0}$ should fall within the convex hull of $\{\boldsymbol{h(Z_i)}, i=1,...,n\}$. 

With the estimated weights, denoted by $\hat{w}_i, i=1,...,n$, now we can estimate $\Lambda_0(t)$ by solving the weighted estimating equation for $V(t)$,
$n^{-1}\sum^n_{i=1} \hat{w}_i \Phi(\boldsymbol{Z_i};V(t),\boldsymbol{\hat{\beta}},S(t))
=n^{-1}\sum^n_{i=1} \hat{w}_i [ \exp\{-V(t)\exp(\boldsymbol{\hat{\beta}}^T \boldsymbol{Z_i})\}-S(t) ]=0$. 
It is easy to show that a unique non-negative solution exists for the above weighted equation when $S(t)>0$. Let  $\widehat{\Lambda}^\textup{w}_0(t)$ denote the solution to this weighted estimating equation. 

Using the Lagrange multiplier, we can show that the weights satisfy
$\hat{w}_i=n^{-1}[1+\boldsymbol{\hat{\gamma}}^T \{\boldsymbol{h(Z_i)} - \boldsymbol{\mu_0}\}]^{-1}, i=1,...,n,$
where $\boldsymbol{\hat{\gamma}}$, the Lagrange multiplier for the constraint $\sum^n_{i=1} w_i \{\boldsymbol{h(Z_i)} - \boldsymbol{\mu_0}\} =  \boldsymbol{0}$, is the solution to
$\sum^n_{i=1} \{\boldsymbol{h(Z_i)} - \boldsymbol{\mu_0}\}/[1+\boldsymbol{\gamma}^T \{\boldsymbol{h(Z_i)} - \boldsymbol{\mu_0}\}]=\boldsymbol{0}.$
This implies that the vector $\{\widehat{\Lambda}^\textup{w}_0(t),\boldsymbol{\hat{\gamma}}\}$ can be written as the solution to the equations 
$\sum^n_{i=1} \boldsymbol{\rho}(\boldsymbol{Z_i};V(t),\boldsymbol{\gamma},\boldsymbol{\mu_0},S(t),\boldsymbol{\hat{\beta}})=\boldsymbol{0}$	 
where
\begin{eqnarray}
\label{ee:rho}
    \boldsymbol{\rho}(\boldsymbol{Z};V(t),\boldsymbol{\gamma},\boldsymbol{\mu},S(t),\boldsymbol{\beta})&=&
    \left( \begin{array}{c}
    	\rho_1(\boldsymbol{Z};V(t),\boldsymbol{\gamma},\boldsymbol{\mu},S(t),\boldsymbol{\beta}) \\
    	\boldsymbol{\rho_2}(\boldsymbol{Z};\boldsymbol{\gamma},\boldsymbol{\mu})
    \end{array}  \right) \nonumber\\
    &=& \left( \begin{array}{c}
    	\frac{\exp\{-V(t)\exp(\boldsymbol{\beta}^T \boldsymbol{Z})\}-S(t)}{1+\boldsymbol{\gamma}^T \{\boldsymbol{h(Z)} - \boldsymbol{\mu}\}}  \\
    	\frac{\boldsymbol{h(Z)} - \boldsymbol{\mu}}{1+\boldsymbol{\gamma}^T \{\boldsymbol{h(Z)} - \boldsymbol{\mu}\}}
    \end{array} \right).
\end{eqnarray}
The weights derived in (\ref{def:elconstraint}) are intended to make the weighted sample representative of the risk factor distribution of the target population. 
Let $\boldsymbol{\gamma_0}$ be the probability limits of $\boldsymbol{\hat{\gamma}}$ (The proof of its existence is provided in Web Appendix D).
Essentially we have three risk factor distributions: 
$f^*(\boldsymbol{Z})$ for $P^*$ from which the individual-level data are sampled,  $f(\boldsymbol{Z})$ for the target population $P$, and $f^{\dag}(\boldsymbol{Z})\equiv f^*(\boldsymbol{Z})/[ 1+ \boldsymbol{\gamma^T_0}\{\boldsymbol{h(Z)-\mu_0}\} ]$ for an artificial population from which the weighted $\{\boldsymbol{Z_i},i=1,...,n\}$ may be considered sampled.  $f^{\dag}(\boldsymbol{Z})$ can be interpreted as the distribution closest to $f^*(\boldsymbol{Z})$, in an empirical likelihood sense, but subject to the restriction that it has the expectation of $\boldsymbol{h(Z)}$ in common with the target population.
 
As it will be shown in Theorem 3 in Section \ref{s:largesample}, under the general case that $\mbox{E}\{\boldsymbol{h(Z)}\} \neq\mbox{E}^*\{\boldsymbol{h(Z)}\}$, 
the estimator $\widehat \Lambda^\textup{w}_0(t)$ obtained from solving (8) converges to $\Lambda^{\dag}_0(t)$, which is the solution to 
$\mbox{E}^{\dag}\{ \Phi(\boldsymbol{Z}; V(t),\boldsymbol{\beta_0},S(t)) \}=0$ where  $\mbox{E}^{\dag}$ is the expectation taken over $f^{\dag}(\boldsymbol{Z})$. Under some conditions, the artificial distribution $f^{\dag}(\boldsymbol{Z})$ is identical to the target distribution $f(\boldsymbol{Z})$. When $f^*(\boldsymbol{Z})$ can be written as $f^*(\boldsymbol{Z})=f(\boldsymbol{Z}) [ 1+ \boldsymbol{\gamma^T_1}\{\boldsymbol{h(Z)-\mu_0}\}]$ for some vector $\boldsymbol{\gamma_1}$, then matching on the expectation $\mbox{E}\{\boldsymbol{h(Z)}\}$ will lead to an artificial distribution $f^{\dag}(\boldsymbol{Z})$ that is identical to $f(\boldsymbol{Z})$, because in this case the probability limit $ \boldsymbol{\gamma_0}$ of  $\boldsymbol{\hat{\gamma}}$ will be equal to $\boldsymbol{\gamma_1}$. Generally speaking, it is unlikely that matching on a few expectations will lead to an artificial population with exactly the same distribution of $\boldsymbol{Z}$ as the target population. However, as more expectations are matched, the artificial risk factor distribution will get closer to the target distribution.

In practice, the true values of $S(t)$ and $\boldsymbol{\mu_0}$ are usually unavailable. Instead, their estimates, denoted as $\hat{S}(t)$ and $\boldsymbol{\hat{\mu}}$, may be available as part of summary-level information that may be obtained from a sample based on the target population. 
We can replace $S(t)$ and $\boldsymbol{\mu_0}$ by consistent estimators $\hat{S}(t)$ and $\boldsymbol{\hat{\mu}}$, respectively, in the estimating equations $\sum^n_{i=1} \boldsymbol{\rho}(\boldsymbol{Z_i};V(t),\boldsymbol{\gamma},\boldsymbol{\mu_0},S(t),\boldsymbol{\hat{\beta}})=\boldsymbol{0}$ to estimate $\Lambda_0(t)$. The variance estimator of $\widehat{\Lambda}^\textup{w}_0(t)$ thus also needs to account this additional variation in $\hat{S}(t)$ and $\boldsymbol{\hat{\mu}}$, which we will consider for inference in Section 3. 

Note that auxiliary variables may also help refine the weights. Let $\boldsymbol{W}$ be an additional $s\times 1$ variable vector recorded in the source study and its summary information is available from the target population.
When the external information of some components of $\boldsymbol{Z}$ is unavailable, $\boldsymbol{W}$ may be served as surrogate if it is correlated with these components. Then the external summary information based on both $\boldsymbol{Z}$ and $\boldsymbol{W}$ provides more auxiliary information than solely on $\boldsymbol{Z}$, in recovering the risk factor distribution of the target population. 
For example, the Charlson comorbidity index predicts the one-year mortality for a patient who may have a range of comorbid conditions, such as heart disease, AIDS, or cancer (a total of more than 20 conditions). Each condition is assigned a score depending on the risk of dying associated with the condition and scores are summed to provide a total score to predict overall mortality. Assume the source cohort has the information of all the conditions and thus the score, but the target only has summary information of partial conditions. Then these partial conditions can be used as surrogates in place of the Charlson comorbidity index for constraining purposes.

\section{LARGE SAMPLE RESULTS}
\label{s:largesample}

In this section we establish the large sample results of the proposed estimators and show the consistency and asymptotic normality of $\widehat{\Lambda}^\textup{u}_0(t)$ and $\widehat{\Lambda}^\textup{w}_0(t)$ under regularity conditions, which are listed in Web Appendix A. Let $n$ be the source study sample size from $P^*$ that we have individual-level data, $m$ be the sample size of the study from the target population $P$ that the summary statistics are based on, and $r=\lim_{n\rightarrow \infty} n/m$.

First, we establish the  consistency and asymptotic normality of unweighted estimator $\widehat{\Lambda}^\textup{u}_0(t)$ in Theorem 1.

\begin{theorem}
	Assume the risk factor distributions are common between $P$ and $P^*$, i.e., $f(\boldsymbol{Z}) = f^*(\boldsymbol{Z})$. Under regularity conditions A1-A3, $\widehat{\Lambda}^\textup{u}_0(t)$ is consistent to $\Lambda_0(t)$, where  $\Lambda_0(t)$ is the true cumulative baseline hazard function in the target population, and $\sqrt{n}\{\widehat{\Lambda}^\textup{u}_0(t)-\Lambda_0(t)\}$ converges to a zero mean normal distribution with variance $ \sigma^2_1(t)$; the formula is provided in Web Appendix B.
\end{theorem}

The $1-\alpha$ confidence interval of $\Lambda_0(t)$ can be constructed as
\begin{eqnarray} 
\label{cl:1}
(\widehat{\Lambda}^\textup{u}_0(t)-z_{1-\alpha/2} \; \hat{\sigma}_1(t)/\sqrt{n},\;\widehat{\Lambda}^\textup{u}_0(t)+z_{1-\alpha/2} \; \hat{\sigma}_1(t)/\sqrt{n}),
\end{eqnarray} 
where $z_{1-\alpha/2}$ is the $1-\alpha/2$ upper quantile of the standard normal distribution, and  $\hat{\sigma}_1(t)$ is the estimator of $\sigma_1(t)$ obtained by replacing the expectations and parameters with corresponding empirical counterparts and parameter estimates, respectively. 

Next, we establish the large sample results for the weighted estimator $\widehat{\Lambda}^\textup{w}_0(t)$ in Theorem 2 when $f(\boldsymbol{Z}) = f^*(\boldsymbol{Z})$. 

\begin{theorem}
	Assume $f(\boldsymbol{Z}) = f^*(\boldsymbol{Z})$. Under regularity conditions A1-A5, we have $\boldsymbol{\gamma_0}=\boldsymbol{0}$ and $\widehat{\Lambda}^\textup{w}_0(t)$ is consistent to $\Lambda_0(t)$. With additional condition A6, $\sqrt{n}\{\widehat{\Lambda}^\textup{w}_0(t)-\Lambda_0(t)\}$ converges to a zero mean normal distribution with variance $\sigma^2_2(t)$; the formula is provided in Web Appendix C.
\end{theorem}
When $n/m$ is low, that is, the sample size from the target population $P$ that generates the summary statistics is much larger than the source cohort from $P^*$, we show that  $\sigma^2_2(t)<\sigma^2_1(t)$, indicating that weighted estimator is asymptotically more efficient than the unweighted estimator. The proof is provided in Web Appendix C. Intuitively, the additional information imposes constraints on the full parameter space to a reduced parameter space, and if this additional information is precise,
it will gain efficiency and the extent will depend on how precise this additional information is.

In the preceding theorems we assume that the two populations have a common distribution of $\boldsymbol{Z}$. This needs not be the case. The proposed weighted estimator is motivated by the presumption that the two populations have different distributions of $\boldsymbol{Z}$ and thus probably different expectations of $\boldsymbol{h(Z)}$.
The large sample results under this situation are given in the following theorem.  

\begin{theorem}
	Under regularity conditions A1-A5,  $\widehat{\Lambda}^\textup{w}_0(t)$ is consistent to $\Lambda^{\dag}_0(t)$, where $\Lambda^{\dag}_0(t)$ is the solution to 
    $\mbox{E}^{\dag}\{ \Phi( \boldsymbol{Z}; V(t),\boldsymbol{\beta_0},S(t)) \}=0$ where  $\mbox{E}^{\dag}$ is the expectation taken over $f^{\dag}(\boldsymbol{Z})$. With additional condition A6, $\sqrt{n}\{\widehat{\Lambda}^\textup{w}_0(t)-\Lambda^{\dag}_0(t)\}$ converges to a zero mean normal distribution with variance $\sigma^2_3(t)$; the formula is provided in Web Appendix D.
\end{theorem}

Based on the asymptotic normality in Theorem 3, the $1-\alpha$ confidence interval of $\Lambda^{\dag}_0(t)$ can be constructed as
\begin{eqnarray} 
\label{cl:2}
(\widehat{\Lambda}^\textup{w}_0(t)-z_{1-\alpha/2} \; \hat{\sigma}_3(t)/\sqrt{n},\;\widehat{\Lambda}^\textup{w}_0(t)+z_{1-\alpha/2} \; \hat{\sigma}_3(t)/\sqrt{n}),
\end{eqnarray} 
where $z_{1-\alpha/2}$ is the $1-\alpha/2$ upper quantile of the standard normal distribution, and  $\hat{\sigma}_3(t)$ is obtained by replacing the expectations with empirical counterparts and parameters with their estimates. This confidence interval can serve as an approximation to the $(1-\alpha)$ confidence interval for $\Lambda_0(t)$.

Let $\boldsymbol{\Sigma_{\mu}}$ be the limiting variance covariance matrix of $\sqrt{m} \boldsymbol{\hat{\mu}}$ and  $\boldsymbol{\Sigma_{\mu,S(t)}}$ the limiting covariance vector of $\sqrt{m} \boldsymbol{\hat{\mu}}$ and $\sqrt{m} \hat{S}(t)$. Variance $\sigma^2_3(t)$ includes $\boldsymbol{\Sigma_{\mu}}$ and $\boldsymbol{\Sigma_{\mu,S(t)}}$ as components.
In practice, however, their estimates are often unavailable except for the diagonal elements of $\boldsymbol{\Sigma_{\mu}}$ (i.e., the variance components). One alternative is to let  all off-diagonal elements of $\boldsymbol{\Sigma_{\mu}}$  and all elements of $\boldsymbol{\Sigma_{\mu, S(t)}}$ be 0. 
By this the resulting estimator, denoted as $\hat{\sigma}^d_3$, would approximate  $\hat{\sigma}_3$ well, if either one of the following two conditions holds: (1) the off-diagonal elements of $\boldsymbol{\Sigma_{\mu}}$ and all elements of $\boldsymbol{\Sigma_{\mu, S(t)}}$ are close to zero; and (2) the ratio of sample sizes, $n/m$, is low.

When the distribution of $\boldsymbol{Z}$ between the two populations differ, the estimator $\widehat{\Lambda}^\textup{w}_0(t)$ using summary information of risk factors has the potential to reduce bias compared to unweighted $\widehat{\Lambda}^\textup{u}_0(t)$ which does not utilize the risk factor summary information, and the confidence interval (\ref{cl:2}) has thus better coverage probability of $\Lambda_0(t)$ than (\ref{cl:1}). Therefore, under the common situation that individual-level data from the target population are not available, the estimator $\widehat{\Lambda}^\textup{w}_0(t)$ with the confidence interval (\ref{cl:2}) would be a more robust solution to external calibration of baseline hazard function.

Sometimes investigators are interested in risk predictions at several different time points, ranging from short term to long term projection. The asymptotic derivations  for a single value of $t$ can be straightforwardly extended to allow for multiple values of $t$ based on the proof in Web Appendix D. 
In addition, the proposed inference procedure can be easily modified to accommodate the scenario where the summary information $(\hat{S}(t), \boldsymbol{\hat{\mu}})$ are obtained from multiple sources..

\section{ABSOLUTE RISK ESTIMATION}
\label{s:cmprsk}

Once we obtain $\widehat\Lambda_0(t)$ and $\boldsymbol{\widehat\beta}$, we can calculate the risk for developing disease for an individual given his/her risk profile. Sometimes the competing risks need to be taken into account. In our real data analysis example of colorectal cancer, as the cancer tends to occur in older ages, individuals may die of other causes before developing colorectal cancer. As such, it is important to account for death when calculating the risk for developing colorectal cancer. For a person who is free of both events of interest and competing risks at age $t_0$ with a risk profile of $\boldsymbol{Z}$, the absolute risk of experiencing the event of interest before age $t_1$ in the presence of competing event is defined as \citep{pfeiffer2017absolute}
\begin{eqnarray*}
\Pr(t_0<T\leq t_1,\Delta=1|T> t_0,\boldsymbol{Z}) & 
=& \int^{t_1}_{t_0} \lambda(u|\boldsymbol{Z}) \exp[-\int^u_{t_0} \{\lambda(s|\boldsymbol{Z})+\lambda_c(s|\boldsymbol{Z})\}ds]du,
\end{eqnarray*}
where $T$ is the time to the first event, $\Delta$ is the event type with values of 1 and 2 indicating the event of interest and competing event, respectively, $\lambda(t|\boldsymbol{Z})$ and $\lambda_c(t|\boldsymbol{Z})$ are the respective cause-specific hazard functions  at age $t$ given $\boldsymbol{Z}$, respectively.

We can use the same method in Section 2 for estimating the baseline hazard function for the event of interest in the presence of competing risks. In equation (\ref{def:ee}), the summary statistic $S(t)$ from the target population may be replaced by the  commonly used Kaplan-Meier estimator
or exponent of the negative Nelson–Aalen estimator, obtained by treating individuals who experience competing events first as censored at that time. The theoretical results in Section \ref{s:largesample}, which assumes the covariates $\boldsymbol{Z}$ having no effects on the time to competing risk, provide an approximated solution to calculating the variance of the baseline hazard function estimator in the presence of competing risk. Subsequently, one can obtain the variance of the absolute risk estimator by the delta method (see Web Appendix D).

\section{SIMULATION STUDIES}
\label{s:simulation}

We conducted extensive simulation to evaluate the finite sample performance of the proposed method under various scenarios. Specifically, we assumed that the effects of risk factors on the hazard function in both the source and target populations followed the Cox proportional hazards models (\ref{def:coxin}) and (\ref{def:coxex}), respectively.  We considered a Weibull distribution for the baseline hazard function $\Lambda_0(t)=(\theta t)^\nu$. For the target population $P$, we assumed $\theta = 0.01$ and $\nu=2$; for the source population $P^*$, we considered four parameter combinations: (A1) $\theta=0.01$, $\nu=2$; (A2) $\theta=0.01$, $\nu=1.5$;  (A3) $\theta=0.008$, $\nu=2$; (A4) $\theta=0.008$, $\nu=1.5$. Under A1, $\Lambda_0(t)$ is the same for both $P$ and $P^*$, and differs with varying degree under other scenarios.

We generated two exposure variables $\boldsymbol{Z}=(Z_1,Z_2)^T$. For $P^*$, $Z_1\sim Bernoulli(0.5)$, $Z_2 \sim N(0,1)$ and $Z_1 \indep Z_2$; for $P$, we considered the following four configurations: (C1) $f(\boldsymbol{Z})=f^*(\boldsymbol{Z})$; (C2) $Z_1 \indep Z_2$, $f(Z_2)$ is same, but $Z_1 \sim Bernoulli(0.8)$; (C3) $f(Z_1)$ is same but  $Z_2|_{Z_1=1} \sim N(0.5,1.2)$ and $Z_2|_{Z_1=0} \sim N(-0.5,0.8)$; and (C4) 
$Z_1\sim Bernoulli(0.8)$, $Z_2|_{Z_1=1} \sim N(0.5,1.2)$ and $Z_2|_{Z_1=0} \sim N(-0.5,0.8)$.
We set the corresponding coefficients $\boldsymbol{\beta_0}=\{\log(2),\log(2)\}$. We generated right censoring time $C=C^*I{(1\leq C^*\leq 100)} + I{(C^*<1)} + 100I{(C^*>100)}$, where $C^*\sim N(40+\zeta Z_1,15)$, $\zeta=-5,0,5$. This yields approximately $64\%$ to $85\%$ censoring rates under various scenarios, representing moderate to high levels of censoring. The observed failure time was rounded to the nearest integer to mimic the real-world data. 
The sample size of the source study was  $n=1,000$, representing moderate source cohort size. To generate the summary information from the target population, we generated a cohort from the target population and obtained the Kaplan-Meier estimator of disease-free probabilities and first- and second-moments of $(Z_1,Z_2)$. Note that the Kaplan-Meier estimator is not consistent given that censoring time is only conditionally independent of failure time given $Z_1$; however, this is a commonly available statistic for disease-free probabilities in practice and hence we used it here. For the sample size of the target cohort, we considered $m = 200, 1,000$ and $100,000$, representing small, moderate and large sample sizes of the external information resources.
We generated 2,000 replicates for each simulation setting. 

We compared six methods: the Breslow estimator $\widehat{\Lambda}^\textup{B}_0(t)$ based on the source cohort data, the unweighted estimator $\widehat{\Lambda}^\textup{u}_0(t)$, weighted estimators $\widehat{\Lambda}^\textup{w}_0(t)$ with four different constraints on $\boldsymbol{Z}$  each having more constraints than the previous one:  $\mbox{E}(Z_1)$,  $\{\mbox{E}(Z_1),\mbox{E}(Z_2)\}$,
$\{\mbox{E}(Z_1),\mbox{E}(Z_2), \mbox{E}(Z^2_2)\}$, and $\{ \mbox{E}(Z_1), 
\mbox{E}(Z_2 I{(Z_1=1)}), \mbox{E}(Z^2_2 I{(Z_1=1)}),\\ \mbox{E}(Z_2 I{(Z_1=0)}), \mbox{E}(Z^2_2 I{(Z_1=0)}) \}$. We denoted the four weighted estimators as $\widehat{\Lambda}^\textup{w}_{0,1}(t)$, $\widehat{\Lambda}^\textup{w}_{0,2}(t)$, $\widehat{\Lambda}^\textup{w}_{0,3}(t)$ and $\widehat{\Lambda}^\textup{w}_{0,4}(t)$, respectively. The Breslow estimator $\widehat{\Lambda}^\textup{B}_0(t)$ is served as a basic model for comparing the performance with other methods, since it represents a conventional  solution when no auxiliary external information is used. 

We assessed the performance of these estimators by the following metrics: percentage of relative bias (PBias),  empirical standard deviation (ESD), asymptotic-based standard error (ASE), square root of mean squared error (sMSE), and coverage probability of 95\% confidence intervals over the true value of $\Lambda_0(t)$ (CP) at selected ages $t=20, 40,$ and  $60$. We also calculated the average cumulative absolute deviation (CAD), $\sum^{T_{max}}_{t=1} |\widehat{\Lambda}_0(t)-\Lambda_0(t)|$, where $T_{max}=60$ and $\widehat{\Lambda}_0(t)$ is an estimator of $\Lambda_0(t)$ from each of the methods, to assess the overall performance across a wide range of time. For the variance estimator of the four weighted estimators, we set all elements of the covariance vector $\boldsymbol{\Sigma_{\mu, S(t)}}$ and the off-diagonal elements of $ \boldsymbol{\Sigma_{\mu}}$  to $0$ since such  information is usually not available in practice.  

The results are very similar for different values of $\zeta$  in the censoring distribution. To save space here we only present results when $\zeta=-5$.
Table \ref{table:simu1} summarizes the results under A1, i.e., $\Lambda^*_0(t)=\Lambda_0(t)$. The target cohort sample size is $m = 100,000$ which yields highly accurate external summary information. Under this scenario, the Breslow estimator $\widehat\Lambda_0^\textup{B}(t)$ obtained from the source cohort is a consistent estimator to $\Lambda_0(t)$ for the target cohort. Indeed, there is little bias for $\widehat\Lambda_0^\textup{B}(t)$ whether or not the risk factor distribution is same between the source and target cohorts (C1--C4). On the other hand, the unweighted estimator $\widehat\Lambda_0^\textup{u}(t)$ that assumes $f(Z_1, Z_2) = f^*(Z_1, Z_2)$ is  unbiased under C1, i.e.,  $f^*(Z_1,Z_2)=f(Z_1,Z_2)$, but can be substantially biased when $f^*(Z_1,Z_2) \ne f(Z_1,Z_2)$ under C2--C4 with relative bias as high as 73\%, even though $\Lambda^*_0(t)=\Lambda_0(t)$. The weighted estimator, especially $\widehat{\Lambda}^\textup{w}_{0,4}(t)$, has little bias with maximum relative bias 3\% and the CAD and sMSE are often the lowest. The coverage probability of 95\% confidence intervals is close to 95\%. As expected, as more constraints are added, the bias of the weighted estimator becomes smaller.  

When $f^*(Z_1,Z_2)=f(Z_1,Z_2)$, the weighted estimators and also the unweighted estimator that leverage the (precise) summary information from target cohort improve the efficiency substantially with efficiency gain over 200\%. Even when $f(Z_1, Z_2)$ is significantly different from $f^*(Z_1,Z_2)$ under C4, the weighted estimators does not lose much efficiency compared to the Breslow estimator. The asymptotic-based variance estimator is close to empirical variance estimator,  suggesting that setting the covariance $\boldsymbol{\Sigma_{\mu, S(t)}}$ and the off-diagonal elements of $ \boldsymbol{\Sigma_{\mu}}$ to $0$ does not bias the variance estimator in any meaningful fashion. In some situations where the off-diagonal elements of $ \boldsymbol{\Sigma_{\mu}}$ is available, we can incorporate such information in the variance estimator; however, the variance estimator is almost identical to the one that sets the off-diagonal elements to 0 (Supplemental Table S1).

\begin{table}[]
	\renewcommand\arraystretch{1.1}
	\setlength{\tabcolsep}{2.5pt}
	\centering
	\scriptsize
	\caption{Summary statistics of simulation results for scenario A1: $\Lambda^*_0(t)=\Lambda_0(t)=(0.01t)^2$, $n=1,000$, and $m=100,000$.}
	\begin{threeparttable}
		\begin{tabular}{lllrrrrrrrrrrrrr}
			\hline
			\hline\\[-2.5 ex]
			&            &       & \multicolumn{6}{l}{(C1)  $Z_1\sim Bernoulli(0.5)$, $Z_2 \sim N(0,1)$,}                        &  & \multicolumn{6}{l}{(C2)  $Z_1\sim Bernoulli(0.8)$, $Z_2 \sim N(0,1)$,}                        \\ 
			&            &       & \multicolumn{6}{l}{\;\;\;\;\;\;\; and $Z_1 \indep Z_2$, i.e., $f^*(Z_1,Z_2)=f(Z_1,Z_2)$}                        &  & \multicolumn{6}{l}{\;\;\;\;\;\;\;  and $Z_1 \indep Z_2$}                        \\ 
			\cline{4-9} \cline{11-16}\\[-2.5 ex]
			$t$   & $\Lambda_0(t)$ &       & $\widehat{\Lambda}^\textup{B}_0(t)$ & $\widehat{\Lambda}^\textup{u}_0(t)$    & $\widehat{\Lambda}^\textup{w}_{0,1}(t)$ & $\widehat{\Lambda}^\textup{w}_{0,2}(t)$ & $\widehat{\Lambda}^\textup{w}_{0,3}(t)$ & $\widehat{\Lambda}^\textup{w}_{0,4}(t)$ &  & $\widehat{\Lambda}^\textup{B}_0(t)$ & $\widehat{\Lambda}^\textup{u}_0(t)$    & $\widehat{\Lambda}^\textup{w}_{0,1}(t)$ & $\widehat{\Lambda}^\textup{w}_{0,2}(t)$ & $\widehat{\Lambda}^\textup{w}_{0,3}(t)$ & $\widehat{\Lambda}^\textup{w}_{0,4}(t)$ \\ [0.5 ex]
			\hline
			\rowcolor{Gray}
			20 & 0.04  & PBias & 0.4\%  & 1.1\%  & 1.1\%  & 1.1\%  & 1.1\%  & 1.0\%  &  & 0.2\%  & 21.7\% & 2.5\%  & 1.8\%  & 1.8\%  & 1.8\%  \\
            &      & ESD   & 0.65   & 0.44   & 0.44   & 0.43   & 0.42   & 0.42   &  & 0.65   & 0.52   & 0.58   & 0.54   & 0.54   & 0.54   \\
            &      & ASE   & 0.65   & 0.47   & 0.46   & 0.43   & 0.42   & 0.42   &  & 0.65   & 0.57   & 0.59   & 0.55   & 0.54   & 0.54   \\
            &      & sMSE  & 0.65   & 0.44   & 0.44   & 0.43   & 0.43   & 0.43   &  & 0.65   & 1.05   & 0.58   & 0.55   & 0.54   & 0.55   \\
            &      & CP    & 94.3\% & 96.3\% & 95.9\% & 95.6\% & 95.1\% & 94.6\% &  & 94.6\% & 67.4\% & 95.9\% & 96.1\% & 95.7\% & 95.6\% \\
			\hline
			\rowcolor{Gray}
			40 & 0.16 & PBias & -0.8\% & 1.0\%  & 1.0\%  & 1.0\%  & 1.0\%  & 1.0\%  &  & -1.0\% & 22.8\% & 2.4\%  & 1.7\%  & 1.7\%  & 1.7\%  \\
            &      & ESD   & 1.89   & 1.53   & 1.52   & 1.47   & 1.47   & 1.47   &  & 1.90   & 1.77   & 2.12   & 1.98   & 1.97   & 1.98   \\
            &      & ASE   & 1.87   & 1.64   & 1.56   & 1.47   & 1.45   & 1.43   &  & 1.87   & 1.94   & 2.12   & 2.00   & 1.98   & 1.98   \\
            &      & sMSE  & 1.90   & 1.54   & 1.53   & 1.48   & 1.48   & 1.47   &  & 1.90   & 4.14   & 2.16   & 1.99   & 1.99   & 2.00   \\
            &      & CP    & 93.7\% & 96.2\% & 95.8\% & 95.3\% & 95.0\% & 94.6\% &  & 93.1\% & 52.1\% & 95.2\% & 95.7\% & 95.7\% & 95.5\% \\
			\hline
			\rowcolor{Gray}
			60 & 0.37 & PBias & -3.2\% & 1.0\%  & 0.9\%  & 0.9\%  & 0.9\%  & 0.9\%  &  & -3.0\% & 24.0\% & 2.3\%  & 1.6\%  & 1.6\%  & 1.7\%  \\
            &      & ESD   & 4.34   & 2.98   & 2.95   & 2.83   & 2.83   & 2.83   &  & 4.33   & 3.44   & 4.47   & 4.12   & 4.12   & 4.14   \\
             &      & ASE   & 4.22   & 3.16   & 2.97   & 2.80   & 2.79   & 2.77   &  & 4.24   & 3.71   & 4.38   & 4.16   & 4.15   & 4.14   \\
            &      & sMSE  & 4.49   & 3.00   & 2.97   & 2.85   & 2.85   & 2.85   &  & 4.47   & 9.43   & 4.55   & 4.16   & 4.16   & 4.19   \\
            &      & CP    & 91.8\% & 95.8\% & 95.1\% & 95.1\% & 95.2\% & 94.9\% &  & 91.3\% & 31.8\% & 93.7\% & 95.7\% & 95.6\% & 95.5\% \\
            \hline
            &      & CAD   & 0.73   & 0.54   & 0.54   & 0.52   & 0.52   & 0.52   &  & 0.74   & 1.76   & 0.77   & 0.71   & 0.71   & 0.72    \\
			\hline\\[-2.5 ex]
			&            &       & \multicolumn{6}{l}{(C3)  $Z_1\sim Bernoulli(0.5)$, }                        &  & \multicolumn{6}{l}{(C4)  $Z_1\sim Bernoulli(0.8)$, }                        \\
			&            &       & \multicolumn{6}{l}{\;\;\;\;\;\;\;  $Z_2|_{Z_1=1} \sim N(0.5,1.2)$,}                        &  & \multicolumn{6}{l}{\;\;\;\;\;\;\;  $Z_2|_{Z_1=1} \sim N(0.5,1.2)$,}                        \\
			&            &       & \multicolumn{6}{l}{\;\;\;\;\;\;\; and $Z_2|_{Z_1=0} \sim N(-0.5,0.8)$}                        &  & \multicolumn{6}{l}{\;\;\;\;\;\;\;  and $Z_2|_{Z_1=0} \sim N(-0.5,0.8)$}                        \\
			\cline{4-9} \cline{11-16} \\[-2.5 ex]
			$t$   & $\Lambda_0(t)$ &       & $\widehat{\Lambda}^\textup{B}_0(t)$ & $\widehat{\Lambda}^\textup{u}_0(t)$    & $\widehat{\Lambda}^\textup{w}_{0,1}(t)$ & $\widehat{\Lambda}^\textup{w}_{0,2}(t)$ & $\widehat{\Lambda}^\textup{w}_{0,3}(t)$ & $\widehat{\Lambda}^\textup{w}_{0,4}(t)$ &  & $\widehat{\Lambda}^\textup{B}_0(t)$ & $\widehat{\Lambda}^\textup{u}_0(t)$    & $\widehat{\Lambda}^\textup{w}_{0,1}(t)$ & $\widehat{\Lambda}^\textup{w}_{0,2}(t)$ & $\widehat{\Lambda}^\textup{w}_{0,3}(t)$ & $\widehat{\Lambda}^\textup{w}_{0,4}(t)$ \\ [0.5 ex]
			
			\hline
			\rowcolor{Gray}
			20 & 0.04 & PBias & -0.3\% & 22.0\% & 22.0\% & 21.9\% & 15.1\% & 0.4\%  &  & 0.3\%  & 73.0\% & 45.5\% & 12.3\% & 4.7\%  & 1.3\%  \\
            &      & ESD   & 0.65   & 0.52   & 0.51   & 0.50   & 0.50   & 0.55   &  & 0.64   & 0.72   & 0.82   & 0.67   & 0.65   & 0.66   \\
            &      & ASE   & 0.64   & 0.57   & 0.55   & 0.51   & 0.50   & 0.54   &  & 0.65   & 0.79   & 0.82   & 0.65   & 0.64   & 0.65   \\
            &      & sMSE  & 0.65   & 1.06   & 1.06   & 1.05   & 0.81   & 0.55   &  & 0.64   & 3.15   & 2.08   & 0.85   & 0.68   & 0.66   \\
            &      & CP    & 93.8\% & 65.6\% & 63.0\% & 58.7\% & 77.8\% & 94.2\% &  & 94.8\% & 0.6\%  & 35.1\% & 91.6\% & 95.2\% & 94.5\% \\
			\hline
			\rowcolor{Gray}
			40 & 0.16 & PBias & -1.3\% & 12.0\% & 12.0\% & 12.0\% & 8.1\%  & 1.4\%  &  & -1.5\% & 63.4\% & 36.0\% & 7.6\%  & 5.2\%  & 2.4\%  \\
            &      & ESD   & 1.89   & 1.65   & 1.64   & 1.60   & 1.57   & 1.69   &  & 1.87   & 2.26   & 2.79   & 2.21   & 2.16   & 2.21   \\
            &      & ASE   & 1.87   & 1.79   & 1.71   & 1.60   & 1.57   & 1.67   &  & 1.86   & 2.46   & 2.75   & 2.20   & 2.17   & 2.21   \\
            &      & sMSE  & 1.90   & 2.57   & 2.56   & 2.53   & 2.05   & 1.70   &  & 1.89   & 10.65  & 6.53   & 2.53   & 2.32   & 2.25   \\
            &      & CP    & 94.1\% & 84.3\% & 81.9\% & 78.8\% & 88.9\% & 95.0\% &  & 94.0\% & 0.2\%  & 44.6\% & 94.4\% & 95.2\% & 95.8\% \\
			\hline
			\rowcolor{Gray}
			60 & 0.37 & PBias & -2.9\% & 2.9\%  & 2.8\%  & 2.8\%  & 1.3\%  & 1.8\%  &  & -3.3\% & 53.7\% & 26.6\% & 2.3\%  & 3.9\%  & 3.0\%  \\
            &      & ESD   & 4.30   & 3.01   & 2.98   & 2.87   & 2.84   & 2.96   &  & 4.21   & 4.06   & 5.49   & 4.22   & 4.18   & 4.29   \\
            &      & ASE   & 4.24   & 3.21   & 3.01   & 2.84   & 2.81   & 2.91   &  & 4.24   & 4.36   & 5.27   & 4.26   & 4.26   & 4.34   \\
            &      & sMSE  & 4.43   & 3.19   & 3.15   & 3.05   & 2.88   & 3.04   &  & 4.38   & 20.06  & 11.17  & 4.30   & 4.42   & 4.43   \\
            &      & CP    & 91.3\% & 95.6\% & 94.8\% & 94.3\% & 95.2\% & 95.1\% &  & 92.2\% & 0.1\%  & 59.6\% & 96.2\% & 95.8\% & 96.3\% \\
            \hline
            &      & CAD   & 0.74   & 0.88   & 0.87   & 0.86   & 0.70   & 0.60   &  & 0.72   & 4.60   & 2.54   & 0.87   & 0.81   & 0.80    \\
			\hline
		\end{tabular}
		\begin{tablenotes}
			\scriptsize
			\item[] NOTE: PBias, the relative bias;  ESD, the empirical standard deviation ($\times 100$) of the 2000 estimates; ASE, the mean of the asymptotic-based standard error ($\times 100$); sMSE, the square root of mean squared error ($\times 100$); CP, the coverage probability of a 95\% confidence interval for $\Lambda_0(t)$;  CAD, the average cumulative absolute deviation between the estimated and true cumulative hazard function across a time range of one through sixty.
		\end{tablenotes}
	\end{threeparttable}
	\label{table:simu1}
\end{table}

\begin{table}[]
\renewcommand\arraystretch{1.1}
\setlength{\tabcolsep}{2.5pt}
\centering
\scriptsize
\caption{Summary statistics of simulation results for scenario A2: $\Lambda_0(t)=(0.01t)^2$, $\Lambda^*_0(t)=(0.01t)^{1.5}$, $n=1,000$, and $m=100,000$.}
\begin{threeparttable}
	\begin{tabular}{lllrrrrrrrrrrrrr}
		\hline
		\hline\\[-2.5 ex]
		&            &       & \multicolumn{6}{l}{(C1)  $Z_1\sim Bernoulli(0.5)$, $Z_2 \sim N(0,1)$,}                        &  & \multicolumn{6}{l}{(C2)  $Z_1\sim Bernoulli(0.8)$, $Z_2 \sim N(0,1)$,}                        \\ 
		&            &       & \multicolumn{6}{l}{\;\;\;\;\;\;\; and $Z_1 \indep Z_2$, i.e., $f^*(Z_1,Z_2)=f(Z_1,Z_2)$}                        &  & \multicolumn{6}{l}{\;\;\;\;\;\;\;  and $Z_1 \indep Z_2$}                        \\ 
		\cline{4-9} \cline{11-16}\\[-2.5 ex]
		$t$   & $\Lambda_0(t)$ &       & $\widehat{\Lambda}^\textup{B}_0(t)$ & $\widehat{\Lambda}^\textup{u}_0(t)$    & $\widehat{\Lambda}^\textup{w}_{0,1}(t)$ & $\widehat{\Lambda}^\textup{w}_{0,2}(t)$ & $\widehat{\Lambda}^\textup{w}_{0,3}(t)$ & $\widehat{\Lambda}^\textup{w}_{0,4}(t)$ &  & $\widehat{\Lambda}^\textup{B}_0(t)$ & $\widehat{\Lambda}^\textup{u}_0(t)$    & $\widehat{\Lambda}^\textup{w}_{0,1}(t)$ & $\widehat{\Lambda}^\textup{w}_{0,2}(t)$ & $\widehat{\Lambda}^\textup{w}_{0,3}(t)$ & $\widehat{\Lambda}^\textup{w}_{0,4}(t)$ \\ [0.5 ex]
		\hline
		\rowcolor{Gray}
		20 & 0.04 & PBias & 120.3\% & 1.1\%  & 1.0\%  & 1.0\%  & 1.0\%  & 1.0\%  &  & 120.4\% & 21.7\% & 2.3\%  & 1.5\%  & 1.5\%  & 1.6\%  \\
        &      & ESD   & 1.10    & 0.37   & 0.37   & 0.36   & 0.36   & 0.35   &  & 1.08    & 0.44   & 0.49   & 0.46   & 0.46   & 0.46   \\
        &      & ASE   & 1.08    & 0.43   & 0.41   & 0.38   & 0.37   & 0.36   &  & 1.08    & 0.51   & 0.52   & 0.47   & 0.47   & 0.46   \\
        &      & sMSE  & 5.18    & 0.37   & 0.37   & 0.36   & 0.36   & 0.36   &  & 5.17    & 1.01   & 0.50   & 0.47   & 0.46   & 0.47   \\
        &      & CP    & 0.1\%   & 97.7\% & 97.1\% & 96.1\% & 96.1\% & 95.5\% &  & 0.0\%   & 59.6\% & 95.5\% & 95.5\% & 95.0\% & 94.9\% \\
		\hline
		\rowcolor{Gray}
		40 & 0.16 & PBias & 55.2\%  & 1.0\%  & 0.9\%  & 0.9\%  & 0.9\%  & 0.9\%  &  & 55.4\%  & 22.8\% & 2.3\%  & 1.4\%  & 1.4\%  & 1.5\%  \\
        &      & ESD   & 2.52    & 1.27   & 1.26   & 1.22   & 1.21   & 1.21   &  & 2.42    & 1.53   & 1.83   & 1.69   & 1.68   & 1.69   \\
        &      & ASE   & 2.44    & 1.47   & 1.38   & 1.27   & 1.26   & 1.24   &  & 2.45    & 1.73   & 1.86   & 1.72   & 1.71   & 1.70   \\
        &      & sMSE  & 9.39    & 1.28   & 1.27   & 1.23   & 1.22   & 1.22   &  & 9.40    & 4.03   & 1.87   & 1.70   & 1.70   & 1.70   \\
        &      & CP    & 1.8\%   & 97.9\% & 96.9\% & 95.7\% & 95.5\% & 95.0\% &  & 1.6\%   & 41.6\% & 94.6\% & 95.3\% & 95.1\% & 95.1\% \\
		\hline
		\rowcolor{Gray}
		60 & 0.37 & PBias & 25.2\%  & 0.8\%  & 0.8\%  & 0.8\%  & 0.8\%  & 0.8\%  &  & 25.2\%  & 23.9\% & 2.2\%  & 1.3\%  & 1.3\%  & 1.4\%  \\
        &      & ESD   & 4.83    & 2.49   & 2.47   & 2.34   & 2.34   & 2.34   &  & 4.82    & 2.99   & 3.91   & 3.54   & 3.54   & 3.55   \\
        &      & ASE   & 4.78    & 2.83   & 2.61   & 2.41   & 2.41   & 2.39   &  & 4.77    & 3.31   & 3.82   & 3.56   & 3.56   & 3.55   \\
        &      & sMSE  & 10.41   & 2.51   & 2.48   & 2.36   & 2.36   & 2.36   &  & 10.39   & 9.25   & 3.99   & 3.57   & 3.57   & 3.58   \\
        &      & CP    & 53.7\%  & 97.5\% & 96.3\% & 95.2\% & 95.4\% & 95.1\% &  & 52.9\%  & 20.3\% & 93.1\% & 95.4\% & 95.4\% & 95.3\% \\
        \hline
        &      & CAD   & 3.89    & 0.46   & 0.45   & 0.44   & 0.43   & 0.43   &  & 3.89    & 1.75   & 0.67   & 0.61   & 0.61   & 0.61      \\
		\hline\\[-2.5 ex]
		&            &       & \multicolumn{6}{l}{(C3)  $Z_1\sim Bernoulli(0.5)$, }                        &  & \multicolumn{6}{l}{(C4)  $Z_1\sim Bernoulli(0.8)$, }                        \\
		&            &       & \multicolumn{6}{l}{\;\;\;\;\;\;\;  $Z_2|_{Z_1=1} \sim N(0.5,1.2)$,}                        &  & \multicolumn{6}{l}{\;\;\;\;\;\;\;  $Z_2|_{Z_1=1} \sim N(0.5,1.2)$,}                        \\
		&            &       & \multicolumn{6}{l}{\;\;\;\;\;\;\; and $Z_2|_{Z_1=0} \sim N(-0.5,0.8)$}                        &  & \multicolumn{6}{l}{\;\;\;\;\;\;\;  and $Z_2|_{Z_1=0} \sim N(-0.5,0.8)$}                        \\
		\cline{4-9} \cline{11-16} \\[-2.5 ex]
		$t$   & $\Lambda_0(t)$ &       & $\widehat{\Lambda}^\textup{B}_0(t)$ & $\widehat{\Lambda}^\textup{u}_0(t)$    & $\widehat{\Lambda}^\textup{w}_{0,1}(t)$ & $\widehat{\Lambda}^\textup{w}_{0,2}(t)$ & $\widehat{\Lambda}^\textup{w}_{0,3}(t)$ & $\widehat{\Lambda}^\textup{w}_{0,4}(t)$ &  & $\widehat{\Lambda}^\textup{B}_0(t)$ & $\widehat{\Lambda}^\textup{u}_0(t)$    & $\widehat{\Lambda}^\textup{w}_{0,1}(t)$ & $\widehat{\Lambda}^\textup{w}_{0,2}(t)$ & $\widehat{\Lambda}^\textup{w}_{0,3}(t)$ & $\widehat{\Lambda}^\textup{w}_{0,4}(t)$ \\ [0.5 ex]
		
		\hline
		\rowcolor{Gray}
		20 & 0.04 & PBias & 119.7\% & 21.8\% & 21.8\% & 21.7\% & 14.9\% & 0.1\%  &  & 119.9\% & 72.9\% & 45.2\% & 11.9\% & 4.4\%  & 0.9\%  \\
        &      & ESD   & 1.07    & 0.44   & 0.44   & 0.42   & 0.42   & 0.46   &  & 1.09    & 0.64   & 0.73   & 0.58   & 0.56   & 0.57   \\
        &      & ASE   & 1.08    & 0.51   & 0.48   & 0.45   & 0.44   & 0.47   &  & 1.08    & 0.71   & 0.72   & 0.56   & 0.55   & 0.56   \\
        &      & sMSE  & 5.14    & 1.02   & 1.02   & 1.01   & 0.76   & 0.46   &  & 5.16    & 3.13   & 2.03   & 0.77   & 0.59   & 0.57   \\
        &      & CP    & 0.0\%   & 59.4\% & 53.8\% & 46.9\% & 73.0\% & 94.8\% &  & 0.0\%   & 0.0\%  & 22.1\% & 88.3\% & 94.6\% & 94.7\% \\
		\hline
		\rowcolor{Gray}
		40 & 0.16 & PBias & 54.9\%  & 11.8\% & 11.8\% & 11.8\% & 7.9\%  & 1.1\%  &  & 54.8\%  & 63.4\% & 35.8\% & 7.3\%  & 5.0\%  & 2.2\%  \\
        &      & ESD   & 2.49    & 1.41   & 1.40   & 1.33   & 1.31   & 1.42   &  & 2.49    & 2.03   & 2.50   & 1.92   & 1.88   & 1.93   \\
        &      & ASE   & 2.45    & 1.61   & 1.51   & 1.39   & 1.36   & 1.44   &  & 2.44    & 2.20   & 2.41   & 1.89   & 1.87   & 1.90   \\
        &      & sMSE  & 9.35    & 2.40   & 2.40   & 2.35   & 1.84   & 1.43   &  & 9.32    & 10.59  & 6.38   & 2.27   & 2.06   & 1.96   \\
        &      & CP    & 2.4\%   & 82.6\% & 79.9\% & 75.9\% & 86.4\% & 95.7\% &  & 2.2\%   & 0.0\%  & 32.5\% & 92.9\% & 95.0\% & 95.0\% \\
		\hline
		\rowcolor{Gray}
		60 & 0.37 & PBias & 25.0\%  & 2.7\%  & 2.7\%  & 2.7\%  & 1.2\%  & 1.6\%  &  & 24.9\%  & 53.7\% & 26.5\% & 2.1\%  & 3.8\%  & 2.9\%  \\
        &      & ESD   & 4.87    & 2.56   & 2.53   & 2.37   & 2.35   & 2.43   &  & 4.81    & 3.70   & 4.99   & 3.72   & 3.70   & 3.80   \\
        &      & ASE   & 4.78    & 2.87   & 2.65   & 2.45   & 2.43   & 2.50   &  & 4.77    & 3.89   & 4.60   & 3.67   & 3.67   & 3.73   \\
        &      & sMSE  & 10.37   & 2.74   & 2.72   & 2.56   & 2.39   & 2.51   &  & 10.30   & 19.99  & 10.89  & 3.80   & 3.95   & 3.95   \\
        &      & CP    & 54.3\%  & 96.3\% & 95.2\% & 94.5\% & 95.9\% & 95.2\% &  & 54.8\%  & 0.0\%  & 48.0\% & 95.3\% & 95.0\% & 95.3\% \\
        \hline
        &      & CAD   & 3.87    & 0.82   & 0.82   & 0.80   & 0.62   & 0.49   &  & 3.86    & 4.60   & 2.52   & 0.77   & 0.72   & 0.70    \\
		\hline
	\end{tabular}
	\begin{tablenotes}
		\scriptsize
		\item[] NOTE: PBias, the relative bias; ESD, the empirical standard deviation ($\times 100$) of the 2000 estimates; ASE, the mean of the asymptotic-based standard error ($\times 100$); sMSE, the square root of mean squared error ($\times 100$); CP, the coverage probability of a 95\% confidence interval for $\Lambda_0(t)$;  CAD, the average cumulative absolute deviation between the estimated and true cumulative hazard function across a time range of one through sixty.
	\end{tablenotes}
\end{threeparttable}
\label{table:simu2}
\end{table}

Table \ref{table:simu2} shows the results when $\Lambda_0(t) \ne \Lambda^*_0(t)$ under A2 where they differ by the shape parameter. The conventional Breslow estimator $\widehat{\Lambda}^\textup{B}_0(t)$ has poor performance with relative bias ranging from 24.9\% to 120.4\%. The unweighted estimator has little bias  when the assumption of $f(Z_1, Z_2) = f^*(Z_1, Z_2)$ holds, but can be substantially biased when it is violated. By further constraining on the moments of risk factors, the weighted estimators reduce the bias, and with adequate constraints, the weighted estimator, e.g., $\widehat{\Lambda}^\textup{w}_{0,4}(t)$, is nearly unbiased. The asymptotic-based variance estimator is close to empirical variance estimator and the coverage probability is close to 95\% nominal level. Under C1 $f(Z_1, Z_2) = f^*(Z_1, Z_2)$ when both unweighted and weighted estimators are unbiased, including additional constraints on the moments of risk factors improves efficiency slightly.

We examined the performance of all the estimators when $\Lambda_0(t)$ differs from the source cohort $\Lambda_0^*(t)$ by the scale parameter (A3) and by both the shape and scale parameters (A4). The results have similar patterns to that under A2 (Supplemental Table S2 and S3).  We also examined the performance when the sample sizes for obtaining auxiliary information from the target cohort are moderate ($m=1,000$, Supplemental Table S4-S7) and low ($m=200$, Supplemental Table S8-S11). Under  $\Lambda_0(t) = \Lambda_0^*(t)$ when all estimators are almost unbiased, the unweighted and weighted estimators have comparable efficiency to the Breslow estimator when $m=1,000$, and are less efficient when $m=200$. However, even when $m=200$, if the baseline hazard functions and risk factor distributions differ between the source and target cohorts, the auxiliary information can help reduce the bias substantially with the weighted estimators (Table S9-S11). 

\begin{figure}
\begin{center}
\includegraphics[width=6.1in]{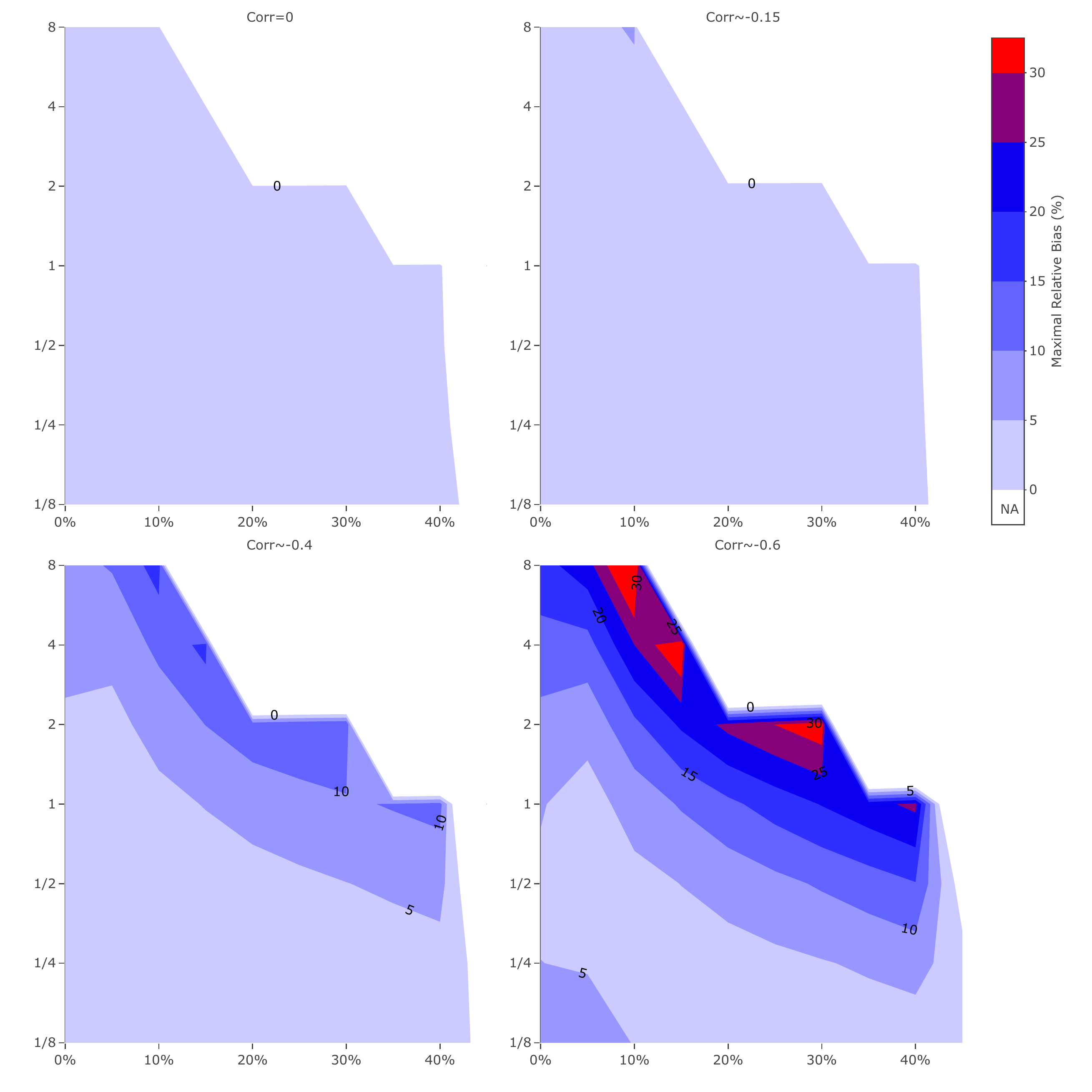}
\end{center}
\caption{ Maximal relative bias of the proposed weighted estimator $\widehat{\Lambda}^\textup{w}_{0,4}(t)$. 
Corr: Pearson correlation coefficient between risk score for event of interest and $T_c$, i.e., $corr(\boldsymbol{\beta^T_0 Z},T_c)$.  Horizontal axis: probability of observing the event of interest; vertical axis: ratio between probability of observing the event of competing risk and probability of observing the event of interest. }
\label{figure_contour}
\end{figure}

We evaluated the performance of the proposed method when the competing risks are present. If the risk score associated with the event of interest is also associated with the competing event, the proposed estimator is no longer unbiased. Thus, we studied the magnitude of the bias of the proposed estimator by varying the values of the following parameters: (1) correlation of the risk score with the competing event; (2) probability of observing the event of interest; and (3) ratio of the probabilities of observing the competing risk and the event of interest. 
Specifically, we generated covariates and time-to-event of interest under the same settings as in the previous simulation. For simplicity, we generated the censoring time $C^* \sim N(40, 15)$. We specified $\Lambda_{0}(t)=\kappa(\theta t)^\nu,\; \kappa=1/16,1/8,1/4,1/2,1,2,4,8,16$ to yield various disease prevalences for the event of interest. Let $T_c$ denote time to a competing event. We let $T_c$ follow the Cox model with the hazard function $\Lambda_c(t|\boldsymbol{Z})=\Lambda_{0c}(t)\exp(\boldsymbol{\beta_c}^T \boldsymbol{Z})$, where $\Lambda_{0c}(t)=\kappa(\theta t)^\nu,$ with $ \kappa=1/16,1/8,1/4,1/2,1,2,4,8,16$, and $\boldsymbol{\beta_c}=(0,0), (\log(1.15),\log(1.15)), (\log(1.41),\log(1.41))$ or $(\log(2),\log(2))$, representing different strengths of association between the risk score (for the event of interest) and $T_c$, and  yielding Pearson correlation coefficients between the two at $0$, $0.15$, $0.4$, and $0.6$, respectively. 
The results were summarized in the four contour plots in Figure \ref{figure_contour}. The contour value is the maximal relative bias  $\max_{t=20,40,60}|\{\overline{\widehat{\Lambda}^\textup{w}_{0,4}(t)}-\Lambda_{0}(t)\}/\Lambda_{0}(t)|$ where $\overline{\widehat{\Lambda}^\textup{w}_{0,4}(t)}$ is the empirical mean of $\widehat{\Lambda}^\textup{w}_{0,4}(t)$ over the 2,000 simulation. 
When the correlation between $T_c$ and risk score for the event of interest is weak ($<0.2$), the proposed weighted estimator is almost unbiased across all scenarios. Note that the correlation coefficients are around $-0.1$ in our real-data examples.  When the correlation is moderate ($\sim-0.4$), the maximal relative bias is below 10\% when the censoring proportion is moderate to high ($>50\%$). When the correlation is high ($\sim-0.6$), the maximal relative bias is still below 10\% when the censoring proportion is high ($>70\%$) and the ratio between probabilities of observing the competing event versus observing the event of interest is not high ($<3$). In addition, we found the coverage probability of 95\% confidence interval using the variance estimator from Theorem 3 is generally above 90\% when the absolute bias is below 5\% (Supplemental Figure S1), reflecting good performance of the inference.

In summary, the proposed weighted estimator incorporating disease incidence and auxiliary risk factor information from target cohort reduces bias substantially compared to the conventional Breslow estimator and the unweighted estimator, both of which can lead to unacceptable large bias, when the baseline hazard function and/or risk factor distribution differ between the two cohorts. The weighted estimator can also gain substantial efficiency  when $f^*(Z_1,Z_2)=f(Z_1,Z_2)$ and the auxiliary external risk factor information is of high accuracy and precision. The proposed estimator is also quite robust across a wide range of realistic scenarios under the competing risks.

\section{APPLICATION}
\label{s:application}

\subsection{Study Cohorts and Risk Factors}

In this section we illustrate an application of our methodology by re-calibrating a colorectal cancer (CRC) risk prediction model, established from the Women's Health Initiative (WHI), to five target cohorts: the intervention and control arms of the Prostate, Lung, Colorectal and Ovarian Cancer Screening Trail (PLCO) study and three countries (England, Scotland, and Wales) from the UK Biobank (UKB). In particular, we will use the WHI individual-level data to build the risk prediction model, and leverage the  summary statistics from the target cohorts, which include overall hazard rates of CRC and means of key risk factors, to re-calibrate the CRC baseline hazard function for the target cohorts. We accommodated the death from other causes as the competing event in the calculation of absolute risk. Below we describe the three studies. 

The WHI observational cohort is a prospective study including 93,676 women aged 50 to 79 years in the U.S. between 1993 and 1998 \citep{study1998design}. The study collects information on socio-demographic and epidemiologic factors using standardized questionnaires, and biological samples at clinic visits. Here we focus on disease risk of CRC for white women. A total of 76,733 subjects are included in the estimation of disease risk, with mean follow-up length of 6.7 years. Among these, 1,073 developed CRC  and 8,815 died from causes other than CRC during the follow-up. The disease outcome is age (in integer years) at diagnosis of CRC, which is subject to left truncation (age at enrollment), right censoring (age at the end of follow-up, loss of follow-up, or death from other causes).

The PLCO study is a large randomized trial conducted to determine the effectiveness of screening in reducing cancer mortality \citep{prorok2000design,gohagan2000prostate}. It enrolled 154,934 participants aged between 55 and 74 years at 10 centers from 1993 to 2001.  To be consistent with WHI, we included only white women (67,254 participants) from PLCO. The mean follow-up length is 11.3 years. During the follow-up 749 women developed CRC and 6,241 died from other causes. As PLCO is a screening trial, we used the intervention and control arms separately as two target cohorts. 
Note that PLCO and WHI are both conducted within the U.S. nationally. To further illustrate recalibration targeted to populations that are more different from the source cohort, we used cohorts assembled from the UK Biobank (UKB) as target cohorts. UKB is a large long-term biobank study in the United Kingdom, aimed to improve the prevention, diagnosis and treatment of a wide range of serious and life-threatening illnesses including cancer, heart diseases, and diabetes \citep{sudlow2015uk}. It recruited 500,000 people aged between 40-69 years in 2006-2010 from across UK except for Northen Ireland and is following the health and well-being of these participants. To be consistent with WHI, only white women (198,058 participants) with enrollment age greater than 50 from UKB were included. The mean follow-up length is 5.8 years; 1,150 developed CRC and 4,623 died from other causes during the follow-up. We analyzed the three countries, England, Scotland and Wales, separately. In total, we have two PLCO-based target cohorts and three UKB-based target cohorts.

The risk factors included in the risk prediction model are based on the model developed by \citet{freedman2009colorectal}. 
These are history of endoscopy (sigmoidoscopy or colonoscopy) in last five years (yes, no); number of first-degree relatives with CRC (0, $\geq$ 1); current leisure-time vigorous activity (0, 0-2, $>$ 2 hours per week); use of aspirin and other nonsteroidal anti-inflammatory drugs (NSAIDs) (nonuser, regular user); vegetable consumption ($<$ , $\geq$  medium portion per day); body mass index (BMI, $<$ 30, $\geq$ 30kg/m$^2$); and estrogen status within the last two years (negative, positive). 
We fit a Cox proportional hazards model, treating the competing risk of other-cause mortaility as censored. Since no women below age 50 were recruited in WHI, we built the model only starting at age 50. For women who enrolled after age 50, they were treated as left truncated at the enrollment age. For example, a woman enrolled at age 60 and developed CRC at age 70 would only be in the at-risk set for developing disease between age 60 and 70. This problem can be appropriately handled by carefully defining the at-risk set for each subject as described in Section \ref{subs:notation}. 

We compared summary information including Kaplan-Meier estimator of CRC-free probability and risk factor distributions across cohorts to determine the need for recalibration. We then applied the proposed methods using the summary information from target cohorts to recalibrate the CRC cumulative baseline hazard function and calculated the t-year absolute risks for individuals in the target cohorts. 

\begin{figure}
\begin{center}
\includegraphics[width=1\textwidth]{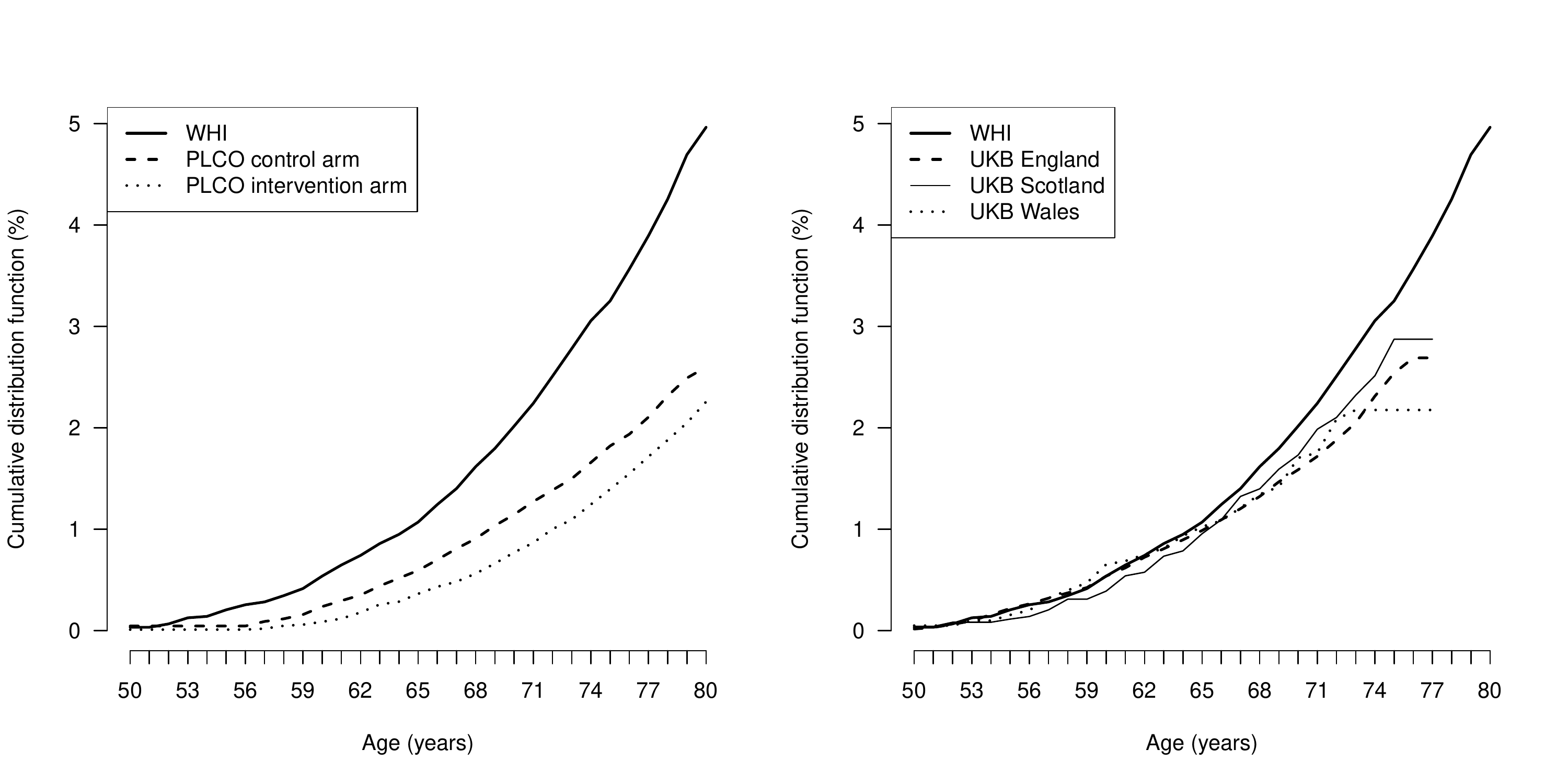}
\end{center}
\caption{Probabilities of developing colorectal cancer for the WHI and five target cohorts obtained by 1 - Kaplan-Meier estimator.
\label{fig:intro1}}
\end{figure}

\subsection{Results}

Figure \ref{fig:intro1} shows the probability of developing colorectal cancer from 50 to 80 years old for the WHI and five target cohorts. It is obvious that these cohorts have very different disease probabilities, justifying the need to leverage CRC risk estimates from the target cohorts for re-calibration.


Table \ref{table:example1} presents the descriptive statistics and hazard ratio estimates from the Cox model. 
We assessed the proportional hazards assumption for all the risk factors \citep{grambsch1994proportional} and the p-values are all greater than 0.05, suggesting there is no evidence the proportionality assumption underlying the Cox model is violated. Generally, having a positive family history in the first-degree relatives and being obese increase CRC risk, while having had endoscopy, more exercise, Aspirin/NSAIDS use, greater vegetable intake and positive estrogen status reduce risk.

The prevalences of risk factors are very similar between WHI and the two arms of PLCO, except for endoscopy history (WHI 56.2\%, PLCO intervention arm 73.9\%, PLCO control arm 10.1\%)and Aspirin/NSAID use (WHI 85.0\%, PLCO $\sim$48\%).  For the three target cohorts from UKB, notably fewer participants have had endoscopy, use NSAIDS and have positive estrogen status but more participants have moderate exercise (0-2 hours/week) than WHI participants.  Such discrepancy justifies the need to re-weight the risk factor distribution for the target PLCO and UKB cohorts.


\begin{table}[]
\renewcommand\arraystretch{1.35}
\setlength{\tabcolsep}{4.5pt}
\centering
\scriptsize
\caption{Prevalence of risk factors and outcome statistics for WHI, PLCO and UKB. Hazard ratio estimates (95\% confidence interval [CI]) for CRC are based on the WHI data.}
\begin{threeparttable}
  \begin{tabular}{lccccccc}
  \hline
  \hline
  & \multicolumn{6}{c}{Prevalence} & \\
  \cline{2-7} 
  &     & PLCO & PLCO & UKB & UKB  & UKB & WHI \\
Risk Factor/Outcome                                  &  WHI   & ctrl & intv & England & Scotland &  Wales & HR (95\% CI)                \\
\hline
No. of subjects                                                    & 76,733  & 33,510    & 33,744    & 175,248      & 14,740        & 8,070      &                   \\
\hline
Endoscopy history                                       & 56.2\% & \cellcolor{gray!25}10.1\%   & \cellcolor{gray!25}73.9\%   & \cellcolor{gray!25}37.7\%      & \cellcolor{gray!25}24.7\%       & \cellcolor{gray!25}23.7\%    & 0.79 (0.70, 0.89) \\
No. of relatives with CRC ($\geq$1)                                       & 14.7\% & 11.1\%   & 11.5\%   & 11.9\%      & 12.4\%       & 10.7\%    & 1.24 (1.06, 1.45) \\
Vigorous leisure exercise 0-2 hr/wk                  & 14.9\% & 23.3\%   & 23.1\%   & \cellcolor{gray!25}33.9\%      & \cellcolor{gray!25}33.6\%       & \cellcolor{gray!25}30.9\%    & 0.99 (0.83, 1.18) \\
Vigorous leisure exercise \textgreater{}2 hr/wk      & 12.1\% & 11.6\%   & 11.6\%   & 14.5\%      & 13.6\%       & 13.0\%    & 0.83 (0.68, 1.03) \\
Aspirin/NSAID use (Regular user)                     & 85.0\% & \cellcolor{gray!25}47.6\%   & \cellcolor{gray!25}48.0\%   & \cellcolor{gray!25}24.6\%      & \cellcolor{gray!25}26.0\%       & \cellcolor{gray!25}24.9\%    & 0.76 (0.65, 0.90) \\
Vegetable intake ($\geq$median portion/day) & 53.0\% & 50.2\%   & 50.7\%   & 46.4\%      & 45.3\%       & 44.6\%    & 0.94 (0.83, 1.06) \\
BMI, kg/$\mbox{m}^2$ ($\geq$30)                       & 22.8\% & 23.6\%   & 24.1\%   & 24.1\%      & 24.2\%       & 28.4\%    & 1.39 (1.21, 1.59) \\
Estrogen status (Positive)                           & 44.9\% & 50.6\%   & 50.7\%   & \cellcolor{gray!25}8.3\%       & \cellcolor{gray!25}6.8\%        & \cellcolor{gray!25}7.6\%     & 0.87 (0.76, 0.99) \\
\hline
Mean follow-up years                                 & 6.7    & 11.3     & 11.2     & 5.7         & 6.9          & 6.7       &                   \\
Other-cause mortality rate (5-year)                  & 3.8\%  & 2.0\%    & 2.1\%    & 1.4\%       & 1.6\%        & 1.4\%     &                   \\
CRC rate (5-year)                                    & 0.7\%  & 0.5\%    & 0.3\%    & 0.5\%       & 0.5\%        & 0.4\%     &                   \\
\hline
No. of risk factors for constraints                  &        & 2        & 2        & 4           & 4            & 4         &    \\ 
\hline
\end{tabular}
  \begin{tablenotes}
		\scriptsize
		\item[] NOTE: PLCO ctrl, PLCO control arm; PLCO intv, PLCO intervention arm. Gray context indicates the prevalence difference compared to WHI is greater than 10\%.
  \end{tablenotes}
\end{threeparttable}
\label{table:example1}
\end{table}

A key aspect of risk model validation is that the model can predict the risk for developing disease accurately. 
To investigate the calibration performance of the proposed methods, for each participant in the targeted cohort, we calculated 5-year absolute risk of developing CRC from the enrollment age based on the risk prediction model, accounting for the competing risks of death from other causes for which the cause-specific hazard functions were estimated from the target cohorts. We stratified participants into five evenly distributed groups according to the levels of their model-based absolute risk estimates. Then for each risk group, we calculated the empirical absolute risk of 5-year CRC risk accounting for competing risks based on the non-parametric estimator \citep{gray1988class}. We expect that a well calibrated model would yield absolute risk estimates, which match well to empirical absolute risks across all levels of risk groups. 

\begin{figure}
\begin{center}
\includegraphics[width=5.8in]{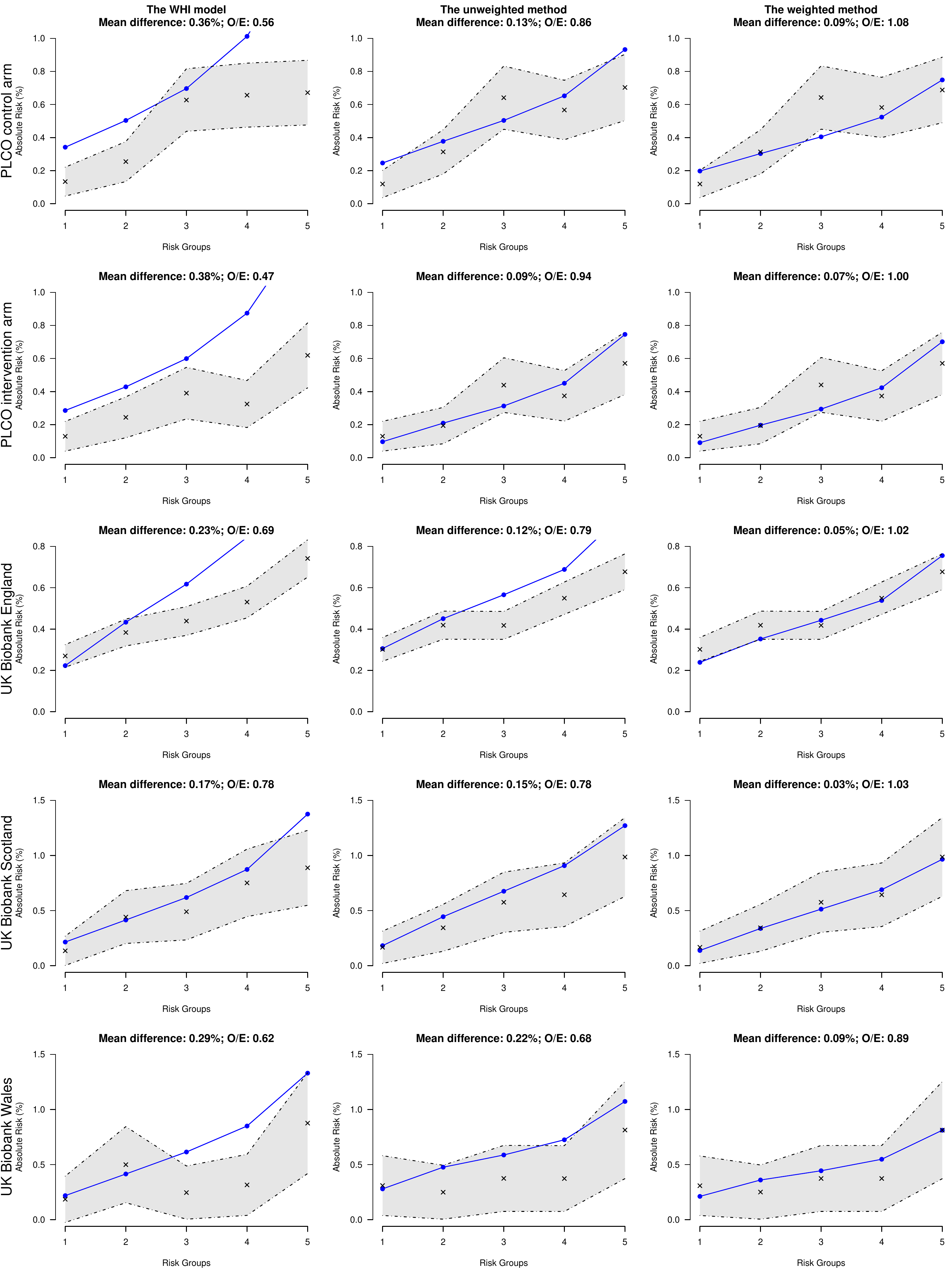}
\end{center}
\caption{Calibration plots of model-based and empirical absolute risk. Solid line: average model-based absolute risk in each risk group; Dotdash line (cross): the 95\% CIs (point estimates) of  empirical absolute risk in each risk group.  \label{fig:example2}}
\end{figure}

Figure \ref{fig:example2} shows the comparison of average model-based and empirical absolute risk estimates in each risk group, for three models: the  WHI model and the re-calibrated models with unweighted $\widehat \Lambda_0^\textup{u}(t)$ and weighted $\widehat\Lambda_0^\textup{w}(t)$ (constraints for selected risk factors for each target cohort are highlighted in gray in Table \ref{table:example1}). The WHI model used the competing hazard function estimated from the WHI data and the recalibrated models used the competing hazards based on the target cohort data.
The absolute risk estimation of the original WHI model is not well calibrated to both PLCO target cohorts. The average model-based absolute risk estimates are far above the upper bounds of the 95\% confidence intervals (CI) of the empirical absolute risk estimates. 
The re-calibrated unweighted estimator improves the calibration substantially. The 95\% CIs of empirical absolute risk estimates generally cover the model-based estimates except for highest and lowest risk groups in the control arm. For the weighted estimator, the 95\% CIs generally covers the average model-based estimates except for the middle risk group in the control arm. We calculated two summary measures of calibration: average absolute difference between model-based and empirical absolute risk estimates across five risk groups and O/E ratio.  The O/E ratio is defined as the overall empirical 5-year absolute risk divided by the model-based 5-year absolute risk. 
The weighted estimator has smaller absolute differences (0.09\% and 0.07\%) and better O/E ratios (1.08 and 1.00) in both PLCO cohorts than the unweighted estimator (absolute difference, 0.13\% and 0.09\%; O/E, 0.86 and 0.94). As the PLCO cohort has a longer follow-up, we also calculated 10-year risk and the the proposed weighted estimator has a very good calibration (Supplemental Figure S2).

For calibration of the UKB cohorts, the weighted estimator again performs the best among the three models, with average absolute differences of 0.05\%, 0.03\% and 0.09\%, and O/E ratios of 1.02, 1.03 and 0.89 for England, Scotland and Wales, respectively. The performance of the unweighted estimator is in between the WHI and weighted estimators. 
For UKB England cohort, the unweighted estimator is poorly calibrated with its estimates falling out of the 95\% CIs of empirical absolute risk in the three higher risk groups and the deviation is substantial in the highest risk group. This highlights the importance of accounting for the potential risk factor distribution difference if one will re-calibrate the model. These results demonstrate that the weighted estimator that leverages the summary information about disease incidence and risk factor distribution has the best performance. 

\subsection{Sensitivity Analyses}

To investigate in-depth the performance of various methods, we used the individual-level data from the PLCO control arm (PLCO-ctrl) to estimate $\Lambda_0(t)$ for CRC by the Breslow estimator with $\boldsymbol{\beta_0}$ obtained from the source WHI data, to serve as the oracle benchmark (i.e., the truth). We compared six different cumulative baseline hazard function estimators. Naturally, the Breslow estimator obtained from the WHI data was served as the base model. For our methods, we had an unweighted and four different weighted estimators. For the unweighted estimator, the baseline hazard function was constrained to yield the same CRC risk in the PLCO-ctrl cohort. For the weighted estimators, we additionally constrained the means of risk factors to be the same as PLCO-ctrl. Specifically, as Aspirin/NSAIDs use and vegetable intake have large differences of prevalence between WHI and PLCO-ctrl  cohorts, we considered using either one  or both to construct constraints to understand the incremental improvement of calibration from incorporating additional summary information. We also constrained on the mean values of all risk factors. 

\begin{figure}
\begin{center}
\includegraphics[width=6.2in]{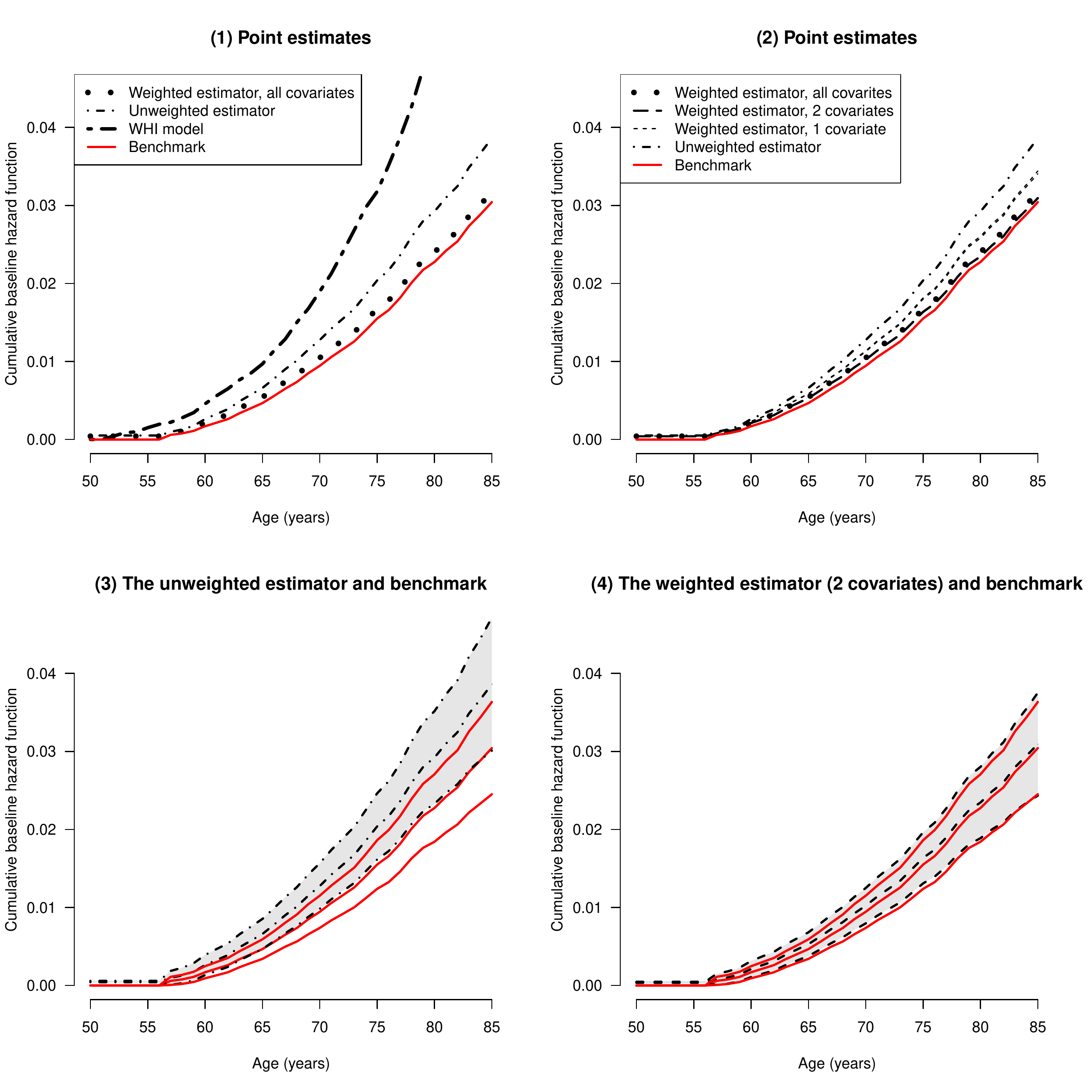}
\end{center}
\caption{Comparison of six cumulative baseline hazard estimators. In subgraphs (1)-(2), the solid line is the benchmark Breslow estimator using the individual-level data of the PLCO control arm. Note that the two weighted estimators that constrain for only one covariate are almost overlapped. In subgraphs (3)-(4), the dot-dashed and dashed lines are point estimates (95\% CIs) of the unweighted method and the weighted method (constraint of Aspirin/NSAIDs use and vegetable intake), respectively. Solid lines are the benchmark Breslow estimator with 95\% CI. 
\label{fig:example1}}
\end{figure}

Figure \ref{fig:example1} shows the benchmark and six estimators, across a wide range of age. From subgraph (1),  the weighted estimator constraining on all risk factors performs best, with its estimates agreeing very well with the benchmark, while the estimator from WHI data performs the worst and the unweighted estimator is in-between. In subgraph (2), the weighted estimator with only one risk factor constraint deviates farther from the benchmark than with two risk factor constraints, but is still closer to the benchmark than the unweighted estimator. The estimator with the constraints of two risk factors is almost overlapped with the one using means of all risk factors, indicating that it is adequate to only put constrains on risk factors that the distributions are significantly different between the source and target cohorts. Subgraphs (3)-(4) show that the benchmark falls within the 95\% confidence intervals of the weighted estimator, but not within that of the unweighted one. As expected, additional valid external information helps improve calibration. Interestingly, the 95\% CI of the weighted estimator is close to that of the benchmark Breslow estimator, suggesting that they are equally efficient. We also calculated the absolute risk estimates incorporating the constraints of all risk factors and they are very similar to the weighted method with fewer constraints in Figure \ref{fig:example2}. 

To assess the performance of the proposed weighted estimator when the competing risk has the same risk factors as the event of interest in the real data, we conducted a real-data based simulation study, mimicking two UKB cohorts and two PLCO cohorts. Specifically, for each cohort, we estimated the prevalences of eight covariates, the proportional hazards functions of the event of interest and competing event conditional on these covariates, and the Kaplan-Meier estimator of the censoring distribution. Some of the risk factors (e.g., BMI and Aspirin/NSAID use) are strongly associated with both the event of interest and competing event. We then generated the covariates, time-to-event of interest, time-to-competing event, and censoring time for each individual with the same sample size as each of the cohorts. The observation time is the earliest age among the three and the indicator records the type corresponding to this earliest age. The simulation results  demonstrate the proposed weighted estimator has robust and efficient performance, with all the  relative bias below 5\% and most of them very close to zero (UK cohorts in Table S12 and PLCO cohorts in Table S13). The coverage probabilities of 95\% CI are all greater than 90\%. In contrast, the Breslow estimator has the largest bias as high as over 200\%. The bias for the unweighted estimator is in between the Breslow and weighted estimators.  

\section{DISCUSSION}
\label{s:discussion}

In this article, we present an empirical likelihood-based weighted estimating equation approach to re-calibrating the risk prediction model developed from one cohort to another, leveraging the summary-level information from the target population.  In the presence of competing risks, the proposed weighted estimating equation yields nearly unbiased estimator and valid inference, even if the risk score $\boldsymbol{\beta}^T \boldsymbol{Z}$ for the event of interest is associated with the competing event. The proposed method offers a practical and robust solution to re-calibration of baseline risk while accommodating competing risks.

Like existing approaches, our proposed method imposes the constrain that the disease incidence rate is consistent to the target population; however, we take an additional step to reweigh the covariate data from the source study towards a more representative sample for the target population. This additional step has made our estimator much more robust than previously proposed approaches \cite{gail1989projecting, liu2014estimating}. Extensive simulation and real data analyses demonstrate that this step is critical, and surprisingly, has not been considered in the literature of risk prediction. While our example cohorts of PLCO and UKB have sufficient individual level data to build its own prediction model, in practice large research cohorts with detailed individual data often do not exist and re-calibration with high-level information is a much more feasible solution. In addition, compared with model developed in these cohorts using full individual level data, our approach, even with only limited cohort information, can yield  absolute risk projection that is both well calibrated to the new cohort and equally efficient. This is very appealing.  

Many risk prediction models have been developed based on various research cohorts. It is of great interest to apply these models to populations beyond the cohorts that were used for building these models. However, the common approach of applying these models directly to the target population can lead to non-ignorable over- or under-estimation of risk due to heterogeneity between source and target cohorts. This phenomenon occurs frequently in practice but has seldom been dealt with statistical rigor. We proposed a practical solution to account for this heterogeneity by synthesizing individual level data from the source cohort and the summary information from the target population. With the rapidly increasing availability of summary data in the public domain, auxiliary summary-level information is readily accessible. For example, cancer incidence rates can be obtained from existing public resources such as the Surveillance Epidemiology and End Results Registry (SEER) data (\url{http://www.seer.cancer.gov}), which is a primary source for cancer statistics representing 26\% of the U.S. population. For casting risk to a specific population, exposure and disease prevalence are often reported in historical publications. Further, the risk factor summary information may be obtained by a small to moderate random sample of the target population without the need of follow-up, which is feasible regarding time and cost to conduct such a study. In this situation, one can plug them directly into the estimating equation (\ref{def:ee}) and obtain the estimate of baseline hazard by solving the equation. If additional summary information (e.g., summary statistics of one or two covariates from a much larger sample size) is available, we can further gain efficiency by using the proposed method through additional constraints.

Regarding which summary statistics for covariates could be most helpful, it depends on the shape  of the covariate distribution, as well as the correlations among covariates. 
In general, mean and variance are sufficient to capture all the information if the covariate is normally distributed. For highly skewed covariates or these with multiple modes, quantiles provide necessarily additional information. If covariates are strongly correlated, conditional statistics can be helpful. It is worth noting that auxiliary summary statistics could be from multiple sources with different precision. After carefully selecting/combining such statistics through balancing their reliability and precision, the proposed method can synthesize them in an optimal way in terms of empirical likelihood.


There are a couple of directions that our proposed method can be extended. First, we use the empirical likelihood approach to impose the equality constraints for re-weighting the samples. When the auxiliary information has a large variance, the proposed approach, though reduces the bias substantially, can lose efficiency. Inequality constraints may allow for some trade-off between bias and efficiency. Second, alternative approaches such as exponential tilting \citep{haberman1984adjustment} and the generalized empirical likelihood \citep{newey2004higher} may generate weights with different properties \citep{little1991models}. 
These extensions deserve further study but are beyond the scope of this article. 





\bibliographystyle{apalike}

\bibliography{Bibliography-External-Calibration}

\end{document}


\if1\blind
{
  \title{\bf Web-based Supplementary Materials for ``Risk Projection for Time-to-event Outcome Leveraging Summary Statistics With Source Individual-level Data'' by Jiayin Zheng, Li Hsu, and Yingye Zheng}
  \author{}
  \date{}
  \maketitle
} \fi

\if0\blind
{
  \bigskip
  \bigskip
  \bigskip
  \begin{center}
    {\LARGE\bf Web-based Supplementary Materials for ``Risk Projection for Time-to-event Outcome Leveraging Summary Statistics With Source Individual-level Data''}
\end{center}
  \medskip
} \fi
	
\newpage

\section*{Web Appendix A:  A List of the Regularity Conditions}
\label{ss:conds}

In the following sections we establish the large sample results of the proposed estimators and show the consistency and asymptotic normality of $\widehat{\Lambda}^\textup{u}_0(t)$ and $\widehat{\Lambda}^\textup{w}_0(t)$ under regularity conditions listed below. We assume the cohort data from $P^*$ and the summary information from $P$ are statistically independent, i.e., $(\hat{S}(t),\boldsymbol{\hat{\mu}}) \indep \{(X_i,\delta_i,\boldsymbol{Z_i}),i=1,...,n\}$.

First we define following notation. \\
Let $s(t)$ be a generic overall survival function.\\
Let $\boldsymbol{\Sigma_{\mu,S(t)}}=\lim_{m\rightarrow \infty}m \mbox{E}^*\{(\boldsymbol{\hat{\mu}}-\boldsymbol{\mu_0})( \hat{S}(t)-S(t))\}$. \\
Define $Q(\boldsymbol{Z;\gamma,\mu})=\log[1+\boldsymbol{\gamma}^T\{\boldsymbol{h(Z)-\mu}\}]\cdot I_{[1+\boldsymbol{\gamma}^T\{\boldsymbol{h(Z)-\mu}\}>0]}-\infty \cdot I_{[1+\boldsymbol{\gamma}^T\{\boldsymbol{h(Z)-\mu}\}\leq 0]}$.
\\
Let $\boldsymbol{\gamma_0}$ be one solution to $\max_{\boldsymbol{\gamma}} \mbox{E}^* \{ Q(\boldsymbol{Z;\gamma,\mu_0}) \}$ if at least one solution exists.
\\
Let $\Lambda^{\dag}_0(t)$ be one solution to $\mbox{E}^*\{ \rho_1(\boldsymbol{Z};V(t),\boldsymbol{\gamma_0},\boldsymbol{\mu_0},S(t),\boldsymbol{\beta_0})\}=0$. \\
Let $\boldsymbol{Q_{\gamma,\gamma^T,0}(Z)}\equiv \partial^2 Q(\boldsymbol{Z;\gamma,\mu_0})/\partial \boldsymbol{\gamma} \partial \boldsymbol{\gamma}^T |_{\boldsymbol{\gamma}=\boldsymbol{\gamma_0}}$.

The regularity conditions are listed below.

\begin{enumerate}
	\item[A1.] The regularity conditions to ensure consistency and asymptotic normality of $\widehat\beta$ under the Cox proportional hazards model: $\{(X_i,\delta_i,\boldsymbol{Z_i}),i=1,...,n\}$ are independently and identically distributed (i.i.d.), and both $L_i$ and $C_i$ are independent of $T_i$ conditional on $\boldsymbol{Z_i}$;
	$\Pr(X>\tau)>0$, where $\tau$ is the maximum follow-up time;
		$\boldsymbol{Z}$ is time-independent and bounded;	and 
	$\boldsymbol{\mbox{E}^*\{n^{-1}I(\beta_0)\}}$ is positive definite, where $\boldsymbol{I(\beta)}=-\partial \boldsymbol{U(\beta)}/\partial \boldsymbol{\beta}$.

	\item[A2.] 
	$\hat{S}(t)$ is consistent and asymptotically normal, and its variance estimator is consistent. For Theorem 2 and 3, $\widehat\mu$'s are consistent and asymptotically normal and their variance estimators are consistent. 
	
	
	\item[A3.] The sample sizes satisfy $n/m\rightarrow r$ as $n\rightarrow \infty$ for $r>0$.
	\item[A4.] 
	Let $\boldsymbol{\mu}\equiv (\mu_1,...,\mu_q)^T$.
	Let $B_{j1}\equiv \{\boldsymbol{Z} | h_j(\boldsymbol{Z})-\mu_j>0\}$, $B_{j2}\equiv \{\boldsymbol{Z} | h_j(\boldsymbol{Z})-\mu_j<0\}$, $j=1,...,q$. 
	For each element in $\{(k_1,...,k_q)|k_j\in \{1,2\}\}$, $\Pr\{\cap^q_{j=1} B_{jk_j}  \}>0$ holds. Further, $\max_{\boldsymbol{\gamma}} \mbox{E}^*\{\boldsymbol{Z;\gamma,\mu_0}) \} < \infty$.
	
	\item[A5.] 
	There exist enough small compact neighborhoods of $\Lambda^{\dag}_0(t)$, $\boldsymbol{\gamma_0}$, $\boldsymbol{\mu_0}$, $S(t)$ and $\boldsymbol{\beta_0}$, denoted as $\Theta_\Lambda$, $\Theta_{\boldsymbol{\gamma}}$, $\Theta_{\boldsymbol{\mu}}$, $\Theta_S$, $\Theta_{\boldsymbol{\beta}}$, respectively, such that $Q(\boldsymbol{Z;\gamma,\mu})$ is continuous at each $(\boldsymbol{\gamma},\boldsymbol{\mu})\in \Theta^{'} \equiv \Theta_{\boldsymbol{\gamma}} \times \Theta_{\boldsymbol{\mu}}$ and $\rho_1(\boldsymbol{Z}; V(t), \boldsymbol{\gamma}, \boldsymbol{\mu}, s(t), \boldsymbol{\beta})$ is continuous at each $(V(t), \boldsymbol{\gamma}, \boldsymbol{\mu}, s(t), \boldsymbol{\beta}) \in \Theta \equiv \Theta_\Lambda \times \Theta_{\boldsymbol{\gamma}} \times \Theta_{\boldsymbol{\mu}} \times \Theta_S \times \Theta_{\boldsymbol{\beta}}$ with probability one for $P^*$. 
	\\
	In addition, there exist functions of $\boldsymbol{Z}$ with finite expectations, dominating the Euclidean $L^2$ norm of the following functions for any $(V(t),\boldsymbol{\gamma},\boldsymbol{\mu},s(t),\boldsymbol{\beta})\in \Theta$: $Q(\boldsymbol{Z;\gamma,\mu})$, $\rho_1(\boldsymbol{Z};V(t),\boldsymbol{\gamma},\boldsymbol{\mu},s(t),\boldsymbol{\beta})$, $\partial^2 Q(\boldsymbol{Z;\gamma,\mu})/\partial \boldsymbol{\gamma} \partial \boldsymbol{\gamma}^T$, $\partial^2 Q(\boldsymbol{Z;\gamma,\mu})/\partial \boldsymbol{\gamma} \partial \boldsymbol{\mu}^T$, and $\partial \rho_1(\boldsymbol{Z};V(t),\boldsymbol{\gamma},\boldsymbol{\mu},s(t),\boldsymbol{\beta}) / \partial R_j, \, j=1,...,5$ where $R_1=V(t), R_2=\boldsymbol{\gamma}, R_3=\boldsymbol{\mu}, R_4=s(t)$ and $R_5=\boldsymbol{\beta}$.
	
	
	
	
	\item[A6.] Assume $\sigma^2_{\rho} \equiv \mbox{E}^*\{ \rho^2_1(\boldsymbol{Z};\Lambda^{\dag}_0(t),\boldsymbol{\gamma_0},\boldsymbol{\mu_0},S(t),\boldsymbol{\beta_0}) \} < \infty$.
	
	
\end{enumerate}

Under the condition A1, the coefficients $\boldsymbol{\hat{\beta}}$ is consistent to $\boldsymbol{\beta_0}$ and asymptotically normal, i.e., $\sqrt{n}(\boldsymbol{\hat{\beta}}-\boldsymbol{\beta_0})\xrightarrow{d} N(\boldsymbol{0},\boldsymbol{\Sigma_{\beta}})$, where $\boldsymbol{0}$ is a $p\times 1$ vector with each component equal to $0$. The estimated variance of $\boldsymbol{\hat{\beta}}$, i.e., $\boldsymbol{\hat{\Sigma}_{\hat{\beta}}}$, satisfies that $n \boldsymbol{\hat{\Sigma}_{\hat{\beta}}}$ is consistent to $\boldsymbol{\Sigma_{\beta}}$.

The condition A2 indicates the following results: the auxiliary external information $\hat{S}(t)$ satisfies $\sqrt{m}(\hat{S}(t)-S(t))\xrightarrow{d} N(0,\sigma^2_{S(t)})$ and the estimated variance of $\hat{S}(t)$, denoted by $\hat{\sigma}^2_{\hat{S}(t)}$, satisfies that $m \hat{\sigma}^2_{\hat{S}(t)}$ is asymptotically consistent to $\sigma^2_{S(t)}$;
$\boldsymbol{\hat{\mu}}$ is consistent and asymptotically normal, and its variance matrix estimator is consistent. Then the auxiliary external information $\boldsymbol{\hat{\mu}}$ satisfies $\sqrt{m}(\boldsymbol{\hat{\mu}}-\boldsymbol{\mu})\xrightarrow{d} N(\boldsymbol{0},\boldsymbol{\Sigma_{\mu}})$ and the estimated variance covariance matrix of $\boldsymbol{\hat{\mu}}$, i.e., $\boldsymbol{\hat{\Sigma}_{\hat{\mu}}}$, satisfies that $m \boldsymbol{\hat{\Sigma}_{\hat{\mu}}}$ is consistent to $\boldsymbol{\Sigma_{\mu}}$.

The condition A4 ensures that with probability approaching one, the zero point vector is within the convex hull of $\{\boldsymbol{h(Z_i)-\mu}, i=1,...,n\}$.

\section*{Web Appendix B: Proof of Theorem 1}
\label{ss:proof1}

\begin{theorem}
	Assume the risk factor distributions are common between $P$ and $P^*$, i.e., $f(\boldsymbol{Z}) = f^*(\boldsymbol{Z})$. Under regularity conditions A1-A3, $\widehat{\Lambda}^{\textup{u}}_0(t)$ is consistent to $\Lambda_0(t)$, where  $\Lambda_0(t)$ is the true cumulative baseline hazard function in the target population, and $\sqrt{n}\{\widehat{\Lambda}^{\textup{u}}_0(t)-\Lambda_0(t)\}$ converges to a zero mean normal distribution with variance 
	\begin{eqnarray*}
	\label{def:limvaree}
	\sigma^2_1(t)=[\textup{E}^*\{\Phi_{\Lambda,0}(t, \boldsymbol{Z})\} ]^{-2}\; \Pi_0,
	\end{eqnarray*}
	where 
    \begin{eqnarray*}
    \Pi_0&= & \textup{E}^*\{\Phi^2_0(t, \boldsymbol{Z})\} + \textup{E}^*\{\boldsymbol{\Phi_{\beta,0}(t, Z)}\}^T \boldsymbol{\Sigma_{\beta}}
    \textup{E}^*\{\boldsymbol{\Phi_{\beta,0}(t, Z)}\} + r \sigma^2_{S(t)} \\
    &&\;\;\;+ 2 \lim_{n\rightarrow \infty}\textup{cov}(\Pi_{0,1},\Pi_{0,2}),\\
\Pi_{0,1}&= & \frac{1}{\sqrt{n}}\sum^n_{i=1}\Phi_0(t, \boldsymbol{Z_i}),\\
    \Pi_{0,2}&= & \textup{E}^*\{\boldsymbol{\Phi_{\beta,0}(t, Z)}\}^T\sqrt{n}(\boldsymbol{\hat{\beta}}-\boldsymbol{\beta_0}),\\
    \Phi_0(t, \boldsymbol{Z}) &\equiv & \Phi( \boldsymbol{Z}; \Lambda_0(t),\boldsymbol{\beta_0},S(t)),\\
	\Phi_{\Lambda,0}(t, \boldsymbol{Z}) &\equiv & \frac{\partial \Phi(\boldsymbol{Z}; V(t),\boldsymbol{\beta_0},S(t))}{\partial V(t)}|_{V=\Lambda_0},\\
\boldsymbol{\Phi_{\beta,0}(t,Z)} &\equiv & \frac{\partial \Phi(\boldsymbol{Z}; \Lambda_0(t),\boldsymbol{\beta},S_0(t))}{\partial \boldsymbol{\beta}}|_{\boldsymbol{\beta}=\boldsymbol{\beta_0}}.
    \end{eqnarray*}
\end{theorem}
\noindent Note that
\begin{eqnarray*}
\lim_{n\rightarrow \infty} \textup{cov} (\Pi_{0,1},\Pi_{0,2}) 
&=&  \mbox{E}^*\{ \Phi_0(t, \boldsymbol{Z}) \boldsymbol{l}^T(X,\delta,\boldsymbol{Z}) \boldsymbol{\Sigma_{\beta}} \} \; \mbox{E}^*\{\boldsymbol{\Phi_{\beta,0}(t, Z)}\}=0,
\end{eqnarray*}
where $\boldsymbol{l}(X,\delta,\boldsymbol{Z})$ is the efficient influence function of $(X,\delta,\boldsymbol{Z})$ for $\boldsymbol{\hat{\beta}}$ obtained by solving the score equation of the partial log-likelihood function under the Cox proportional hazards model.

\begin{proof}

$\Phi(\boldsymbol{Z};V(t),\boldsymbol{\beta},s(t))$ is continuous at each $(V(t),\boldsymbol{\beta},s(t))\in \Theta^{''}$ for any $\boldsymbol{Z}$, and bounded in $\Theta^{''}$, where $\Theta^{''}$ is any compact set satisfying $0 \leq s(t)\leq 1$ and $V(t) \geq 0$. From Lemma 2.4 in \citet{newey1994large}, $\mbox{E}^*\{\Phi(\boldsymbol{Z};V(t),\boldsymbol{\beta},s(t))\}$ is continuous and $\Phi_n(V(t),\boldsymbol{\beta},s(t))\equiv n^{-1}\sum^{n}_{i=1} \Phi(\boldsymbol{Z_i};V(t),\boldsymbol{\beta},s(t))$ converges uniformly in probability to $\mbox{E}^*\{\Phi(\boldsymbol{Z};V(t),\boldsymbol{\beta},s(t))\}$ in $\Theta^{''}$. With $\boldsymbol{\hat{\beta}}\xrightarrow{p} \boldsymbol{\beta_0}$ and $\hat{S}(t)\xrightarrow{p} S(t)$, given a large positive number $M$,
\begin{eqnarray*}
&&\sup_{0 \leq V(t) \leq M} \|\Phi_n(V(t),\boldsymbol{\hat{\beta}},\hat{S}(t)) - \mbox{E}^*\{\Phi(\boldsymbol{Z};V(t),\boldsymbol{\beta_0},S(t))\}\| \\
&\leq& \sup_{0 \leq V(t) \leq M} \|\Phi_n(V(t),\boldsymbol{\hat{\beta}},\hat{S}(t)) - \mbox{E}^*\{\Phi(\boldsymbol{Z};V(t),\boldsymbol{\hat{\beta}},\hat{S}(t))\}\| \\
&+& \sup_{0 \leq V(t) \leq M} \|\mbox{E}^*\{\Phi(\boldsymbol{Z};V(t),\boldsymbol{\beta_0},S(t))\} - \mbox{E}^*\{\Phi(\boldsymbol{Z};V(t),\boldsymbol{\hat{\beta}},\hat{S}(t))\}\|.
\end{eqnarray*}
Each of the last two terms converges to zero in probability, from the uniform consistency of $\Phi_n(V(t),\boldsymbol{\beta},s(t))$ and the uniform continuity of $\mbox{E}^*\{\Phi(\boldsymbol{Z};V(t),\boldsymbol{\beta},s(t))\}$ over $\Theta^{''}$ by Heine-Cantor theorem. 
Since $\mbox{E}^*\{\Phi(\boldsymbol{Z};V(t),\boldsymbol{\beta_0},S(t))\}$ is strictly decreasing with respect to $V(t)$, $\mbox{E}^*\{\Phi(\boldsymbol{Z};0,\boldsymbol{\beta_0},S(t))\}\geq 0$ and $\mbox{E}^*\{\Phi(\boldsymbol{Z};\infty,\boldsymbol{\beta_0},S(t))\} \leq 0$, there exists a unique solution of $\mbox{E}^*\{\Phi(\boldsymbol{Z};V(t),\boldsymbol{\beta_0},S(t))\}$ at $\Lambda_0(t)$.
We have 
\begin{eqnarray*}
&&\| \mbox{E}^*\{\Phi(\boldsymbol{Z};\widehat{\Lambda}^\textup{u}_0(t),\boldsymbol{\beta_0},S(t))\} - \mbox{E}^*\{\Phi(\boldsymbol{Z};\Lambda_0(t),\boldsymbol{\beta_0},S(t))\} \| \\
&=& \| \mbox{E}^*\{\Phi(\boldsymbol{Z};\widehat{\Lambda}^\textup{u}_0(t),\boldsymbol{\beta_0},S(t))\} \| \\
&=&\| \Phi_n(\widehat{\Lambda}^\textup{u}_0(t),\boldsymbol{\hat{\beta}},\hat{S}(t)) - \mbox{E}^*\{\Phi(\boldsymbol{Z};\widehat{\Lambda}^\textup{u}_0(t),\boldsymbol{\beta_0},S(t))\} \| \xrightarrow{p} 0. 
\end{eqnarray*}
Since $\mbox{E}^*\{\Phi(\boldsymbol{Z};V(t),\boldsymbol{\beta_0},S(t))\}$ is a strictly decreasing and continuous function of $V$, by continuous mapping theorem, we have $\widehat{\Lambda}^\textup{u}_0(t) \xrightarrow{p} \Lambda_0(t)$. The consistency of $\widehat{\Lambda}^\textup{u}_0(t)$ is proved.

By Taylor's expansion, we have
\begin{eqnarray*}
\widehat{\Lambda}^\textup{u}_0(t) - \Lambda_0(t) = - ( \Phi_{n, \bar\Lambda} )^{-1} \{ \Phi_{n,0} + \boldsymbol{\Phi_{n, \bar\beta}} (\boldsymbol{\hat{\beta}} - \boldsymbol{\beta_0}) - (\hat{S}(t) - S(t)) \}
\end{eqnarray*}
where 
\begin{eqnarray*}
\Phi_{n,0} &\equiv& n^{-1} \sum^n_{i=1} \Phi_0(t,\boldsymbol{Z_i}), \\
\Phi_{n, \bar\Lambda} &\equiv& n^{-1} \sum^n_{i=1} \partial \Phi(\boldsymbol{Z_i};V(t),\boldsymbol{\bar{\beta}},\bar{S}(t))/\partial V(t)|_{V=\bar{\Lambda}_0}, \\
\boldsymbol{\Phi_{n, \bar\beta}} &\equiv& n^{-1} \sum^n_{i=1} \partial \Phi(\boldsymbol{Z_i};\bar{\Lambda}_0(t),\boldsymbol{\beta},\bar{S}(t))/\partial \boldsymbol{\beta}|_{\boldsymbol{\beta}=\boldsymbol{\bar{\beta}}}, 
\end{eqnarray*}
and $(\bar{\Lambda}_0(t), \boldsymbol{\bar{\beta}},\bar{S})$ is a mean value on the line joining $(\widehat{\Lambda}^\textup{u}_0(t), \boldsymbol{\hat{\beta}}, \hat{S})$ and $(\Lambda_0(t), \boldsymbol{\beta_0}, S)$. Based the boundedness of $\boldsymbol{Z}$, it is easy to show that $\Phi_{n, \bar\Lambda}$ and $\boldsymbol{\Phi_{n, \bar\beta}}$ converge in probability to $\mbox{E}^*\{\Phi_{\Lambda,0}(t,\boldsymbol{Z})\}$ and $\mbox{E}^*\{\boldsymbol{\Phi_{\beta,0}(t,Z)}\}$, respectively. From the asymptotic normality of $\sqrt{n}\Phi_{n,0}$, $\sqrt{n}(\boldsymbol{\hat{\beta}} - \boldsymbol{\beta_0})$ and $\sqrt{m}(\hat{S}(t) - S(t))$, we have $\sqrt{n}(\widehat{\Lambda}^\textup{u}_0(t) - \Lambda_0(t))$ is asymptotically normal with mean $0$ and variances 
$[\mbox{E}^*\{\Phi_{\Lambda,0}(t, \boldsymbol{Z})\} ]^{-2}\; \Pi_0 $.

\end{proof}

\section*{Web Appendix C: Proof of Theorem 2}
\label{ss:proof2}

\begin{theorem}
    Assume $f(\boldsymbol{Z}) = f^*(\boldsymbol{Z})$. Under regularity conditions A1-A5, we have $\boldsymbol{\gamma_0}=\boldsymbol{0}$ and $\widehat{\Lambda}^\textup{w}_0(t)$ is consistent to $\Lambda_0(t)$. With additional condition A6, $\sqrt{n}\{\widehat{\Lambda}^\textup{w}_0(t)-\Lambda_0(t)\}$ converges to a zero mean normal distribution with variance
	\begin{eqnarray*}
	\label{def:limvarwee1}
	\sigma^2_2(t)=[\textup{E}^*\{\Phi_{\Lambda,0}(t, \boldsymbol{Z})\}]^{-2} \;	\Pi_1,
	\end{eqnarray*}
	where 
	\begin{eqnarray*}
		\Pi_1&=&\Pi_0
		-\textup{E}^*\{ \boldsymbol{\rho_{1,\gamma,0}}(t,\boldsymbol{Z}) \}^T 
		[\textup{E}^*\{ \boldsymbol{Q_{\gamma,\gamma^T,0}(Z)} \}]^{-1} \textup{E}^*\{ \boldsymbol{\rho_{1,\gamma,0}}(t,\boldsymbol{Z}) \} + \boldsymbol{\Pi_{1,4}}^T \; r \boldsymbol{\Sigma_{\mu}} \; \boldsymbol{\Pi_{1,4}}\\
		&+& 2 \textup{cov}(\Pi_{0,1},\Pi_{1,3}) + 2 \lim_{n\rightarrow \infty} \textup{cov}(\Pi_{0,2},\Pi_{1,3})
		- 2 \boldsymbol{\Pi^T_{1,4}} \; r \boldsymbol{\Sigma_{\mu, S(t)}} ,
	\end{eqnarray*}
    with
    \begin{eqnarray*}
        \Pi_{1,3}&=&-\textup{E}^*\{ \boldsymbol{\rho_{1,\gamma,0}}(t,\boldsymbol{Z}) \}^T
	    [\textup{E}^*\{\boldsymbol{Q_{\gamma,\gamma^T, 0}(Z)} \}]^{-1}
	    \frac{\sqrt{n}}{n} \sum^n_{i=1} \boldsymbol{Q_{\gamma, 0}(Z_i)},\\
        \boldsymbol{\Pi_{1,4}} &=& \textup{E}^*\{ \boldsymbol{\rho_{1,\mu,0}}(t,\boldsymbol{Z}) \} 
        -\textup{E}^*\{ \boldsymbol{\rho_{1,\gamma,0}}(t,\boldsymbol{Z}) \}^T
	    [\textup{E}^*\{\boldsymbol{Q_{\gamma,\gamma^T,0}(Z)} \}]^{-1}
	    [\textup{E}^*\{\boldsymbol{Q_{\gamma,\mu^T,0}(Z)} \}]^{-1},\\
	    \boldsymbol{\rho_{1,\gamma,0}}(t,\boldsymbol{Z}) &\equiv& \frac{\partial \rho_1(\boldsymbol{Z};\Lambda_0(t),\boldsymbol{\gamma},\boldsymbol{\mu_0},S(t),\boldsymbol{\beta_0}) }{ \partial \boldsymbol{\gamma}} |_{\boldsymbol{\gamma}=\boldsymbol{\gamma_0}},\\
	    \boldsymbol{\rho_{1,\mu,0}}(t,\boldsymbol{Z}) &\equiv&  \frac{\partial \rho_1(\boldsymbol{Z};\Lambda_0(t),\boldsymbol{\gamma_0},\boldsymbol{\mu},S(t),\boldsymbol{\beta_0}) }{ \partial \boldsymbol{\mu}} |_{\boldsymbol{\mu}=\boldsymbol{\mu_0}},\\
	    \boldsymbol{Q_{\gamma,\gamma^T,0}(Z)} &\equiv& \frac{\partial^2 Q(\boldsymbol{Z;\gamma,\mu_0})}{\partial \boldsymbol{\gamma} \partial \boldsymbol{\gamma}^T }|_{\boldsymbol{\gamma}=\boldsymbol{\gamma_0}},\\
	    \boldsymbol{Q_{\gamma,\mu^T, 0}(Z)} &\equiv& \frac{ \partial^2 Q(\boldsymbol{Z;\gamma,\mu})} {\partial \boldsymbol{\gamma} \partial \boldsymbol{\mu}^T} |_{\boldsymbol{\gamma}=\boldsymbol{\gamma_0},\boldsymbol{\mu}=\boldsymbol{\mu_0}},\\
	    \boldsymbol{Q_{\gamma,0}(Z)} &\equiv& \frac{\partial Q(\boldsymbol{Z;\gamma,\mu_0})}{\partial \boldsymbol{\gamma}}|_{\boldsymbol{\gamma}=\boldsymbol{\gamma_0}}.
    \end{eqnarray*}
\end{theorem}

\begin{proof}
As this theorem can be seen as a special case of Theorem 3, we omit the proof here.
\end{proof}

It is easy to show 
\begin{eqnarray*}
	\textup{cov}(\Pi_{0,1},\Pi_{1,3})&=&\mbox{E}^*\{ \boldsymbol{\rho_{1,\gamma,0}}(t,\boldsymbol{Z}) \}^T 
		[\mbox{E}^*\{ \boldsymbol{Q_{\gamma,\gamma^T,0}(Z)} \}]^{-1}
		\mbox{E}^*\{ \boldsymbol{\rho_{1,\gamma,0}}(t,\boldsymbol{Z}) \},\\
	\lim_{n\rightarrow \infty} \textup{cov}(\Pi_{0,2},\Pi_{1,3})&=&
	\mbox{E}^*\{ \boldsymbol{\rho_{1,\gamma,0}}(t,\boldsymbol{Z}) \}^T 
		[\mbox{E}^*\{ \boldsymbol{Q_{\gamma,\gamma^T,0}(Z)} \}]^{-1}\\
	    && \cdot  \mbox{E}^*\{ \boldsymbol{Q_{\gamma,0}(Z)}  \boldsymbol{l}^T(X,\delta,\boldsymbol{Z})\}
	    \cdot \boldsymbol{\Sigma_{\beta}} \mbox{E}^*\{\boldsymbol{\rho_{1,\beta,0}}(t,\boldsymbol{Z})\}=0,
\end{eqnarray*}
where $\boldsymbol{\rho_{1,\beta,0}}(t,\boldsymbol{Z}) \equiv \partial \rho_1(\boldsymbol{Z};\Lambda_0(t),\boldsymbol{\gamma_0},\boldsymbol{\mu_0},S(t),\boldsymbol{\beta}) / \partial \boldsymbol{\beta} |_{\boldsymbol{\beta}=\boldsymbol{\beta_0}}$. From the asymptotic variance expression, we have $\sigma^2_2(t)<\sigma^2_1(t)$ when $\boldsymbol{\Pi_{1,4}}^T \; r \boldsymbol{\Sigma_{\mu}} \; \boldsymbol{\Pi_{1,4}}
+  \textup{cov}(\Pi_{0,1},\Pi_{1,3})
- 2 \boldsymbol{\Pi^T_{1,4}} \; r \boldsymbol{\Sigma_{\mu, S(t)}}<0$. When $r$ is low,  both $\boldsymbol{\Pi^T_{1,4}} \; r \boldsymbol{\Sigma_{\mu, S(t)}}$ and $\boldsymbol{\Pi_{1,4}}^T \; r \boldsymbol{\Sigma_{\mu}} \; \boldsymbol{\Pi_{1,4}}$ are close to zero. We show in the following section that $\mbox{E}^*\{ \boldsymbol{Q_{\gamma,\gamma^T,0}(Z)} \}$ is negative definite. Its inverse is also negative definite and then $\textup{cov}(\Pi_{0,1},\Pi_{1,3})<0$. Thus $\sigma^2_2(t)<\sigma^2_1(t)$.

\section*{Web Appendix D: Proof of Theorem 3}
\label{ss:proof3}

\begin{theorem}
    Under regularity conditions A1-A5,   we have  $\widehat{\Lambda}^\textup{w}_0(t)$ is consistent to $\Lambda^{\dag}_0(t)$. With additional condition A6, $\sqrt{n}\{\widehat{\Lambda}^\textup{w}_0(t)-\Lambda^{\dag}_0(t)\}$ converges to a zero mean normal distribution with variance
	\begin{eqnarray*}
	\label{def:limvarwee2}
	\sigma^2_3(t)=[\textup{E}^*\{ \rho_{1,\Lambda, 0}(t,\boldsymbol{Z}) \}]^{-2} \;	\Pi_2,
	\end{eqnarray*}
	where 
	\begin{eqnarray*}
		\Pi_2&=& \textup{E}^*\{{\rho^2_{1,0}}(t,\boldsymbol{Z})\} -\textup{E}^*\{ \boldsymbol{\rho_{1,\gamma,0}}(t,\boldsymbol{Z}) \}^T 
		[\textup{E}^*\{ \boldsymbol{Q_{\gamma,\gamma^T,0}(Z)} \}]^{-1} \textup{E}^*\{ \boldsymbol{\rho_{1,\gamma,0}}(t,\boldsymbol{Z}) \}\\
		&+& \boldsymbol{\Pi_{1,4}^T} \; r \boldsymbol{\Sigma_{\mu}} \; \boldsymbol{\Pi_{1,4}} 
		+ \textup{E}^*\{\boldsymbol{\rho_{1,\beta,0}}(t,\boldsymbol{Z})\}^T 
		\boldsymbol{\Sigma_{\beta}}    \textup{E}^*\{\boldsymbol{\rho_{1,\beta,0}}(t,\boldsymbol{Z})\} 
		+ r \sigma^2_{S(t)} [\textup{E}^*\{  \rho_{1,S,0}(t,\boldsymbol{Z}) \}]^2\\
        &+& 2 \lim_{n\rightarrow \infty}\textup{cov}(\Pi_{1,1},\Pi_{1,2})
		+ 2 \textup{cov}(\Pi_{1,1},\Pi_{1,3}) + 2 \lim_{n\rightarrow \infty} \textup{cov}(\Pi_{1,2},\Pi_{1,3}) \\
		&+& 2 \boldsymbol{\Pi^T_{1,4}} \; r \boldsymbol{\Sigma_{\mu, S(t)}} \textup{E}^*\{  \rho_{1,S,0}(t,\boldsymbol{Z})\},
	\end{eqnarray*}
    with
    \begin{eqnarray*}
            \rho_{1,0}(t,\boldsymbol{Z}) &\equiv& \rho_1(\boldsymbol{Z};\Lambda^{\dag}_0(t),\boldsymbol{\gamma_0},\boldsymbol{\mu_0},S(t),\boldsymbol{\beta_0}),\\
            \rho_{1,\Lambda,0}(t,\boldsymbol{Z}) &\equiv& \frac{\partial \rho_1(\boldsymbol{Z};V(t),\boldsymbol{\gamma_0},\boldsymbol{\mu_0},S(t),\boldsymbol{\beta_0}) }{ \partial V (t)} |_{V=\Lambda^{\dag}_0},\\
            \boldsymbol{\rho_{1,\beta,0}(t,Z)} &\equiv& \frac{\partial \rho_1(\boldsymbol{Z};\Lambda^{\dag}_0(t),\boldsymbol{\gamma_0},\boldsymbol{\mu_0},S(t),\boldsymbol{\beta}) }{ \partial \boldsymbol{\beta}} |_{\boldsymbol{\beta}=\boldsymbol{\beta_0}},\\
            \rho_{1,S,0}(t,\boldsymbol{Z}) &\equiv&
            \frac{\partial \rho_1(\boldsymbol{Z};\Lambda^{\dag}_0(t),\boldsymbol{\gamma_0},\boldsymbol{\mu_0},s(t),\boldsymbol{\beta_0})}{\partial s(t)}  |_{s=S},\\
            \Pi_{1,1} &=& \frac{1}{\sqrt{n}}\sum^n_{i=1} \rho_{1,0}(t,\boldsymbol{Z_i}),\\
            \Pi_{1,2} &=& \textup{E}^*\{\boldsymbol{\rho_{1,\beta,0}}(t,\boldsymbol{Z_i})\}^T\sqrt{n}(\boldsymbol{\hat{\beta}}-\boldsymbol{\beta_0}).
    \end{eqnarray*}
\end{theorem}

\begin{proof}

First we prove the consistency. We estimate the weights $w_i$ by maximizing $\sum^n_{i=1} \log(w_i)$ subject to the constraints $\sum^n_{i=1} w_i = 1$ and $\sum^n_{i=1} w_i \{\boldsymbol{h(Z_i)} - \boldsymbol{\hat{\mu}}\} =  \boldsymbol{0}$. Under condition A4, the solution exists with probability approaching one and can be written as $\hat{w}_i=n^{-1}[1+\boldsymbol{\hat{\gamma}}^T \{\boldsymbol{h(Z_i)} - \boldsymbol{\hat{\mu}}\}]^{-1}$, where $\boldsymbol{\hat{\gamma}} = \mbox{argmax}_{\gamma} \sum^{n}_{i=1} \log[ 1+\boldsymbol{\hat{\gamma}}^T \{\boldsymbol{h(Z_i)} - \boldsymbol{\hat{\mu}}\} ]$ by convex duality \citep{owen2001empirical}. 

Let $Q_n(\boldsymbol{\gamma},\boldsymbol{\mu}) \equiv n^{-1}\sum^n_{i=1} Q(\boldsymbol{Z_i;\gamma,\mu})$.
Under the condition A5, from Lemma 2.4 in \citet{newey1994large}, we have $\mbox{E}^*\{ Q(\boldsymbol{Z;\gamma,\mu})\}$ is (uniformly) continuous and $\sup_{(\boldsymbol{\gamma},\boldsymbol{\mu})\in \Theta^{'}} \| Q_n(\boldsymbol{\gamma},\boldsymbol{\mu}) -  \mbox{E}^*\{ Q(\boldsymbol{Z;\gamma,\mu})\}\| \xrightarrow{p} 0$.

Now we prove the existence of a unique solution to $\max_{\boldsymbol{\gamma}} \mbox{E}^* \{ Q(\boldsymbol{Z;\gamma,\mu_0}) \}$. Since $\boldsymbol{\gamma} \mapsto Q(\boldsymbol{Z;\gamma,\mu_0})$ is concave, so is $\boldsymbol{\gamma} \mapsto \mbox{E}^*\{ Q(\boldsymbol{Z;\gamma,\mu_0}) \}$. Define \\
$\Omega\equiv \{ \boldsymbol{\gamma} | \| \mbox{E}^*\{Q(\boldsymbol{Z;\gamma,\mu_0})\} \| <\infty \}$. From the condition A4, $\mbox{E}^*\{\partial^2 Q(\boldsymbol{Z;\gamma,\mu_0})/\partial \boldsymbol{\gamma} \partial \boldsymbol{\gamma}^T\}$ is negative definite on $\Omega$. Thus $\boldsymbol{\gamma} \mapsto \mbox{E}^* \{Q(\boldsymbol{Z;\gamma,\mu_0})\}$ is strictly concave on $\Omega$. As $\mbox{E}^* \{Q(\boldsymbol{Z;0,\mu_0})\}=0$, 
$\Omega$ is non-empty and $\boldsymbol{0}$ is an interior point in $\Omega$. It is easy to see that $\Omega$ is a convex and open set. From the condition A4 and the strict concaveness of $\boldsymbol{\gamma} \mapsto \mbox{E}^*\{ Q(\boldsymbol{Z;\gamma,\mu_0}) \}$ on the open convex set $\Omega$, there exits a unique interior solution to $\max_{\boldsymbol{\gamma}} \mbox{E}^* \{ Q(\boldsymbol{Z;\gamma,\mu_0}) \}$, denoted as $\boldsymbol{\gamma_0}$. Therefore $\boldsymbol{\gamma_0}$ must satisfy \\
$\mbox{E}^*\{ \partial Q(\boldsymbol{Z;\gamma,\mu_0})/\partial \boldsymbol{\gamma} |_{ \boldsymbol{\gamma}=\boldsymbol{\gamma_0}}\}=\boldsymbol{0}$. 

From the uniform consistency of $Q_n(\boldsymbol{\gamma},\boldsymbol{\mu})$ and the uniform continuity of $\mbox{E}^*\{ Q(\boldsymbol{Z;\gamma,\mu})\}$ on $\Theta^{'}$, we have 
\begin{eqnarray*}
&& \sup_{\boldsymbol{\gamma} \in \Theta_{\boldsymbol{\gamma}}} \| Q_n(\boldsymbol{\gamma},\boldsymbol{\hat{\mu}}) -  \mbox{E}^*\{ Q(\boldsymbol{Z;\gamma,\mu_0})\} \| \\
&\leq& \sup_{\boldsymbol{\gamma} \in \Theta_{\boldsymbol{\gamma}}} \| Q_n(\boldsymbol{\gamma},\boldsymbol{\hat{\mu}}) -  \mbox{E}^*\{ Q(\boldsymbol{Z;\gamma,\hat{\mu}})\} \| 
+  \sup_{\boldsymbol{\gamma} \in \Theta_{\boldsymbol{\gamma}}} \| \mbox{E}^*\{ Q(\boldsymbol{Z;\gamma,\hat{\mu}})\} - \mbox{E}^*\{ Q(\boldsymbol{Z;\gamma,\mu_0})\} \| \\
&\xrightarrow{p}& 0.
\end{eqnarray*} 
Then we have
\begin{eqnarray*}
\lefteqn{\mbox{E}^*\{ Q(\boldsymbol{Z;\hat{\gamma},\mu_0})\} - \mbox{E}^*\{ Q(\boldsymbol{Z;\gamma_0,\mu_0})\}} \\
&=& Q_n(\boldsymbol{\hat{\gamma}},\boldsymbol{\hat{\mu}}) + o_p(1) - \mbox{E}^*\{ Q(\boldsymbol{Z;\gamma_0,\mu_0})\} \\
&\geq& Q_n(\boldsymbol{\gamma_0},\boldsymbol{\hat{\mu}}) + o_p (1) - \mbox{E}^*\{ Q(\boldsymbol{Z;\gamma_0,\mu_0})\} \\
&=& \mbox{E}^*\{ Q(\boldsymbol{Z;\gamma_0,\hat{\mu}})\} + o_p (1) - \mbox{E}^*\{ Q(\boldsymbol{Z;\gamma_0,\mu_0})\} \\
&=& o_p (1).
\end{eqnarray*}
Now $\mbox{E}^*\{ Q(\boldsymbol{Z;\hat{\gamma},\mu_0})\} \geq \mbox{E}^*\{ Q(\boldsymbol{Z;\gamma_0,\mu_0})\} + o_p (1)$. As $\boldsymbol{\gamma_0}$ is the unique solution of $\max_{\boldsymbol{\gamma}} \mbox{E}^* \{ Q(\boldsymbol{Z;\gamma,\mu_0}) \}$ and $\Theta_{\boldsymbol{\gamma}}$ is compact, we have $\boldsymbol{\hat{\gamma}} \xrightarrow{p} \boldsymbol{\gamma_0}$.

Now we prove that there exists a unique solution to $ \rho_{01}(t,V) \equiv \\ \mbox{E}^*\{ \rho_1(\boldsymbol{Z};V(t),\boldsymbol{\gamma_0},\boldsymbol{\mu_0},S(t),\boldsymbol{\beta_0}) \}=0$. Let the solution be $\Lambda^{\dag}_0(t)$. $ \rho_{01}(t,V)$ is continuous and strictly decreasing on any non-negative compact set, since $1+\boldsymbol{\gamma_0}^T \{\boldsymbol{h(Z)} - \boldsymbol{\mu_0}\}>0$ with probability one for $P^*$,  $\Phi(\boldsymbol{Z};V(t),\boldsymbol{\beta_0},S(t))$ is bounded, continuous and strictly decreasing for $V(t)$, $\lim_{V(t) \rightarrow 0} \Phi(\boldsymbol{Z};V(t),\boldsymbol{\beta_0},S(t))>0$, $\lim_{V(t) \rightarrow \infty} \Phi(\boldsymbol{Z};V(t),\boldsymbol{\beta_0},S(t))<0$. Then there is a unique solution to $ \rho_{01}(t,V)=0$, and $(\boldsymbol{\hat{\gamma}}, \widehat{\Lambda}^\textup{w}_0(t))$ exist with probability approaching one.

Let $\rho_{n1}(V(t),\boldsymbol{\gamma},\boldsymbol{\mu},s(t),\boldsymbol{\beta}) \equiv n^{-1}\sum^n_{i=1} \rho_1(\boldsymbol{Z_i};V(t),\boldsymbol{\gamma},\boldsymbol{\mu},s(t),\boldsymbol{\beta})$.
Under the condition A5, using Lemma 2.4 in \citet{newey1994large}, we have $\mbox{E}^*\{ \rho_1(\boldsymbol{Z};V(t),\boldsymbol{\gamma},\boldsymbol{\mu},s(t),\boldsymbol{\beta}) \}$ is (uniformly) continuous and
\begin{eqnarray*}
\sup_{(V(t),\boldsymbol{\gamma},\boldsymbol{\mu},s(t),\boldsymbol{\beta})\in \Theta} \| \rho_{n1}(V(t),\boldsymbol{\gamma},\boldsymbol{\mu},s(t),\boldsymbol{\beta}) -  \mbox{E}^*\{ \rho_1(\boldsymbol{Z};V(t),\boldsymbol{\gamma},\boldsymbol{\mu},s(t),\boldsymbol{\beta}) \}\| \xrightarrow{p} 0.
\end{eqnarray*}

By the uniform consistency of $\rho_{n1}(V(t),\boldsymbol{\gamma},\boldsymbol{\mu},s(t),\boldsymbol{\beta})$ on $\Theta$, $\boldsymbol{\hat{\gamma}} \xrightarrow{p} \boldsymbol{\gamma_0}$, $\boldsymbol{\hat{\beta}}\xrightarrow{p} \boldsymbol{\beta_0}$, $\boldsymbol{\hat{\mu}}\xrightarrow{p} \boldsymbol{\mu_0}$ and $\hat{S}(t)\xrightarrow{p} S(t)$, we have
\begin{eqnarray*}
&&\sup_{V(t) \in \Theta_\Lambda} 
\|\rho_{n1}(V(t),\boldsymbol{\hat{\gamma}},\boldsymbol{\hat{\mu}},\hat{S}(t),\boldsymbol{\hat{\beta}})
- \rho_{01}(t,V)\| \\
&\leq& \sup_{V(t) \in \Theta_\Lambda} 
\|\rho_{n1}(V(t),\boldsymbol{\hat{\gamma}},\boldsymbol{\hat{\mu}},\hat{S}(t),\boldsymbol{\hat{\beta}})
- \mbox{E}^*\{\rho_1(\boldsymbol{Z};V(t),\boldsymbol{\hat{\gamma}},\boldsymbol{\hat{\mu}},\hat{S}(t),\boldsymbol{\hat{\beta}})\}\| \\
&+& \sup_{V(t) \in \Theta_\Lambda} 
\|\mbox{E}^*\{\rho_1(\boldsymbol{Z};V(t),\boldsymbol{\hat{\gamma}},\boldsymbol{\hat{\mu}},\hat{S}(t),\boldsymbol{\hat{\beta}})\}
- \rho_{01}(t,V)\|.
\end{eqnarray*}
Each of the last two terms converges to $0$ in probability, from the uniform consistency of $\rho_{n1}(V(t),\boldsymbol{\gamma},\boldsymbol{\mu},s(t),\boldsymbol{\beta})$ and the uniform continuity of $\mbox{E}^*\{ \rho_1(\boldsymbol{Z};V(t),\boldsymbol{\gamma},\boldsymbol{\mu},s(t),\boldsymbol{\beta}) \}$ over $\Theta$ by Heine-Cantor theorem. 
Then we have $\| \rho_{01}(t,\widehat{\Lambda}^\textup{w}_0) - \rho_{01}(t,\Lambda^{\dag}_0) \| = \| \rho_{01}(t,\widehat{\Lambda}^\textup{w}_0) \| =\| \rho_{01}(t,\widehat{\Lambda}^\textup{w}_0) - \rho_{n1}(\widehat{\Lambda}^\textup{w}_0(t),\boldsymbol{\hat{\gamma}},\boldsymbol{\hat{\mu}},\hat{S}(t),\boldsymbol{\hat{\beta}}) \| \xrightarrow{p} 0$. From the existence of a unique solution $\Lambda^{\dag}_0(t)$ to $ \rho_{01}(t,V)=0$, and the compactness of $\Theta_{\Lambda}$, we have $\widehat{\Lambda}^\textup{w}_0(t) \xrightarrow{p} \Lambda^{\dag}_0(t)$. The consistency of $\widehat{\Lambda}^\textup{w}_0(t)$ is proved.

Define $f^{\dag}(\boldsymbol{Z}) \equiv f^*(\boldsymbol{Z}) / [1+\boldsymbol{\gamma^T_0} \{ \boldsymbol{h(Z)} - \boldsymbol{\mu_0} \}]$. Since $1+\boldsymbol{\gamma^T_0} \{ \boldsymbol{h(Z)} - \boldsymbol{\mu_0} \}>0,\;a.s.$, and
\begin{eqnarray*}
1&=& \int f^*(\boldsymbol{Z})d\boldsymbol{Z} \equiv \int [1+\boldsymbol{\gamma^T_0} \{ \boldsymbol{h(Z)} - \boldsymbol{\mu_0} \}] f^{\dag}(\boldsymbol{Z})d\boldsymbol{Z} \\
&=& \int f^{\dag}(\boldsymbol{Z})d\boldsymbol{Z} + \int \boldsymbol{\gamma^T_0} \{ \boldsymbol{h(Z)} - \boldsymbol{\mu_0} \} f^{\dag}(\boldsymbol{Z})d\boldsymbol{Z} \\
&=& \int f^{\dag}(\boldsymbol{Z})d\boldsymbol{Z} + \boldsymbol{\gamma^T_0} \int  \frac{ \{ \boldsymbol{h(Z)} - \boldsymbol{\mu_0} \}}{1+\boldsymbol{\gamma^T_0} \{ \boldsymbol{h(Z)} - \boldsymbol{\mu_0} \}} f^*(\boldsymbol{Z})d\boldsymbol{Z} \\
&=& \int f^{\dag}(\boldsymbol{Z})d\boldsymbol{Z} + \boldsymbol{\gamma^T_0} \mbox{E}^* \{\partial Q(\boldsymbol{Z;\gamma,\mu_0}) / \partial \boldsymbol{\gamma} |_{ \boldsymbol{\gamma}=\boldsymbol{\gamma_0} } \} \\
&=& \int f^{\dag}(\boldsymbol{Z})d\boldsymbol{Z}+ \boldsymbol{\gamma^T_0} \boldsymbol{0} =\int f^{\dag}(\boldsymbol{Z})d\boldsymbol{Z}
\end{eqnarray*}
$f^{\dag}(\boldsymbol{Z})$ is a valid probability density function for $\boldsymbol{Z}$. 
In addition,
\begin{eqnarray*}
&&\mbox{E}^* \{ \rho_1(\boldsymbol{Z};V(t),\boldsymbol{\gamma_0},\boldsymbol{\mu_0},S(t),\boldsymbol{\beta_0}) \} \\
&=& \int \rho_1(\boldsymbol{Z};V(t),\boldsymbol{\gamma_0},\boldsymbol{\mu_0},S(t),\boldsymbol{\beta_0}) f^*(\boldsymbol{Z})d\boldsymbol{Z} \\
&=& \int \frac{\Phi(\boldsymbol{Z};V(t),\boldsymbol{\beta_0},S(t))}{1+\boldsymbol{\gamma^T_0} \{ \boldsymbol{h(Z)} - \boldsymbol{\mu_0} \}} f^*(\boldsymbol{Z})d\boldsymbol{Z}\\
&=& \int \Phi(\boldsymbol{Z};V(t),\boldsymbol{\beta_0},S(t)) \frac{f^*(\boldsymbol{Z})d\boldsymbol{Z}}{1+\boldsymbol{\gamma^T_0} \{ \boldsymbol{h(Z)} - \boldsymbol{\mu_0} \}} \\
&=& \int \Phi(\boldsymbol{Z};V(t),\boldsymbol{\beta_0},S(t)) f^{\dag}(\boldsymbol{Z})d\boldsymbol{Z} \\
&=& \mbox{E}^{\dag} \{ \Phi(\boldsymbol{Z};V(t),\boldsymbol{\beta_0},S(t)) \},
\end{eqnarray*}
where $\mbox{E}^{\dag}$ is the expectation of $\boldsymbol{Z}$ with respect to the probability density function $f^{\dag}(\boldsymbol{Z})$. Then $\Lambda^{\dag}_0(t)$ is also the unique solution to $\mbox{E}^{\dag} \{ \Phi(\boldsymbol{Z};V(t),\boldsymbol{\beta_0},S(t)) \}=0$.
Note that under the conditions of Theorem 2, $\boldsymbol{\gamma_0}=\boldsymbol{0}$ and $\Lambda^{\dag}_0(t)=\Lambda_0(t)$.

Now we prove the asymptotic normality.

By Taylor-series expansion, we have 
\begin{eqnarray*}
0 &=& \rho_{n1}(\widehat{\Lambda}^\textup{w}_0(t),\boldsymbol{\hat{\gamma}},\boldsymbol{\hat{\mu}},\hat{S}(t),\boldsymbol{\hat{\beta}}) \\
&=& \rho_{n1}(\Lambda^{\dag}_0(t),\boldsymbol{\gamma_0},\boldsymbol{\mu_0},S(t),\boldsymbol{\beta_0})
+ \rho_{n1,\bar{\Lambda}} (\widehat{\Lambda}^\textup{w}_0(t)-\Lambda^{\dag}_0(t))
+ \boldsymbol{\rho_{n1,\bar{\gamma}}} (\boldsymbol{\hat{\gamma}} - \boldsymbol{\gamma_0}) \\
&&+ \boldsymbol{\rho_{n1,\bar{\mu}}} (\boldsymbol{\hat{\mu}}-\boldsymbol{\mu_0})
+ \rho_{n1,\bar{S}} \{\hat{S}(t)-S(t)\}
+ \boldsymbol{\rho_{n1,\bar{\beta}}} (\boldsymbol{\hat{\beta}} - \boldsymbol{\beta_0}),\\
\boldsymbol{0} &=&  \partial Q_n(\boldsymbol{\gamma},\boldsymbol{\mu})/ \partial \boldsymbol{\gamma}
|_{(\boldsymbol{\gamma},\boldsymbol{\mu})=(\boldsymbol{\hat{\gamma}},\boldsymbol{\hat{\mu}})} \\
&=& \partial Q_n(\boldsymbol{\gamma},\boldsymbol{\mu})/ \partial \boldsymbol{\gamma}
|_{(\boldsymbol{\gamma},\boldsymbol{\mu})=(\boldsymbol{\gamma_0},\boldsymbol{\mu_0})}
+ \boldsymbol{Q_{n,\bar{\gamma},\bar{\gamma}^T}} (\boldsymbol{\hat{\gamma}} - \boldsymbol{\gamma_0})
+ \boldsymbol{Q_{n,\bar{\gamma},\bar{\mu}^T}} (\boldsymbol{\hat{\mu}}-\boldsymbol{\mu_0})
\end{eqnarray*}
where
\begin{eqnarray*}
\rho_{n1,\bar{\Lambda}} &=& \partial \rho_{n1}(V(t),\boldsymbol{\gamma},\boldsymbol{\mu},s(t),\boldsymbol{\beta}) / \partial V(t) |_{(V,\boldsymbol{\gamma},\boldsymbol{\mu},s,\boldsymbol{\beta})=(\bar{\Lambda}_0,\boldsymbol{\bar{\gamma}},\boldsymbol{\bar{\mu}},\bar{S},\boldsymbol{\bar{\beta}})}, \\
\boldsymbol{\rho_{n1,\bar{\gamma}}} &=& \partial \rho_{n1}(V(t),\boldsymbol{\gamma},\boldsymbol{\mu},s(t),\boldsymbol{\beta}) / \partial \boldsymbol{\gamma} |_{(V,\boldsymbol{\gamma},\boldsymbol{\mu},s,\boldsymbol{\beta})=(\bar{\Lambda}_0,\boldsymbol{\bar{\gamma}},\boldsymbol{\bar{\mu}},\bar{S},\boldsymbol{\bar{\beta}})},\\
\boldsymbol{\rho_{n1,\bar{\mu}}} &=& \partial \rho_{n1}(V(t),\boldsymbol{\gamma},\boldsymbol{\mu},s(t),\boldsymbol{\beta}) / \partial \boldsymbol{\mu} |_{(V,\boldsymbol{\gamma},\boldsymbol{\mu},s,\boldsymbol{\beta})=(\bar{\Lambda}_0,\boldsymbol{\bar{\gamma}},\boldsymbol{\bar{\mu}},\bar{S},\boldsymbol{\bar{\beta}})}, \\
\rho_{n1,\bar{S}} &=& \partial \rho_{n1}(V(t),\boldsymbol{\gamma},\boldsymbol{\mu},s(t),\boldsymbol{\beta}) / \partial s(t) |_{(V,\boldsymbol{\gamma},\boldsymbol{\mu},s,\boldsymbol{\beta})=(\bar{\Lambda}_0,\boldsymbol{\bar{\gamma}},\boldsymbol{\bar{\mu}},\bar{S},\boldsymbol{\bar{\beta}})}, \\
\boldsymbol{\rho_{n1,\bar{\beta}}} &=& \partial \rho_{n1}(V(t),\boldsymbol{\gamma},\boldsymbol{\mu},s(t),\boldsymbol{\beta}) / \partial \boldsymbol{\beta} |_{(V,\boldsymbol{\gamma},\boldsymbol{\mu},s,\boldsymbol{\beta})=(\bar{\Lambda}_0,\boldsymbol{\bar{\gamma}},\boldsymbol{\bar{\mu}},\bar{S},\boldsymbol{\bar{\beta}})},\\
\boldsymbol{Q_{n,\bar{\gamma},\bar{\gamma}^T}} &=& \partial^2 Q_n(\boldsymbol{\gamma},\boldsymbol{\mu})/ \partial \boldsymbol{\gamma}  \partial \boldsymbol{\gamma^T} |_{(\boldsymbol{\gamma},\boldsymbol{\mu})=(\boldsymbol{\bar{\gamma}},\boldsymbol{\bar{\mu}})}, \\
\boldsymbol{Q_{n,\bar{\gamma},\bar{\mu}^T}} &=&  \partial^2 Q_n(\boldsymbol{\gamma},\boldsymbol{\mu})/ \partial \boldsymbol{\gamma}  \partial \boldsymbol{\mu^T} |_{(\boldsymbol{\gamma},\boldsymbol{\mu})=(\boldsymbol{\bar{\gamma}},\boldsymbol{\bar{\mu}})}, 
\end{eqnarray*}
$(\bar{\Lambda}_0(t),\boldsymbol{\bar{\gamma}},\boldsymbol{\bar{\mu}},\bar{S}(t),\boldsymbol{\bar{\beta}})$ is a mean value on the line joining $(\Lambda^{\dag}_0(t),\boldsymbol{\gamma_0},\boldsymbol{\mu_0},S(t),\boldsymbol{\beta_0})$ and $(\widehat{\Lambda}^\textup{w}_0(t),\boldsymbol{\hat{\gamma}},\boldsymbol{\hat{\mu}},\hat{S}(t),\boldsymbol{\hat{\beta}})$. Then we have 
\begin{eqnarray*}
\boldsymbol{\hat{\gamma}} - \boldsymbol{\gamma_0} &=& 
- \boldsymbol{Q^{-1}_{n,\bar{\gamma},\bar{\gamma}^T}}
\{
\partial Q_n(\boldsymbol{\gamma},\boldsymbol{\mu})/ \partial \boldsymbol{\gamma}
|_{(\boldsymbol{\gamma},\boldsymbol{\mu})=(\boldsymbol{\gamma_0},\boldsymbol{\mu_0})}
+ \boldsymbol{Q_{n,\bar{\gamma},\bar{\mu}^T}} (\boldsymbol{\hat{\mu}}-\boldsymbol{\mu_0})
\},\\
\widehat{\Lambda}^\textup{w}_0(t)-\Lambda^{\dag}_0(t) &=&
- \rho^{-1}_{n1,\bar{\Lambda}} 
\{
\rho_{n1}(\Lambda^{\dag}_0(t),\boldsymbol{\gamma_0},\boldsymbol{\mu_0},S(t),\boldsymbol{\beta_0})
+ \boldsymbol{\rho_{n1,\bar{\gamma}}} (\boldsymbol{\hat{\gamma}} - \boldsymbol{\gamma_0}) \\
&&+ \boldsymbol{\rho_{n1,\bar{\mu}}} (\boldsymbol{\hat{\mu}}-\boldsymbol{\mu_0})
+ \rho_{n1,\bar{S}} \{\hat{S}(t)-S(t)\}
+ \boldsymbol{\rho_{n1,\bar{\beta}}} (\boldsymbol{\hat{\beta}} - \boldsymbol{\beta_0})
\}.
\end{eqnarray*}
Reorganizing the two equations, we have
\begin{eqnarray}
\label{Taylor1}
&&\widehat{\Lambda}^\textup{w}_0(t)-\Lambda^{\dag}_0(t) \\
&=&
- \rho^{-1}_{n1,\bar{\Lambda}} 
[
\rho_{n1}(\Lambda^{\dag}_0(t),\boldsymbol{\gamma_0},\boldsymbol{\mu_0},S(t),\boldsymbol{\beta_0})
+ \boldsymbol{\rho_{n1,\bar{\mu}}} (\boldsymbol{\hat{\mu}}-\boldsymbol{\mu_0})
+ \rho_{n1,\bar{S}} \{\hat{S}(t)-S(t)\} \nonumber\\
&& 
- \boldsymbol{\rho_{n1,\bar{\gamma}}} \boldsymbol{Q^{-1}_{n,\bar{\gamma},\bar{\gamma}^T}}
\{
\partial Q_n(\boldsymbol{\gamma},\boldsymbol{\mu})/ \partial \boldsymbol{\gamma}
|_{(\boldsymbol{\gamma},\boldsymbol{\mu})=(\boldsymbol{\gamma_0},\boldsymbol{\mu_0})}
+ \boldsymbol{Q_{n,\bar{\gamma},\bar{\mu}^T}} (\boldsymbol{\hat{\mu}}-\boldsymbol{\mu_0})
\}
+ \boldsymbol{\rho_{n1,\bar{\beta}}} (\boldsymbol{\hat{\beta}} - \boldsymbol{\beta_0})
]. \nonumber
\end{eqnarray}
By the condition A5, using Lemma 2.4 in \citet{newey1994large}, we have 
\begin{eqnarray*}
\rho_{n1,\bar{\Lambda}} &\xrightarrow{p}& \mbox{E}^*\{ \rho_{1,\Lambda,0}(t,\boldsymbol{Z}) \}, \\
\boldsymbol{\rho_{n1,\bar{\gamma}}} &\xrightarrow{p}& \mbox{E}^*\{ \boldsymbol{\rho_{1,\gamma,0}}(t,\boldsymbol{Z}) \},\\
\boldsymbol{\rho_{n1,\bar{\mu}}} &\xrightarrow{p}& \mbox{E}^*\{ \boldsymbol{\rho_{1,\mu,0}}(t,\boldsymbol{Z}) \}, \\
\rho_{n1,\bar{S}} &\xrightarrow{p}& \mbox{E}^*\{ \rho_{1,S,0}(t,\boldsymbol{Z}) \}, \\
\boldsymbol{\rho_{n1,\bar{\beta}}} &\xrightarrow{p}& \mbox{E}^*\{ \boldsymbol{\rho_{1,\beta,0}}(t,\boldsymbol{Z}) \},\\
\boldsymbol{Q_{n,\bar{\gamma},\bar{\gamma}^T}} &\xrightarrow{p} &
\mbox{E}^*\{\boldsymbol{Q_{\gamma,\gamma^T,0}(Z)} \}, \\
\boldsymbol{Q_{n,\bar{\gamma},\bar{\mu}^T}} &\xrightarrow{p} &
\mbox{E}^*\{\boldsymbol{Q_{\gamma,\mu^T,0}(Z)} \}.
\end{eqnarray*}
$\mbox{E}^*\{\boldsymbol{Q_{\gamma,\gamma^T,0}(Z)} \}$ is negative definite and $\mbox{E}^*\{ \rho_{1,\Lambda,0}(t,\boldsymbol{Z}) \}<0$. With the condition A1 and A2, we have 
$\sqrt{n}(\boldsymbol{\hat{\beta}} - \boldsymbol{\beta_0})$, $\sqrt{m}\{\hat{S}(t)-S(t)\}$ and $\sqrt{m}(\boldsymbol{\hat{\mu}}-\boldsymbol{\mu_0})$ are asymptotically normal. Under the condition A5 and A6, using the central limit theorem, $\sqrt{n}\rho_{n1}(\Lambda^{\dag}_0(t),\boldsymbol{\gamma_0},\boldsymbol{\mu_0},S(t),\boldsymbol{\beta_0})$ and  $\sqrt{n} \partial Q_n(t,\boldsymbol{\gamma},\boldsymbol{\mu})/ \partial \boldsymbol{\gamma}
|_{(\boldsymbol{\gamma},\boldsymbol{\mu})=(\boldsymbol{\gamma_0},\boldsymbol{\mu_0})}$ are asymptotically normal. 
With the condition A3, from (\ref{Taylor1}), $\sqrt{n}\{\widehat{\Lambda}^\textup{w}_0(t)-\Lambda^{\dag}_0(t)\}$ is asymptotically normal with mean $0$ and variance 
$[\mbox{E}^*\{ \rho_{1,\Lambda, 0}(t,\boldsymbol{Z}) \}]^{-2} \;	\Pi_2$.

\end{proof}
It is easy to show 
\begin{eqnarray*}
        \lim_{n\rightarrow \infty} \textup{cov}(\Pi_{1,1},\Pi_{1,2}) &=& \lim_{n\rightarrow \infty} \mbox{E}^*\{ n^{-1}\sum^n_{i=1}\rho_{1,0}(t,\boldsymbol{Z_i}) \boldsymbol{l}^T(X_i,\delta_i,\boldsymbol{Z_i}) \boldsymbol{\Sigma_{\beta}} \} \; \mbox{E}^*\{\boldsymbol{\rho_{1,\beta,0}}(t,\boldsymbol{Z})\} \\
        &=&  \mbox{E}^*\{ \rho_{1,0}(t,\boldsymbol{Z}) \boldsymbol{l}^T(X,\delta,\boldsymbol{Z})  \} \boldsymbol{\Sigma_{\beta}}\; \mbox{E}^*\{\boldsymbol{\rho_{1,\beta,0}}(t,\boldsymbol{Z})\}=0, \\
	\textup{cov}(\Pi_{1,1},\Pi_{1,3})&=&\mbox{E}^*\{ \boldsymbol{\rho_{1,\gamma,0}}(t,\boldsymbol{Z}) \}^T 
		[\mbox{E}^*\{ \boldsymbol{Q_{\gamma,\gamma^T,0}(Z)} \}]^{-1}
		\mbox{E}^*\{ \boldsymbol{\rho_{1,\gamma,0}}(t,\boldsymbol{Z}) \},\\
	\lim_{n\rightarrow \infty} \textup{cov}(\Pi_{1,2},\Pi_{1,3})&=&
	\mbox{E}^*\{ \boldsymbol{\rho_{1,\gamma,0}}(t,\boldsymbol{Z}) \}^T
	    [\mbox{E}^*\{\boldsymbol{Q_{\gamma,\gamma^T,0}(Z)} \}]^{-1}\\
	    && \cdot  \mbox{E}^*\{ \boldsymbol{Q_{\gamma,0}(Z)}  \boldsymbol{l}^T(X,\delta,\boldsymbol{Z})\}
	    \cdot \boldsymbol{\Sigma_{\beta}} \mbox{E}^*\{\boldsymbol{\rho_{1,\beta,0}}(t,\boldsymbol{Z})\}=0.
\end{eqnarray*}

Now we briefly show the derivation of the asymptotic-based variance estimator of the absolute risk. 
The estimate of the absolute risk $\Pr(t_0<T\leq t_1,\Delta=1|T> t_0, \boldsymbol{Z})  = \int^{t_1}_{t_0} \lambda(u|\boldsymbol{Z}) \exp[-\int^u_{t_0} \{\lambda(s|\boldsymbol{Z})+\lambda_c(s|\boldsymbol{Z})\}ds]du$ can be written as a sum over the distinct times $u_1, u_2, ..., u_L$ for event of interest occurring within $[t_0, t_1]$, as
\begin{eqnarray*} 
    &&\widehat{\Pr}(t_0<T\leq t_1,\Delta=1|T> t_0,\boldsymbol{Z})  \\
    &=& \sum_{l=1}^L \widehat{\lambda}(u_l|\boldsymbol{Z}) \exp{\left[ - \{\widehat{\Lambda}(u_l|\boldsymbol{Z}) - \widehat{\Lambda}(t_0|\boldsymbol{Z})\} - \{\widehat{\Lambda}_c(u_l|\boldsymbol{Z})-\widehat{\Lambda}_c(t_0|\boldsymbol{Z})\}\right]} \\
    &=& \sum_{l=1}^L \widehat{\lambda}_0(u_l)\exp{(\boldsymbol{\widehat{\beta}}^T \boldsymbol{Z})} \exp{\left[ - \{\widehat{\Lambda}_0(u_l) - \widehat{\Lambda}_0(t_0)\} \exp{(\boldsymbol{\widehat{\beta}}^T \boldsymbol{Z})}  - \{\widehat{\Lambda}_c(u_l|\boldsymbol{Z})-\widehat{\Lambda}_c(t_0|\boldsymbol{Z})\}\right]} \\
    &=& \sum_{l=1}^L \{\widehat{\Lambda}_0(u_{l+1})-\widehat{\Lambda}_0(u_l)\}\exp{(\boldsymbol{\widehat{\beta}}^T \boldsymbol{Z})}  \cdot \exp{\left[ - \{\widehat{\Lambda}_0(u_l) - \widehat{\Lambda}_0(t_0)\} \exp{(\boldsymbol{\widehat{\beta}}^T \boldsymbol{Z})}  - \{\widehat{\Lambda}_c(u_l|\boldsymbol{Z})-\widehat{\Lambda}_c(t_0|\boldsymbol{Z})\}\right]}
\end{eqnarray*}
where $\widehat{\Lambda}_c(t|\boldsymbol{Z})$ is the hazard function estimate of the competing event at age $t$ given $\boldsymbol{Z}$ with respect to the target population.

Due to practical limitation of data availability, one often ignores covariates and sets the competing risk to be independent with covariates in calculating the absolute risk. This is often a good approximation, because allowing covariates in modelling the competing risk may have little impact on estimates of the absolute risk of the event of interest (see discussion in Chapter 5.4 in \citet{pfeiffer2017absolute} and Section 5 in \citet{liu2014estimating}). Under this approximation, the absolute risk estimate can be written as
\begin{eqnarray*} 
    &&\widehat{\Pr}(t_0<T\leq t_1,\Delta=1|T> t_0,\boldsymbol{Z})  \\
    &=& \sum_{l=1}^L \{\widehat{\Lambda}_0(u_{l+1})-\widehat{\Lambda}_0(u_l)\}\exp{(\boldsymbol{\widehat{\beta}}^T \boldsymbol{Z})}  \cdot \exp{\left[ - \{\widehat{\Lambda}_0(u_l) - \widehat{\Lambda}_0(t_0)\} \exp{(\boldsymbol{\widehat{\beta}}^T \boldsymbol{Z})}  - \{\widehat{\Lambda}_c(u_l)-\widehat{\Lambda}_c(t_0)\}\right]} 
\end{eqnarray*}
where $\widehat{\Lambda}_c(t)$ is the overall hazard function estimate of the competing event at age $t$  with respect to the target population. 
Let us denote $\boldsymbol{\Xi} = (\boldsymbol{\beta_0}^T,\Lambda_0(u_l),\Lambda_c(u_l),l=1,...,L)^T$ and $\boldsymbol{\widehat{\Xi}} = (\boldsymbol{\hat{\beta}^T},\widehat{\Lambda}_0(u_l),\widehat{\Lambda}_c(u_l),l=1,...,L)^T$.
The absolute risk estimator $\widehat{\Pr}(t_0<T\leq t_1,\Delta=1|T> t_0,\boldsymbol{Z})$ is a function of $(\boldsymbol{Z};\boldsymbol{\widehat{\Xi}})$, denoted as $f(\boldsymbol{Z};\boldsymbol{\widehat{\Xi}})$. By the first order expansion, we have $f(\boldsymbol{Z};\boldsymbol{\widehat{\Xi}})\approx f(\boldsymbol{Z};\boldsymbol{\Xi}) + f_{\boldsymbol{\Xi}}^{T} (\boldsymbol{Z};\boldsymbol{\widehat{\Xi}})(\boldsymbol{\widehat{\Xi}}-\boldsymbol{\Xi})$ where $f_{\boldsymbol{\Xi}} (\boldsymbol{Z};\boldsymbol{\widehat{\Xi}})$ is the derivative vector of the function $f(\boldsymbol{Z};\boldsymbol{\Xi})$ with respect to $\boldsymbol{\Xi}$ at $\boldsymbol{\Xi}=\boldsymbol{\widehat{\Xi}}$. By the delta method, we have the variance estimator of $f(\boldsymbol{Z};\boldsymbol{\widehat{\Xi}})$ as $f_{\boldsymbol{\Xi}}^{T} (\boldsymbol{Z};\boldsymbol{\widehat{\Xi}})  
\widehat{var}(\boldsymbol{\widehat{\Xi}}-\boldsymbol{\Xi})
f_{\boldsymbol{\Xi}}(\boldsymbol{Z};\boldsymbol{\widehat{\Xi}})$, where $\widehat{var}(\boldsymbol{\widehat{\Xi}}-\boldsymbol{\Xi})$ is the estimate of variance-covariance matrix of $\boldsymbol{\widehat{\Xi}}-\boldsymbol{\Xi}$. To calculate $\widehat{var}(\boldsymbol{\widehat{\Xi}}-\boldsymbol{\Xi})$, we need the variances $\mbox{var}(\boldsymbol{\hat{\beta}})$ and $\mbox{var}(\widehat{\Lambda}_c(u_l)), l=1,...,L$ that are readily available, and $\mbox{var}(\widehat{\Lambda}_0(u_l)), l=1,...,L$ that can be obtained by the above asymptotic results. We also need the covariances $\mbox{cov}(\boldsymbol{\hat{\beta}},\widehat{\Lambda}_0(u_l)), l=1,...,L$ and $\mbox{cov}(\widehat{\Lambda}_0(u_j)),\widehat{\Lambda}_0(u_l)), j,l=1,...,L$ that can be easily derived with a similar process as in the proof of Theorem 3. Note that $\mbox{cov}(\widehat{\Lambda}_0(u_j)),\widehat{\Lambda}_c(u_l))$ and $\mbox{cov}(\widehat{\Lambda}_c(u_j)),\widehat{\Lambda}_c(u_l)), j,l=1,...,L$ are generally not available and we can set them as $0$ in the calculation.

In some application, where the hazard of competing event is high and strongly associated with covariates of event of interest, it is essential to include covariates in modelling the competing risk for absolute risk calculation. Since the competing event is not uncommon, the information of covariate-specific competing risk model may be available for the target population. By incorporating these information, the absolute risk and its variance, can be similarly derived as described above.

\bibliographystyle{apalike}
\bibliography{Bibliography-External-Calibration-sup}

\section*{Web Appendix E: the Connection to the Attributable Hazard Function}

In this section we show that for a general hazard function, the proposed estimating equation can be seen as an integral form of the attributable hazard function.

Let $\lambda(t|\boldsymbol{Z(t)})$ be a general hazard function given a time-varying risk profile $\boldsymbol{Z(t)}$ at time $t$. Let $\boldsymbol{\bar{Z}}$ represent the entire risk profile history $\{\boldsymbol{Z(t)},t\geq 0\}$. The survival function given $\boldsymbol{\bar{Z}}$ is then $S(t|\boldsymbol{\bar{Z}})=\exp\{-\int^t_0 \lambda(t|\boldsymbol{Z(u)})du\}$ and the overall survival function is $S(t)=\exp\{-\int^t_0 \lambda(u)du\}$, where the overall hazard function $\lambda(t)=\mbox{E}_{\boldsymbol{\bar{Z}}}\{S(u|\boldsymbol{\bar{Z}}) \lambda(t|\boldsymbol{Z(t)})\}/\mbox{E}_{\boldsymbol{\bar{Z}}}\{S(u|\boldsymbol{\bar{Z}}) \}=\{\int_{\boldsymbol{\bar{Z}}} S(u|\boldsymbol{\bar{Z}}) \lambda(t|\boldsymbol{Z(t)}) f(\boldsymbol{\bar{Z}}) d\boldsymbol{\bar{Z}}\}/\{ \int_{\boldsymbol{\bar{Z}}} S(u|\boldsymbol{\bar{Z}}) f(\boldsymbol{\bar{Z}}) d\boldsymbol{\bar{Z}} \}$ and $f(\boldsymbol{\bar{Z}})$ is the probability density function of $\boldsymbol{\bar{Z}}$. Let $\lambda_0(t)\equiv \lambda(t|\boldsymbol{Z(t)}=\boldsymbol{0})$ be the baseline hazard function and the time-varying hazard ratio function $\eta(t,\boldsymbol{Z(t)}) \equiv  \lambda(t|\boldsymbol{Z(t)}) / \lambda_0(t)$. Our proposed estimating equation has the general form of $\mbox{E}_{\boldsymbol{\bar{Z}}}\{S(t|\boldsymbol{\bar{Z}})\}=S(t)$. Taking the derivative of both sides with respect to $t$, we have
\begin{eqnarray}
\label{deriv}
&&\frac{\partial E_{\boldsymbol{\bar{Z}}}\{S(t|\boldsymbol{\bar{Z}})\}}{\partial t}=\frac{\partial S(t)}{\partial t} \notag\\
&\Leftrightarrow& E_{\boldsymbol{\bar{Z}}}\{ \frac{\partial S(t|\boldsymbol{\bar{Z}}) }{\partial t}\} = -\lambda(t)S(t) \notag\\
&\Leftrightarrow& \int_{\boldsymbol{\bar{Z}}} \exp\{-\int^t_0 \lambda(t|\boldsymbol{Z(u)})du\} \{-\lambda(t|\boldsymbol{Z(t)})\} f(\boldsymbol{\bar{Z}}) d\boldsymbol{\bar{Z}} = -\lambda(t)S(t) \notag\\
&\Leftrightarrow& \int_{\boldsymbol{\bar{Z}}} \exp\{-\int^t_0 \lambda(t|\boldsymbol{Z(u)})du\} \lambda_0(t) \eta(t,\boldsymbol{Z(t)}) f(\boldsymbol{\bar{Z}}) d\boldsymbol{\bar{Z}} = \lambda(t)S(t) \notag\\
&\Leftrightarrow& \frac{\lambda_0(t)}{\lambda(t)} = \frac{S(t)}{\int_{\boldsymbol{\bar{Z}}} \exp\{-\int^t_0 \lambda(t|\boldsymbol{Z(u)})du\} \eta(t,\boldsymbol{Z(t)}) f(\boldsymbol{\bar{Z}}) d\boldsymbol{\bar{Z}}}
= \frac{S(t)}{\mbox{E}_{\boldsymbol{\bar{Z}}}\{S(t|\boldsymbol{\bar{Z}}) \eta(t|\boldsymbol{Z(t)})\}} \notag\\
\,
\end{eqnarray}
Note that the attributable hazard function
\begin{eqnarray*}
AR(t)&=&\frac{\lambda(t)-\lambda_0(t)}{\lambda(t)}=1-\frac{\lambda_0(t)}{\lambda(t)} \\
&=&1-\frac{\lambda_0(t)}{\mbox{E}_{\boldsymbol{\bar{Z}}}\{S(t|\boldsymbol{\bar{Z}}) \lambda(t|\boldsymbol{Z(t)})\}/\mbox{E}_{\boldsymbol{\bar{Z}}}\{S(t|\boldsymbol{\bar{Z}}) \}} \\
&=&1-\frac{\lambda_0(t)}{\mbox{E}_{\boldsymbol{\bar{Z}}}\{S(t|\boldsymbol{\bar{Z}}) \lambda(t|\boldsymbol{Z(t)})\}/S(t)} \\
&=&1-\frac{S(t)}{\mbox{E}_{\boldsymbol{\bar{Z}}}\{S(t|\boldsymbol{\bar{Z}}) \eta(t|\boldsymbol{Z(t)})\}}
\end{eqnarray*}
and then $1-AR(t)={S(t)}/{\mbox{E}_{\boldsymbol{\bar{Z}}}\{S(t|\boldsymbol{\bar{Z}}) \eta(t|\boldsymbol{Z(t)})\}}$. Thus (\ref{deriv}) is equivalent to ${\lambda_0(t)}/{\lambda(t)}=1-AR(t)$.

\section*{Web Appendix F: Short list of the Women's Health Initiative (WHI)
Investigators}

Program Office: (National Heart, Lung, and Blood Institute, Bethesda, Maryland)
Jacques Rossouw, Shari Ludlam, Joan McGowan, Leslie Ford, and Nancy Geller

Clinical Coordinating Center: (Fred Hutchinson Cancer Research Center, Seattle, WA) Garnet
Anderson, Ross Prentice, Andrea LaCroix, and Charles Kooperberg

Investigators and Academic Centers: (Brigham and Women's Hospital, Harvard Medical
School, Boston, MA) JoAnn E. Manson; (MedStar Health Research Institute/Howard University,
Washington, DC) Barbara V. Howard; (Stanford Prevention Research Center, Stanford, CA)
Marcia L. Stefanick; (The Ohio State University, Columbus, OH) Rebecca Jackson; (University
of Arizona, Tucson/Phoenix, AZ) Cynthia A. Thomson; (University at Buffalo, Buffalo, NY)
Jean Wactawski-Wende; (University of Florida, Gainesville/Jacksonville, FL) Marian Limacher;
(University of Iowa, Iowa City/Davenport, IA) Jennifer Robinson; (University of Pittsburgh,
Pittsburgh, PA) Lewis Kuller; (Wake Forest University School of Medicine, Winston-Salem,
NC) Sally Shumaker; (University of Nevada, Reno, NV) Robert Brunner

Women’s Health Initiative Memory Study: (Wake Forest University School of Medicine,
Winston-Salem, NC) Mark Espeland

For a list of all the investigators who have contributed to WHI science, please visit:
https://www.whi.org/researchers/Documents\%20\%20Write\%20a\%20Paper/WHI\%20
Investigator\%20Long\%20List.pdf

\section*{Web Appendix G: Supplementary Results for the Simulation Studies}

In this section, we provide the simulation results under a variety of settings that can not be accommodated in the main article due to space constraint. This includes Table S1 - S3 for when the sample size used to calculate the summary statistics is large (m = 100,000), Table S4 - S7 when m = 1,000, and Table S8 - S11 when m = 200, representing the precision of summary statistics information high, moderate, and low, respectively. Under each sample size configuration, we also evaluated the performance of various methods when the baseline hazard function from the target population differs from the source study that was used for developing the risk prediction model with varying degree.   The results are similar to what is presented in the main article. We also calculated the asymptotic-based standard error assuming known off-diagonal elements of $\boldsymbol{\hat{\Sigma}_{\hat{\mu}}}$, to accommodate the situation that covariance between the moments may be available.
The corresponding asymptotic-based standard error and 95\%coverage probability are denoted by ASEd and CPd. These estimates are nearly identical to those assuming covariance equal to 0.

\begin{figure}
\begin{center}
\includegraphics[width=5.8in]{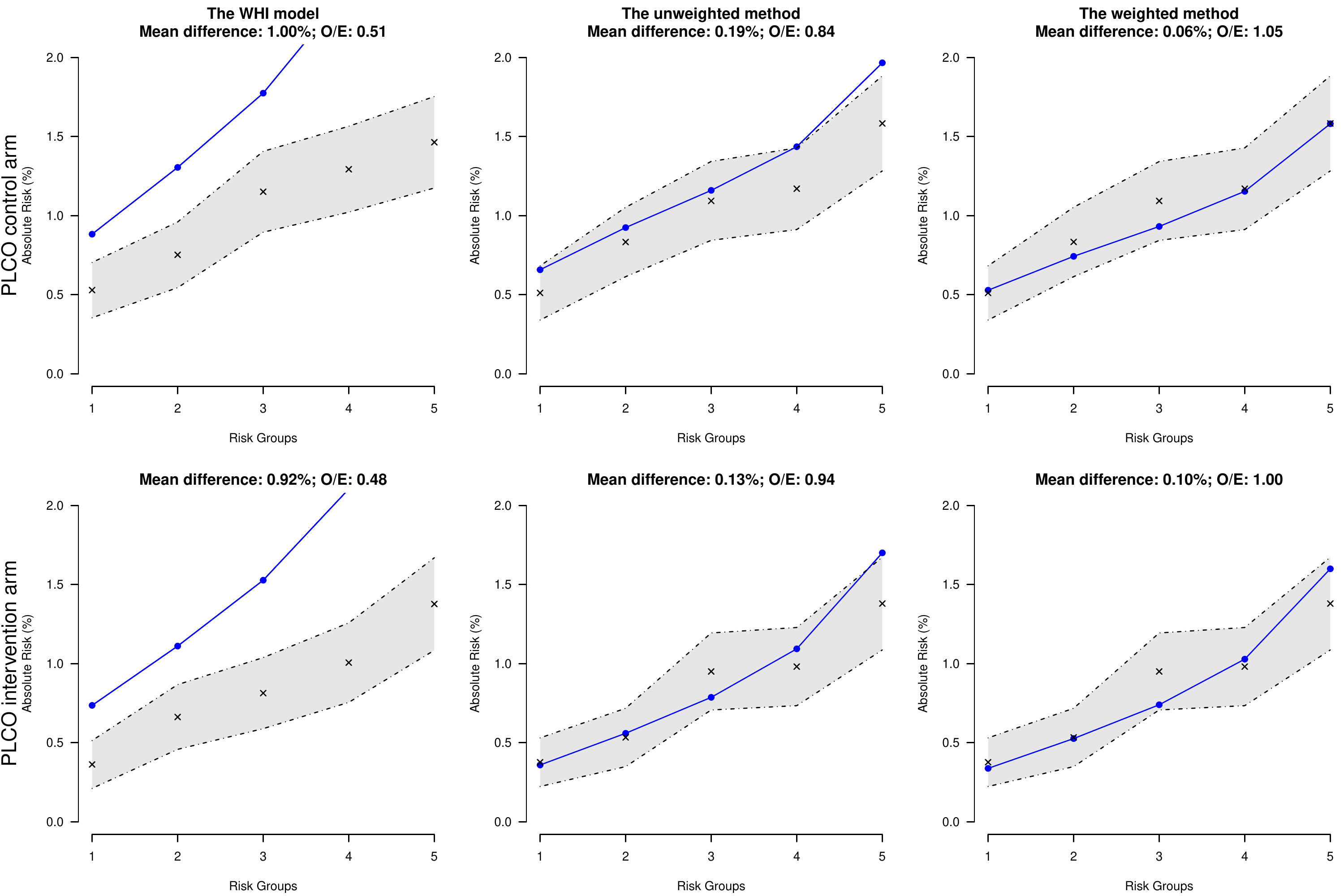}
\end{center}
\caption{Calibration plots of model-based and empirical 10-year absolute risk. Solid line: average model-based absolute risk in each risk group; Dotdash line (cross): the 95\% CIs (point estimates) of  empirical absolute risk in each risk group.  }
\label{figure_exsup}
\end{figure}

In addition, we provide the results for the calibration of 10-year absolute risk in the real example in Figure \ref{figure_exsup}  and the results of real data-based simulation in the presence of competing risks in Table \ref{table:simu.real1} and \ref{table:simu.real2}.

\begin{table}[]
	\renewcommand\arraystretch{1.1}
	\setlength{\tabcolsep}{2.5pt}
	\centering
	\scriptsize
	\caption{Summary statistics of simulation results for the scenario of $m=100,000$, (A1)/(A2), and $n=1,000$.}
	\begin{threeparttable}
\begin{tabular}{lllrrrrrrrrrrrrr}
			\hline
			\hline
			\\
			&            &   &&&&     \multicolumn{4}{l}{(A1) $\Lambda^*_0(t)=\Lambda_0(t)=(0.01t)^2$}         &&&               &      &                  \\
			\\
			\hline\\[-2.5 ex]
			&            &       & \multicolumn{6}{l}{(C1)  $Z_1\sim Bernoulli(0.5)$, $Z_2 \sim N(0,1)$,}                        &  & \multicolumn{6}{l}{(C2)  $Z_1\sim Bernoulli(0.8)$, $Z_2 \sim N(0,1)$,}                        \\ 
			&            &       & \multicolumn{6}{l}{\;\;\;\;\;\;\; and $Z_1 \indep Z_2$, i.e., $f^*(Z_1,Z_2)=f(Z_1,Z_2)$}                        &  & \multicolumn{6}{l}{\;\;\;\;\;\;\;  and $Z_1 \indep Z_2$}                        \\ 
			\cline{4-9} \cline{11-16}\\[-2.5 ex]
			$t$   & $\Lambda_0(t)$ &       & $\widehat{\Lambda}^\textup{b}_0(t)$ & $\widehat{\Lambda}^\textup{u}_0(t)$    & $\widehat{\Lambda}^\textup{w}_{0,1}(t)$ & $\widehat{\Lambda}^\textup{w}_{0,2}(t)$ & $\widehat{\Lambda}^\textup{w}_{0,3}(t)$ & $\widehat{\Lambda}^\textup{w}_{0,4}(t)$ &  & $\widehat{\Lambda}^\textup{b}_0(t)$ & $\widehat{\Lambda}^\textup{u}_0(t)$    & $\widehat{\Lambda}^\textup{w}_{0,1}(t)$ & $\widehat{\Lambda}^\textup{w}_{0,2}(t)$ & $\widehat{\Lambda}^\textup{w}_{0,3}(t)$ & $\widehat{\Lambda}^\textup{w}_{0,4}(t)$ \\
\hline
		\end{tabular}
	\end{threeparttable}
	\label{table:simu1}
\end{table}

\begin{table}[]
	\renewcommand\arraystretch{1.1}
	\setlength{\tabcolsep}{2.5pt}
	\centering
	\scriptsize
	\caption{Summary statistics of simulation results for the scenario of $m=100,000$, (A3) $\Lambda_0(t)=(0.01t)^2 \; and\; \Lambda^*_0(t)=(0.008t)^2$, and $n=1,000$.}
	\begin{threeparttable}
		\begin{tabular}{lllrrrrrrrrrrrrr}
			\hline
			\hline\\[-2.5 ex]
			&            &       & \multicolumn{6}{l}{(C1)  $Z_1\sim Bernoulli(0.5)$, $Z_2 \sim N(0,1)$,}                        &  & \multicolumn{6}{l}{(C2)  $Z_1\sim Bernoulli(0.8)$, $Z_2 \sim N(0,1)$,}                        \\ 
			&            &       & \multicolumn{6}{l}{\;\;\;\;\;\;\; and $Z_1 \indep Z_2$, i.e., $f^*(Z_1,Z_2)=f(Z_1,Z_2)$}                        &  & \multicolumn{6}{l}{\;\;\;\;\;\;\;  and $Z_1 \indep Z_2$}                        \\ 
			\cline{4-9} \cline{11-16}\\[-2.5 ex]
			$t$   & $\Lambda_0(t)$ &       & $\widehat{\Lambda}^\textup{b}_0(t)$ & $\widehat{\Lambda}^\textup{u}_0(t)$    & $\widehat{\Lambda}^\textup{w}_{0,1}(t)$ & $\widehat{\Lambda}^\textup{w}_{0,2}(t)$ & $\widehat{\Lambda}^\textup{w}_{0,3}(t)$ & $\widehat{\Lambda}^\textup{w}_{0,4}(t)$ &  & $\widehat{\Lambda}^\textup{b}_0(t)$ & $\widehat{\Lambda}^\textup{u}_0(t)$    & $\widehat{\Lambda}^\textup{w}_{0,1}(t)$ & $\widehat{\Lambda}^\textup{w}_{0,2}(t)$ & $\widehat{\Lambda}^\textup{w}_{0,3}(t)$ & $\widehat{\Lambda}^\textup{w}_{0,4}(t)$ \\ [0.5 ex]
			\hline
			\rowcolor{Gray}
			20 & 0.04        & PBias & -35.7\% & 1.3\%  & 1.3\%   & 1.3\%   & 1.2\%   & 1.2\%   &  & -36.0\% & 21.8\% & 2.8\%   & 1.9\%   & 1.9\%   & 2.0\%   \\
			&            & ESD   & 0.51    & 0.51   & 0.51    & 0.50    & 0.50    & 0.50    &  & 0.50    & 0.60   & 0.67    & 0.64    & 0.64    & 0.64    \\
			&            & ASEd  & 0.51    & 0.54   & 0.53    & 0.50    & 0.50    & 0.49    &  & 0.50    & 0.65   & 0.68    & 0.65    & 0.64    & 0.64    \\
\hline
		\end{tabular}
	\end{threeparttable}
	\label{table:simu3}
\end{table}

\begin{table}[]
	\renewcommand\arraystretch{1.1}
	\setlength{\tabcolsep}{2.5pt}
	\centering
	\scriptsize
	\caption{Summary statistics of simulation results for the scenario of $m=100,000$, (A4) $\Lambda_0(t)=(0.01t)^2 \; and\; \Lambda^*_0(t)=(0.008t)^{1.5}$, and $n=1,000$.}
	\begin{threeparttable}
		\begin{tabular}{lllrrrrrrrrrrrrr}
			\hline
			\hline\\[-2.5 ex]
			&            &       & \multicolumn{6}{l}{(C1)  $Z_1\sim Bernoulli(0.5)$, $Z_2 \sim N(0,1)$,}                        &  & \multicolumn{6}{l}{(C2)  $Z_1\sim Bernoulli(0.8)$, $Z_2 \sim N(0,1)$,}                        \\ 
			&            &       & \multicolumn{6}{l}{\;\;\;\;\;\;\; and $Z_1 \indep Z_2$, i.e., $f^*(Z_1,Z_2)=f(Z_1,Z_2)$}                        &  & \multicolumn{6}{l}{\;\;\;\;\;\;\;  and $Z_1 \indep Z_2$}                        \\ 
			\cline{4-9} \cline{11-16}\\[-2.5 ex]
			$t$   & $\Lambda_0(t)$ &       & $\widehat{\Lambda}^\textup{b}_0(t)$ & $\widehat{\Lambda}^\textup{u}_0(t)$    & $\widehat{\Lambda}^\textup{w}_{0,1}(t)$ & $\widehat{\Lambda}^\textup{w}_{0,2}(t)$ & $\widehat{\Lambda}^\textup{w}_{0,3}(t)$ & $\widehat{\Lambda}^\textup{w}_{0,4}(t)$ &  & $\widehat{\Lambda}^\textup{b}_0(t)$ & $\widehat{\Lambda}^\textup{u}_0(t)$    & $\widehat{\Lambda}^\textup{w}_{0,1}(t)$ & $\widehat{\Lambda}^\textup{w}_{0,2}(t)$ & $\widehat{\Lambda}^\textup{w}_{0,3}(t)$ & $\widehat{\Lambda}^\textup{w}_{0,4}(t)$ \\ [0.5 ex]
			\hline
			\rowcolor{Gray}
			20 & 0.04        & PBias & 58.3\%  & 0.6\%  & 0.6\%   & 0.7\%   & 0.7\%   & 0.7\%   &  & 58.2\%  & 21.8\% & 2.5\%   & 1.7\%   & 1.7\%   & 1.7\%   \\
			&            & ESD   & 0.89    & 0.41   & 0.41    & 0.40    & 0.40    & 0.40    &  & 0.89    & 0.49   & 0.55    & 0.52    & 0.51    & 0.52    \\
			&            & ASEd  & 0.89    & 0.46   & 0.44    & 0.42    & 0.41    & 0.40    &  & 0.89    & 0.55   & 0.57    & 0.53    & 0.52    & 0.52    \\
\hline
		\end{tabular}
	\end{threeparttable}
	\label{table:simu4}
\end{table}

\begin{table}[]
	\renewcommand\arraystretch{1.1}
	\setlength{\tabcolsep}{2.5pt}
	\centering
	\scriptsize
	\caption{Summary statistics of simulation results for the scenario of $m=1,000$, (A1) $\Lambda^*_0(t)=\Lambda_0(t)=(0.01t)^2$, and $n=1,000$.}
	\begin{threeparttable}
		\begin{tabular}{lllrrrrrrrrrrrrr}
			\hline
			\hline\\[-2.5 ex]
			&            &       & \multicolumn{6}{l}{(C1)  $Z_1\sim Bernoulli(0.5)$, $Z_2 \sim N(0,1)$,}                        &  & \multicolumn{6}{l}{(C2)  $Z_1\sim Bernoulli(0.8)$, $Z_2 \sim N(0,1)$,}                        \\ 
			&            &       & \multicolumn{6}{l}{\;\;\;\;\;\;\; and $Z_1 \indep Z_2$, i.e., $f^*(Z_1,Z_2)=f(Z_1,Z_2)$}                        &  & \multicolumn{6}{l}{\;\;\;\;\;\;\;  and $Z_1 \indep Z_2$}                        \\ 
			\cline{4-9} \cline{11-16}\\[-2.5 ex]
			$t$   & $\Lambda_0(t)$ &       & $\widehat{\Lambda}^\textup{b}_0(t)$ & $\widehat{\Lambda}^\textup{u}_0(t)$    & $\widehat{\Lambda}^\textup{w}_{0,1}(t)$ & $\widehat{\Lambda}^\textup{w}_{0,2}(t)$ & $\widehat{\Lambda}^\textup{w}_{0,3}(t)$ & $\widehat{\Lambda}^\textup{w}_{0,4}(t)$ &  & $\widehat{\Lambda}^\textup{b}_0(t)$ & $\widehat{\Lambda}^\textup{u}_0(t)$    & $\widehat{\Lambda}^\textup{w}_{0,1}(t)$ & $\widehat{\Lambda}^\textup{w}_{0,2}(t)$ & $\widehat{\Lambda}^\textup{w}_{0,3}(t)$ & $\widehat{\Lambda}^\textup{w}_{0,4}(t)$ \\ [0.5 ex]
			\hline
			\rowcolor{Gray}
			20 & 0.04        & PBias & 0.2\%   & 0.9\%  & 0.9\%   & 0.8\%   & 0.8\%   & 0.8\%   &  & 0.1\%   & 22.0\% & 2.4\%   & 1.6\%   & 1.7\%   & 1.7\%   \\
			&            & ESD   & 0.66    & 0.65   & 0.65    & 0.63    & 0.63    & 0.63    &  & 0.65    & 0.75   & 0.73    & 0.71    & 0.70    & 0.70    \\
			&            & ASEd  & 0.65    & 0.69   & 0.68    & 0.67    & 0.67    & 0.66    &  & 0.65    & 0.80   & 0.75    & 0.73    & 0.72    & 0.72    \\
\hline
		\end{tabular}
	\end{threeparttable}
	\label{table:simu5}
\end{table}

\begin{table}[]
	\renewcommand\arraystretch{1.1}
	\setlength{\tabcolsep}{2.5pt}
	\centering
	\scriptsize
	\caption{Summary statistics of simulation results for the scenario of $m=1,000$, (A2) $\Lambda_0(t)=(0.01t)^2 \; and\; \Lambda^*_0(t)=(0.01t)^{1.5}$, and $n=1,000$.}
	\begin{threeparttable}
		\begin{tabular}{lllrrrrrrrrrrrrr}
			\hline
			\hline\\[-2.5 ex]
			&            &       & \multicolumn{6}{l}{(C1)  $Z_1\sim Bernoulli(0.5)$, $Z_2 \sim N(0,1)$,}                        &  & \multicolumn{6}{l}{(C2)  $Z_1\sim Bernoulli(0.8)$, $Z_2 \sim N(0,1)$,}                        \\ 
			&            &       & \multicolumn{6}{l}{\;\;\;\;\;\;\; and $Z_1 \indep Z_2$, i.e., $f^*(Z_1,Z_2)=f(Z_1,Z_2)$}                        &  & \multicolumn{6}{l}{\;\;\;\;\;\;\;  and $Z_1 \indep Z_2$}                        \\ 
			\cline{4-9} \cline{11-16}\\[-2.5 ex]
			$t$   & $\Lambda_0(t)$ &       & $\widehat{\Lambda}^\textup{b}_0(t)$ & $\widehat{\Lambda}^\textup{u}_0(t)$    & $\widehat{\Lambda}^\textup{w}_{0,1}(t)$ & $\widehat{\Lambda}^\textup{w}_{0,2}(t)$ & $\widehat{\Lambda}^\textup{w}_{0,3}(t)$ & $\widehat{\Lambda}^\textup{w}_{0,4}(t)$ &  & $\widehat{\Lambda}^\textup{b}_0(t)$ & $\widehat{\Lambda}^\textup{u}_0(t)$    & $\widehat{\Lambda}^\textup{w}_{0,1}(t)$ & $\widehat{\Lambda}^\textup{w}_{0,2}(t)$ & $\widehat{\Lambda}^\textup{w}_{0,3}(t)$ & $\widehat{\Lambda}^\textup{w}_{0,4}(t)$ \\ [0.5 ex]
			\hline
			\rowcolor{Gray}
			20 & 0.04        & PBias & 121.0\% & 2.1\%  & 2.0\%   & 1.9\%   & 1.9\%   & 1.9\%   &  & 120.0\% & 21.8\% & 2.1\%   & 1.5\%   & 1.5\%   & 1.5\%   \\
			&            & ESD   & 1.07    & 0.65   & 0.64    & 0.63    & 0.62    & 0.62    &  & 1.11    & 0.73   & 0.70    & 0.67    & 0.67    & 0.67    \\
			&            & ASEd  & 1.08    & 0.66   & 0.65    & 0.64    & 0.64    & 0.63    &  & 1.08    & 0.76   & 0.70    & 0.67    & 0.67    & 0.67    \\
\hline
		\end{tabular}
	\end{threeparttable}
	\label{table:simu6}
\end{table}

\begin{table}[]
	\renewcommand\arraystretch{1.1}
	\setlength{\tabcolsep}{2.5pt}
	\centering
	\scriptsize
	\caption{Summary statistics of simulation results for the scenario of $m=1,000$, (A3) $\Lambda_0(t)=(0.01t)^2 \; and\; \Lambda^*_0(t)=(0.008t)^2$, and $n=1,000$.}
	\begin{threeparttable}
		\begin{tabular}{lllrrrrrrrrrrrrr}
			\hline
			\hline\\[-2.5 ex]
			&            &       & \multicolumn{6}{l}{(C1)  $Z_1\sim Bernoulli(0.5)$, $Z_2 \sim N(0,1)$,}                        &  & \multicolumn{6}{l}{(C2)  $Z_1\sim Bernoulli(0.8)$, $Z_2 \sim N(0,1)$,}                        \\ 
			&            &       & \multicolumn{6}{l}{\;\;\;\;\;\;\; and $Z_1 \indep Z_2$, i.e., $f^*(Z_1,Z_2)=f(Z_1,Z_2)$}                        &  & \multicolumn{6}{l}{\;\;\;\;\;\;\;  and $Z_1 \indep Z_2$}                        \\ 
			\cline{4-9} \cline{11-16}\\[-2.5 ex]
			$t$   & $\Lambda_0(t)$ &       & $\widehat{\Lambda}^\textup{b}_0(t)$ & $\widehat{\Lambda}^\textup{u}_0(t)$    & $\widehat{\Lambda}^\textup{w}_{0,1}(t)$ & $\widehat{\Lambda}^\textup{w}_{0,2}(t)$ & $\widehat{\Lambda}^\textup{w}_{0,3}(t)$ & $\widehat{\Lambda}^\textup{w}_{0,4}(t)$ &  & $\widehat{\Lambda}^\textup{b}_0(t)$ & $\widehat{\Lambda}^\textup{u}_0(t)$    & $\widehat{\Lambda}^\textup{w}_{0,1}(t)$ & $\widehat{\Lambda}^\textup{w}_{0,2}(t)$ & $\widehat{\Lambda}^\textup{w}_{0,3}(t)$ & $\widehat{\Lambda}^\textup{w}_{0,4}(t)$ \\ [0.5 ex]
			\hline
			\rowcolor{Gray}
			20 & 0.04        & PBias & -35.9\% & 1.6\%  & 1.5\%   & 1.5\%   & 1.6\%   & 1.5\%   &  & -36.3\% & 21.2\% & 1.8\%   & 1.1\%   & 1.2\%   & 1.2\%   \\
			&            & ESD   & 0.52    & 0.71   & 0.71    & 0.70    & 0.70    & 0.70    &  & 0.50    & 0.80   & 0.79    & 0.77    & 0.76    & 0.76    \\
			&            & ASEd  & 0.50    & 0.74   & 0.73    & 0.72    & 0.72    & 0.71    &  & 0.50    & 0.85   & 0.82    & 0.80    & 0.80    & 0.79    \\
\hline
		\end{tabular}
	\end{threeparttable}
	\label{table:simu7}
\end{table}

\begin{table}[]
	\renewcommand\arraystretch{1.1}
	\setlength{\tabcolsep}{2.5pt}
	\centering
	\scriptsize
	\caption{Summary statistics of simulation results for the scenario of $m=1,000$, (A4) $\Lambda_0(t)=(0.01t)^2 \; and\; \Lambda^*_0(t)=(0.008t)^{1.5}$, and $n=1,000$.}
	\begin{threeparttable}
		\begin{tabular}{lllrrrrrrrrrrrrr}
			\hline
			\hline\\[-2.5 ex]
			&            &       & \multicolumn{6}{l}{(C1)  $Z_1\sim Bernoulli(0.5)$, $Z_2 \sim N(0,1)$,}                        &  & \multicolumn{6}{l}{(C2)  $Z_1\sim Bernoulli(0.8)$, $Z_2 \sim N(0,1)$,}                        \\ 
			&            &       & \multicolumn{6}{l}{\;\;\;\;\;\;\; and $Z_1 \indep Z_2$, i.e., $f^*(Z_1,Z_2)=f(Z_1,Z_2)$}                        &  & \multicolumn{6}{l}{\;\;\;\;\;\;\;  and $Z_1 \indep Z_2$}                        \\ 
			\cline{4-9} \cline{11-16}\\[-2.5 ex]
			$t$   & $\Lambda_0(t)$ &       & $\widehat{\Lambda}^\textup{b}_0(t)$ & $\widehat{\Lambda}^\textup{u}_0(t)$    & $\widehat{\Lambda}^\textup{w}_{0,1}(t)$ & $\widehat{\Lambda}^\textup{w}_{0,2}(t)$ & $\widehat{\Lambda}^\textup{w}_{0,3}(t)$ & $\widehat{\Lambda}^\textup{w}_{0,4}(t)$ &  & $\widehat{\Lambda}^\textup{b}_0(t)$ & $\widehat{\Lambda}^\textup{u}_0(t)$    & $\widehat{\Lambda}^\textup{w}_{0,1}(t)$ & $\widehat{\Lambda}^\textup{w}_{0,2}(t)$ & $\widehat{\Lambda}^\textup{w}_{0,3}(t)$ & $\widehat{\Lambda}^\textup{w}_{0,4}(t)$ \\ [0.5 ex]
			\hline
			\rowcolor{Gray}
			20 & 0.04        & PBias & 57.7\%  & 0.7\%  & 0.6\%   & 0.7\%   & 0.7\%   & 0.7\%   &  & 57.3\%  & 21.4\% & 1.9\%   & 1.4\%   & 1.5\%   & 1.5\%   \\
			&            & ESD   & 0.88    & 0.66   & 0.65    & 0.64    & 0.64    & 0.64    &  & 0.89    & 0.75   & 0.73    & 0.70    & 0.69    & 0.70    \\
			&            & ASEd  & 0.88    & 0.68   & 0.67    & 0.66    & 0.66    & 0.65    &  & 0.88    & 0.78   & 0.74    & 0.71    & 0.71    & 0.71    \\
\hline
		\end{tabular}
	\end{threeparttable}
	\label{table:simu8}
\end{table}

\begin{table}[]
	\renewcommand\arraystretch{1.1}
	\setlength{\tabcolsep}{2.5pt}
	\centering
	\scriptsize
	\caption{Summary statistics of simulation results for the scenario of $m=200$, (A1) $\Lambda^*_0(t)=\Lambda_0(t)=(0.01t)^2$, and $n=1,000$.}
	\begin{threeparttable}
		\begin{tabular}{lllrrrrrrrrrrrrr}
			\hline
			\hline\\[-2.5 ex]
			&            &       & \multicolumn{6}{l}{(C1)  $Z_1\sim Bernoulli(0.5)$, $Z_2 \sim N(0,1)$,}                        &  & \multicolumn{6}{l}{(C2)  $Z_1\sim Bernoulli(0.8)$, $Z_2 \sim N(0,1)$,}                        \\ 
			&            &       & \multicolumn{6}{l}{\;\;\;\;\;\;\; and $Z_1 \indep Z_2$, i.e., $f^*(Z_1,Z_2)=f(Z_1,Z_2)$}                        &  & \multicolumn{6}{l}{\;\;\;\;\;\;\;  and $Z_1 \indep Z_2$}                        \\ 
			\cline{4-9} \cline{11-16}\\[-2.5 ex]
			$t$   & $\Lambda_0(t)$ &       & $\widehat{\Lambda}^\textup{b}_0(t)$ & $\widehat{\Lambda}^\textup{u}_0(t)$    & $\widehat{\Lambda}^\textup{w}_{0,1}(t)$ & $\widehat{\Lambda}^\textup{w}_{0,2}(t)$ & $\widehat{\Lambda}^\textup{w}_{0,3}(t)$ & $\widehat{\Lambda}^\textup{w}_{0,4}(t)$ &  & $\widehat{\Lambda}^\textup{b}_0(t)$ & $\widehat{\Lambda}^\textup{u}_0(t)$    & $\widehat{\Lambda}^\textup{w}_{0,1}(t)$ & $\widehat{\Lambda}^\textup{w}_{0,2}(t)$ & $\widehat{\Lambda}^\textup{w}_{0,3}(t)$ & $\widehat{\Lambda}^\textup{w}_{0,4}(t)$ \\ [0.5 ex]
			\hline
			\rowcolor{Gray}
			20 & 0.04        & PBias & 0.5\%   & 1.9\%  & 1.9\%   & 1.7\%   & 1.8\%   & 1.8\%   &  & 0.6\%   & 23.4\% & 3.4\%   & 2.7\%   & 3.0\%   & 3.0\%   \\
			&            & ESD   & 0.64    & 1.21   & 1.21    & 1.19    & 1.18    & 1.18    &  & 0.64    & 1.36   & 1.19    & 1.16    & 1.15    & 1.15    \\
			&            & ASEd  & 0.65    & 1.23   & 1.22    & 1.23    & 1.23    & 1.23    &  & 0.65    & 1.38   & 1.21    & 1.21    & 1.21    & 1.21    \\
\hline
		\end{tabular}
	\end{threeparttable}
	\label{table:simu9}
\end{table}

\begin{table}[]
	\renewcommand\arraystretch{1.1}
	\setlength{\tabcolsep}{2.5pt}
	\centering
	\scriptsize
	\caption{Summary statistics of simulation results for the scenario of $m=200$, (A2) $\Lambda_0(t)=(0.01t)^2 \; and\; \Lambda^*_0(t)=(0.01t)^{1.5}$, and $n=1,000$.}
	\begin{threeparttable}
		\begin{tabular}{lllrrrrrrrrrrrrr}
			\hline
			\hline\\[-2.5 ex]
			&            &       & \multicolumn{6}{l}{(C1)  $Z_1\sim Bernoulli(0.5)$, $Z_2 \sim N(0,1)$,}                        &  & \multicolumn{6}{l}{(C2)  $Z_1\sim Bernoulli(0.8)$, $Z_2 \sim N(0,1)$,}                        \\ 
			&            &       & \multicolumn{6}{l}{\;\;\;\;\;\;\; and $Z_1 \indep Z_2$, i.e., $f^*(Z_1,Z_2)=f(Z_1,Z_2)$}                        &  & \multicolumn{6}{l}{\;\;\;\;\;\;\;  and $Z_1 \indep Z_2$}                        \\ 
			\cline{4-9} \cline{11-16}\\[-2.5 ex]
			$t$   & $\Lambda_0(t)$ &       & $\widehat{\Lambda}^\textup{b}_0(t)$ & $\widehat{\Lambda}^\textup{u}_0(t)$    & $\widehat{\Lambda}^\textup{w}_{0,1}(t)$ & $\widehat{\Lambda}^\textup{w}_{0,2}(t)$ & $\widehat{\Lambda}^\textup{w}_{0,3}(t)$ & $\widehat{\Lambda}^\textup{w}_{0,4}(t)$ &  & $\widehat{\Lambda}^\textup{b}_0(t)$ & $\widehat{\Lambda}^\textup{u}_0(t)$    & $\widehat{\Lambda}^\textup{w}_{0,1}(t)$ & $\widehat{\Lambda}^\textup{w}_{0,2}(t)$ & $\widehat{\Lambda}^\textup{w}_{0,3}(t)$ & $\widehat{\Lambda}^\textup{w}_{0,4}(t)$ \\ [0.5 ex]
			\hline
			\rowcolor{Gray}
			20 & 0.04        & PBias & 120.5\% & 1.2\%  & 1.2\%   & 0.9\%   & 1.0\%   & 0.9\%   &  & 121.0\% & 22.6\% & 2.6\%   & 2.0\%   & 2.1\%   & 2.2\%   \\
			&            & ESD   & 1.07    & 1.24   & 1.23    & 1.20    & 1.20    & 1.20    &  & 1.12    & 1.35   & 1.17    & 1.14    & 1.14    & 1.14    \\
			&            & ASEd  & 1.08    & 1.20   & 1.20    & 1.20    & 1.21    & 1.21    &  & 1.08    & 1.35   & 1.17    & 1.17    & 1.17    & 1.17    \\
\hline
		\end{tabular}
	\end{threeparttable}
	\label{table:simu10}
\end{table}

\begin{table}[]
	\renewcommand\arraystretch{1.1}
	\setlength{\tabcolsep}{2.5pt}
	\centering
	\scriptsize
	\caption{Summary statistics of simulation results for the scenario of $m=200$, (A3) $\Lambda_0(t)=(0.01t)^2 \; and\; \Lambda^*_0(t)=(0.008t)^2$, and $n=1,000$.}
	\begin{threeparttable}
		\begin{tabular}{lllrrrrrrrrrrrrr}
			\hline
			\hline\\[-2.5 ex]
			&            &       & \multicolumn{6}{l}{(C1)  $Z_1\sim Bernoulli(0.5)$, $Z_2 \sim N(0,1)$,}                        &  & \multicolumn{6}{l}{(C2)  $Z_1\sim Bernoulli(0.8)$, $Z_2 \sim N(0,1)$,}                        \\ 
			&            &       & \multicolumn{6}{l}{\;\;\;\;\;\;\; and $Z_1 \indep Z_2$, i.e., $f^*(Z_1,Z_2)=f(Z_1,Z_2)$}                        &  & \multicolumn{6}{l}{\;\;\;\;\;\;\;  and $Z_1 \indep Z_2$}                        \\ 
			\cline{4-9} \cline{11-16}\\[-2.5 ex]
			$t$   & $\Lambda_0(t)$ &       & $\widehat{\Lambda}^\textup{b}_0(t)$ & $\widehat{\Lambda}^\textup{u}_0(t)$    & $\widehat{\Lambda}^\textup{w}_{0,1}(t)$ & $\widehat{\Lambda}^\textup{w}_{0,2}(t)$ & $\widehat{\Lambda}^\textup{w}_{0,3}(t)$ & $\widehat{\Lambda}^\textup{w}_{0,4}(t)$ &  & $\widehat{\Lambda}^\textup{b}_0(t)$ & $\widehat{\Lambda}^\textup{u}_0(t)$    & $\widehat{\Lambda}^\textup{w}_{0,1}(t)$ & $\widehat{\Lambda}^\textup{w}_{0,2}(t)$ & $\widehat{\Lambda}^\textup{w}_{0,3}(t)$ & $\widehat{\Lambda}^\textup{w}_{0,4}(t)$ \\ [0.5 ex]
			\hline
			\rowcolor{Gray}
			20 & 0.04        & PBias & -36.2\% & 1.9\%  & 1.8\%   & 1.7\%   & 1.9\%   & 1.9\%   &  & -36.6\% & 21.5\% & 1.6\%   & 1.0\%   & 1.2\%   & 1.2\%   \\
			&            & ESD   & 0.51    & 1.26   & 1.25    & 1.23    & 1.23    & 1.22    &  & 0.49    & 1.36   & 1.21    & 1.17    & 1.17    & 1.17    \\
			&            & ASEd  & 0.50    & 1.25   & 1.25    & 1.26    & 1.26    & 1.26    &  & 0.50    & 1.40   & 1.24    & 1.24    & 1.24    & 1.24    \\
\hline
		\end{tabular}
	\end{threeparttable}
	\label{table:simu11}
\end{table}

\begin{table}[]
	\renewcommand\arraystretch{1.1}
	\setlength{\tabcolsep}{2.5pt}
	\centering
	\scriptsize
	\caption{Summary statistics of simulation results for the scenario of $m=200$, (A4) $\Lambda_0(t)=(0.01t)^2 \; and\; \Lambda^*_0(t)=(0.008t)^{1.5}$, and $n=1,000$.}
	\begin{threeparttable}
		\begin{tabular}{lllrrrrrrrrrrrrr}
			\hline
			\hline\\[-2.5 ex]
			&            &       & \multicolumn{6}{l}{(C1)  $Z_1\sim Bernoulli(0.5)$, $Z_2 \sim N(0,1)$,}                        &  & \multicolumn{6}{l}{(C2)  $Z_1\sim Bernoulli(0.8)$, $Z_2 \sim N(0,1)$,}                        \\ 
			&            &       & \multicolumn{6}{l}{\;\;\;\;\;\;\; and $Z_1 \indep Z_2$, i.e., $f^*(Z_1,Z_2)=f(Z_1,Z_2)$}                        &  & \multicolumn{6}{l}{\;\;\;\;\;\;\;  and $Z_1 \indep Z_2$}                        \\ 
			\cline{4-9} \cline{11-16}\\[-2.5 ex]
			$t$   & $\Lambda_0(t)$ &       & $\widehat{\Lambda}^\textup{b}_0(t)$ & $\widehat{\Lambda}^\textup{u}_0(t)$    & $\widehat{\Lambda}^\textup{w}_{0,1}(t)$ & $\widehat{\Lambda}^\textup{w}_{0,2}(t)$ & $\widehat{\Lambda}^\textup{w}_{0,3}(t)$ & $\widehat{\Lambda}^\textup{w}_{0,4}(t)$ &  & $\widehat{\Lambda}^\textup{b}_0(t)$ & $\widehat{\Lambda}^\textup{u}_0(t)$    & $\widehat{\Lambda}^\textup{w}_{0,1}(t)$ & $\widehat{\Lambda}^\textup{w}_{0,2}(t)$ & $\widehat{\Lambda}^\textup{w}_{0,3}(t)$ & $\widehat{\Lambda}^\textup{w}_{0,4}(t)$ \\ [0.5 ex]
			\hline
			\rowcolor{Gray}
			20 & 0.04        & PBias & 57.4\%  & 1.6\%  & 1.6\%   & 1.4\%   & 1.6\%   & 1.6\%   &  & 58.2\%  & 23.0\% & 3.0\%   & 2.4\%   & 2.7\%   & 2.7\%   \\
			&            & ESD   & 0.90    & 1.20   & 1.20    & 1.18    & 1.17    & 1.17    &  & 0.91    & 1.39   & 1.22    & 1.18    & 1.18    & 1.18    \\
			&            & ASEd  & 0.88    & 1.22   & 1.21    & 1.22    & 1.22    & 1.22    &  & 0.89    & 1.37   & 1.20    & 1.20    & 1.20    & 1.20    \\
\hline
		\end{tabular}
	\end{threeparttable}
	\label{table:simu12}
\end{table}

\begin{figure}
\begin{center}
\includegraphics[width=6.5in]{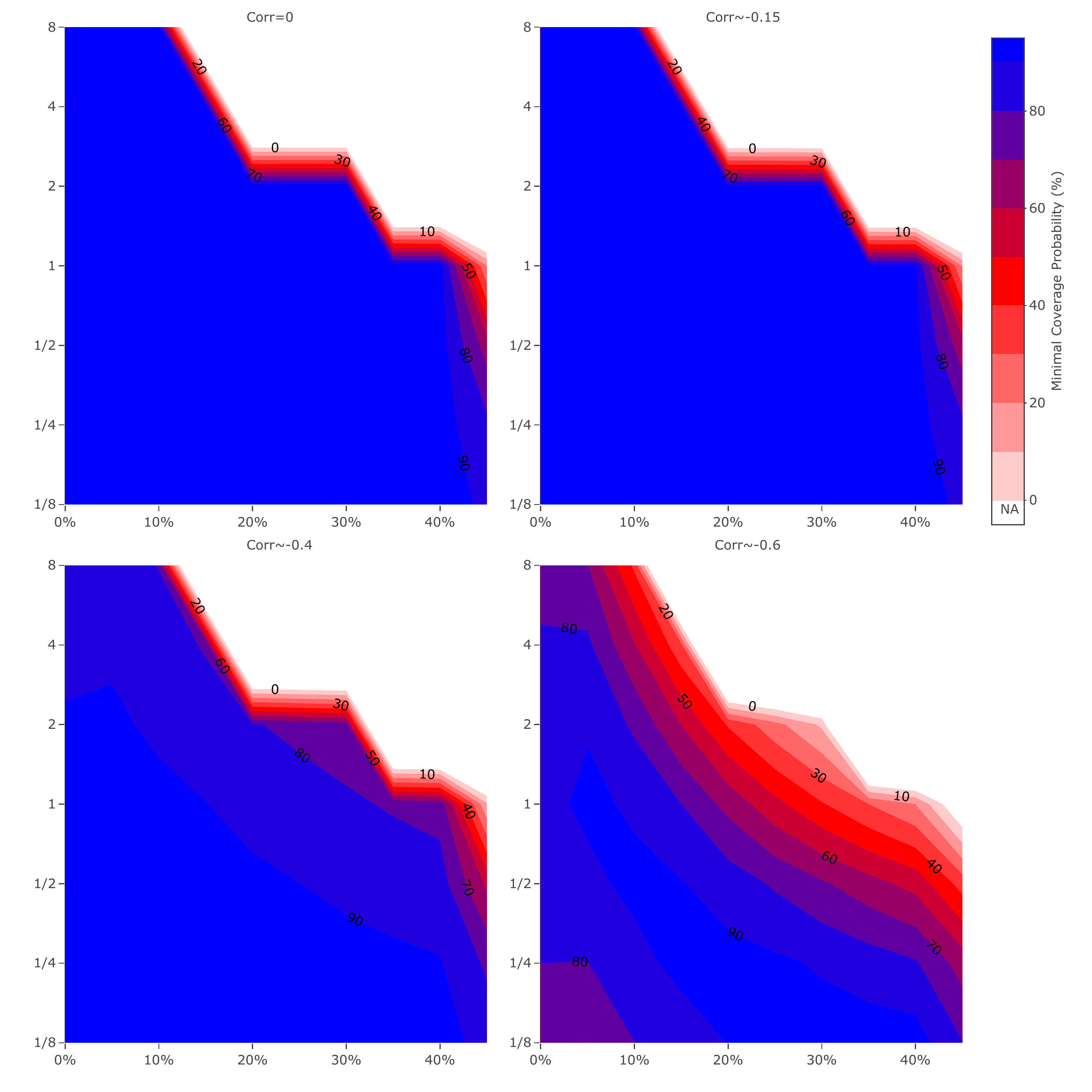}
\end{center}
\caption{ Minimal coverage probability of the proposed weighted estimator $\widehat{\Lambda}^\textup{w}_{0,4}(t)$. Minimal coverage probability is defined as the minimum of coverage probabilities at the three time points, 20, 40, and 60.
Corr: Pearson correlation coefficient between risk score for event of interest and $T_c$, i.e., $corr(\boldsymbol{\beta^T_0 Z},T_c)$.  Horizontal axis: probability of observing the event of interest; vertical axis: ratio between probability of observing the event of competing risk and probability of observing the event of interest. }
\label{figure_contour}
\end{figure}

\begin{table}[]
\renewcommand\arraystretch{1.1}
\setlength{\tabcolsep}{3.5pt}
\centering
\scriptsize
\caption{Real data based simulation results in the presence of competing risks (UKB cohorts as the target populations).}
\begin{threeparttable}
	\begin{tabular}{lclrrrrrclrrrr}
		\hline
		\hline\\[-2.5 ex]
		&             \multicolumn{6}{c}{UKB England}                    &    &   \multicolumn{6}{c}{UKB Scotland}                        \\ 
		\cline{2-7} \cline{9-14}\\[-2.5 ex]
		$t$   & $\Lambda_0(t)(10^{-3}) $  &       & $\widehat{\Lambda}^\textup{b}_0(t)$ & $\widehat{\Lambda}^\textup{u}_0(t)$    &  $\widehat{\Lambda}^\textup{w}_{0,p}(t)$ & $\widehat{\Lambda}^\textup{w}_{0,a}(t)$ & & $\Lambda_0(t) (10^{-3}) $  &  & $\widehat{\Lambda}^\textup{b}_0(t)$ & $\widehat{\Lambda}^\textup{u}_0(t)$    &  $\widehat{\Lambda}^\textup{w}_{0,p}(t)$ & $\widehat{\Lambda}^\textup{w}_{0,a}(t)$ \\ [0.5 ex]
		\hline
		\rowcolor{Gray}
		5 & 1.2 & PBias & -45.4\% & 17.7\% & -0.6\% & -0.1\% &  & 0.4 & PBias & 61.3\% & 16.9\% & 1.3\% & 2.4\% \\
 &  & Bias & -0.47 & 0.19 & -0.01 & 0.00 &  &  & Bias & 0.22 & 0.06 & 0.00 & 0.01 \\
 &  & ESD & 0.11 & 0.14 & 0.10 & 0.10 &  &  & ESD & 0.11 & 0.18 & 0.15 & 0.16 \\
 &  & ASE & 0.10 & 0.14 & 0.09 & 0.09 &  &  & ASE & 0.10 & 0.17 & 0.15 & 0.15 \\
 &  & sMSE & 0.49 & 0.23 & 0.10 & 0.10 &  &  & sMSE & 0.24 & 0.19 & 0.15 & 0.16 \\
 &  & CP & 2.8\% & 76.4\% & 93.5\% & 93.9\% &  &  & CP & 45.2\% & 93.3\% & 91.1\% & 91.1\% \\
		\hline
		\rowcolor{Gray}
		10 & 3.6 & PBias & -5.5\% & 17.6\% & -0.7\% & -0.2\% &  & 2.2 & PBias & 60.0\% & 14.7\% & -0.9\% & 0.2\% \\
 &  & Bias & -0.18 & 0.59 & -0.02 & -0.01 &  &  & Bias & 1.20 & 0.30 & -0.02 & 0.00 \\
 &  & ESD & 0.37 & 0.39 & 0.25 & 0.24 &  &  & ESD & 0.37 & 0.49 & 0.41 & 0.41 \\
 &  & ASE & 0.36 & 0.38 & 0.24 & 0.24 &  &  & ASE & 0.36 & 0.48 & 0.39 & 0.40 \\
 &  & sMSE & 0.41 & 0.71 & 0.25 & 0.24 &  &  & sMSE & 1.26 & 0.58 & 0.41 & 0.41 \\
 &  & CP & 88.2\% & 69.3\% & 93.8\% & 94.5\% &  &  & CP & 4.4\% & 94.2\% & 93.4\% & 93.9\% \\
		\hline
		\rowcolor{Gray}
		15 & 7.1 & PBias & 30.7\% & 17.6\% & -0.6\% & -0.1\% &  & 5.9 & PBias & 59.3\% & 14.8\% & -0.8\% & 0.3\% \\
 &  & Bias & 2.06 & 1.18 & -0.04 & -0.01 &  &  & Bias & 3.26 & 0.81 & -0.05 & 0.01 \\
 &  & ESD & 0.91 & 0.76 & 0.48 & 0.46 &  &  & ESD & 0.88 & 1.03 & 0.81 & 0.82 \\
 &  & ASE & 0.88 & 0.74 & 0.45 & 0.45 &  &  & ASE & 0.88 & 1.02 & 0.79 & 0.80 \\
 &  & sMSE & 2.25 & 1.41 & 0.49 & 0.46 &  &  & sMSE & 3.38 & 1.31 & 0.81 & 0.82 \\
 &  & CP & 33.2\% & 66.6\% & 93.3\% & 94.4\% &  &  & CP & 1.1\% & 91.6\% & 93.4\% & 93.9\% \\
        \hline
		\rowcolor{Gray}
		20 & 11.3 & PBias & 64.2\% & 17.1\% & -1.1\% & -0.6\% &  & 11.9 & PBias & 59.0\% & 14.6\% & -1.0\% & 0.1\% \\
 &  & Bias & 7.01 & 1.87 & -0.12 & -0.06 &  &  & Bias & 6.66 & 1.65 & -0.11 & 0.01 \\
 &  & ESD & 1.77 & 1.26 & 0.81 & 0.77 &  &  & ESD & 1.75 & 1.98 & 1.53 & 1.55 \\
 &  & ASE & 1.73 & 1.22 & 0.76 & 0.75 &  &  & ASE & 1.73 & 1.98 & 1.52 & 1.54 \\
 &  & sMSE & 7.23 & 2.25 & 0.82 & 0.77 &  &  & sMSE & 6.88 & 2.57 & 1.53 & 1.55 \\
 &  & CP & 0.4\% & 69.2\% & 92.9\% & 93.9\% &  &  & CP & 0.7\% & 91.8\% & 93.8\% & 94.5\% \\
        \hline
         &  & CAD & 29.6 & 14.7 & 4.8 & 4.6 &  &  & CAD & 38.6 & 12.9 & 8.3 & 8.4 \\
		\hline
	\end{tabular}
	\begin{tablenotes}
		\scriptsize
		\item[] NOTE: $\widehat{\Lambda}^\textup{b}_0(t)$, the WHI estimator; $\widehat{\Lambda}^\textup{u}_0(t)$, the unweighted estimator; $\widehat{\Lambda}^\textup{p}_0(t)$, the weighted estimator with the same constrain as in the example; $\widehat{\Lambda}^\textup{a}_0(t)$, the weighted estimator with the constrain of all risk factor prevalences;  PBias, the relative bias; ESD, the empirical standard deviation ($\times 1000$) of the 2000 estimates; ASE, the mean of the asymptotic-based standard error ($\times 1000$); sMSE, the square root of mean squared error ($\times 1000$); CP, the coverage probability of a 95\% confidence interval for $\Lambda_0(t)$;  CAD, the average cumulative absolute deviation between the estimates and the true values across a time range of one through twenty ($\times 1000$).
	\end{tablenotes}
\end{threeparttable}
\label{table:simu.real1}
\end{table}

\begin{table}[]
\renewcommand\arraystretch{1.1}
\setlength{\tabcolsep}{3.5pt}
\centering
\scriptsize
\caption{Real data based simulation results in the presence of competing risks (PLCO cohorts as the target populations).}
\begin{threeparttable}
	\begin{tabular}{lclrrrrrclrrrr}
		\hline
		\hline\\[-2.5 ex]
		&             \multicolumn{6}{c}{PLCO intervention cohort}                    &    &   \multicolumn{6}{c}{PLCO control cohort}                        \\ 
		\cline{2-7} \cline{9-14}\\[-2.5 ex]
		$t$   & $\Lambda_0(t)(10^{-3}) $  &       & $\widehat{\Lambda}^\textup{b}_0(t)$ & $\widehat{\Lambda}^\textup{u}_0(t)$    &  $\widehat{\Lambda}^\textup{w}_{0,p}(t)$ & $\widehat{\Lambda}^\textup{w}_{0,a}(t)$ & & $\Lambda_0(t) (10^{-3}) $  &  & $\widehat{\Lambda}^\textup{b}_0(t)$ & $\widehat{\Lambda}^\textup{u}_0(t)$    &  $\widehat{\Lambda}^\textup{w}_{0,p}(t)$ & $\widehat{\Lambda}^\textup{w}_{0,a}(t)$ \\ [0.5 ex]
		\hline
		\rowcolor{Gray}
		5 & 0.1 & PBias & 771.6\% & 23.7\% & 2.1\% & 3.9\% &  & 0.3 & PBias & 47.4\% & -0.1\% & 1.1\% & 2.5\% \\
 &  & Bias & 0.51 & 0.02 & 0.00 & 0.00 &  &  & Bias & 0.18 & 0.00 & 0.00 & 0.01 \\
 &  & ESD & 0.11 & 0.05 & 0.05 & 0.05 &  &  & ESD & 0.10 & 0.13 & 0.12 & 0.13 \\
 &  & ASE & 0.10 & 0.05 & 0.04 & 0.04 &  &  & ASE & 0.10 & 0.12 & 0.12 & 0.12 \\
 &  & sMSE & 0.52 & 0.05 & 0.05 & 0.05 &  &  & sMSE & 0.21 & 0.13 & 0.12 & 0.13 \\
 &  & CP & 0.0\% & 86.8\% & 77.0\% & 78.2\% &  &  & CP & 61.2\% & 91.7\% & 92.2\% & 92.7\% \\
		\hline
		\rowcolor{Gray}
		10 & 0.8 & PBias & 515.2\% & 16.2\% & 1.2\% & 2.1\% &  & 1.6 & PBias & 58.2\% & -2.3\% & -1.1\% & 0.3\% \\
 &  & Bias & 2.67 & 0.08 & 0.01 & 0.01 &  &  & Bias & 1.17 & -0.05 & -0.02 & 0.01 \\
 &  & ESD & 0.36 & 0.16 & 0.14 & 0.14 &  &  & ESD & 0.36 & 0.33 & 0.30 & 0.31 \\
 &  & ASE & 0.36 & 0.16 & 0.13 & 0.13 &  &  & ASE & 0.36 & 0.32 & 0.30 & 0.30 \\
 &  & sMSE & 2.70 & 0.18 & 0.14 & 0.14 &  &  & sMSE & 1.23 & 0.33 & 0.31 & 0.31 \\
 &  & CP & 0.0\% & 95.3\% & 93.5\% & 93.6\% &  &  & CP & 5.0\% & 92.2\% & 93.3\% & 93.8\% \\
		\hline
		\rowcolor{Gray}
		15 & 2.7 & PBias & 402.6\% & 13.9\% & -1.1\% & -0.2\% &  & 4.1 & PBias & 64.6\% & -2.3\% & -1.1\% & 0.3\% \\
 &  & Bias & 7.02 & 0.24 & -0.02 & 0.00 &  &  & Bias & 3.43 & -0.12 & -0.06 & 0.02 \\
 &  & ESD & 0.88 & 0.33 & 0.28 & 0.29 &  &  & ESD & 0.88 & 0.65 & 0.56 & 0.58 \\
 &  & ASE & 0.88 & 0.33 & 0.27 & 0.28 &  &  & ASE & 0.88 & 0.64 & 0.55 & 0.56 \\
 &  & sMSE & 7.07 & 0.41 & 0.29 & 0.29 &  &  & sMSE & 3.54 & 0.66 & 0.56 & 0.58 \\
 &  & CP & 0.0\% & 92.5\% & 92.9\% & 93.7\% &  &  & CP & 0.6\% & 92.4\% & 94.0\% & 94.5\% \\
        \hline
		\rowcolor{Gray}
		20 & 6.3 & PBias & 335.5\% & 14.1\% & -0.9\% & 0.0\% &  & 7.9 & PBias & 69.4\% & -2.4\% & -1.2\% & 0.2\% \\
 &  & Bias & 13.83 & 0.58 & -0.04 & 0.00 &  &  & Bias & 7.35 & -0.25 & -0.12 & 0.03 \\
 &  & ESD & 1.74 & 0.61 & 0.52 & 0.52 &  &  & ESD & 1.72 & 1.12 & 0.92 & 0.95 \\
 &  & ASE & 1.73 & 0.61 & 0.50 & 0.51 &  &  & ASE & 1.73 & 1.12 & 0.91 & 0.95 \\
 &  & sMSE & 13.94 & 0.84 & 0.52 & 0.52 &  &  & sMSE & 7.55 & 1.15 & 0.92 & 0.95 \\
 &  & CP & 0.0\% & 89.1\% & 93.2\% & 94.2\% &  &  & CP & 0.2\% & 92.6\% & 93.8\% & 94.5\% \\
        \hline
		\rowcolor{Gray}
		\hline
		25 & 12.1 & PBias & 289.4\% & 13.9\% & -1.1\% & -0.2\% &  & 13.4 & PBias & 73.1\% & -2.8\% & -1.6\% & -0.2\% \\
 &  & Bias & 23.27 & 1.12 & -0.08 & -0.01 &  &  & Bias & 13.21 & -0.50 & -0.28 & -0.03 \\
 &  & ESD & 2.97 & 1.07 & 0.91 & 0.90 &  &  & ESD & 2.98 & 1.84 & 1.47 & 1.53 \\
 &  & ASE & 2.97 & 1.08 & 0.87 & 0.89 &  &  & ASE & 2.98 & 1.82 & 1.45 & 1.50 \\
 &  & sMSE & 23.46 & 1.55 & 0.91 & 0.90 &  &  & sMSE & 13.54 & 1.91 & 1.50 & 1.53 \\
 &  & CP & 0.0\% & 87.1\% & 93.5\% & 94.6\% &  &  & CP & 0.0\% & 92.2\% & 93.5\% & 94.4\% \\
		\rowcolor{Gray}
        \hline
        30 & 20.7 & PBias & 255.5\% & 13.5\% & -1.5\% & -0.6\% &  & 20.6 & PBias & 76.2\% & -2.9\% & -1.7\% & -0.3\% \\
 &  & Bias & 35.47 & 1.87 & -0.20 & -0.08 &  &  & Bias & 21.31 & -0.82 & -0.49 & -0.10 \\
 &  & ESD & 4.73 & 1.85 & 1.57 & 1.55 &  &  & ESD & 4.71 & 2.84 & 2.28 & 2.37 \\
 &  & ASE & 4.71 & 1.84 & 1.49 & 1.52 &  &  & ASE & 4.71 & 2.83 & 2.26 & 2.35 \\
 &  & sMSE & 35.78 & 2.63 & 1.58 & 1.55 &  &  & sMSE & 21.82 & 2.96 & 2.33 & 2.37 \\
 &  & CP & 0.0\% & 87.4\% & 92.3\% & 93.4\% &  &  & CP & 0.0\% & 92.3\% & 93.1\% & 94.4\% \\
         &  & CAD & 320.9 & 17.2 & 10.5 & 10.4 &  &  & CAD & 177.1 & 22.4 & 18.1 & 18.4 \\
		\hline
	\end{tabular}
	\begin{tablenotes}
		\scriptsize
		\item[] NOTE: $\widehat{\Lambda}^\textup{b}_0(t)$, the WHI estimator; $\widehat{\Lambda}^\textup{u}_0(t)$, the unweighted estimator; $\widehat{\Lambda}^\textup{p}_0(t)$, the weighted estimator with the same constrain as in the example; $\widehat{\Lambda}^\textup{a}_0(t)$, the weighted estimator with the constrain of all risk factor prevalences;  PBias, the relative bias; ESD, the empirical standard deviation ($\times 1000$) of the 2000 estimates; ASE, the mean of the asymptotic-based standard error ($\times 1000$); sMSE, the square root of mean squared error ($\times 1000$); CP, the coverage probability of a 95\% confidence interval for $\Lambda_0(t)$;  CAD, the average cumulative absolute deviation between the estimates and the true values across a time range of one through thirty ($\times 1000$).
	\end{tablenotes}
\end{threeparttable}
\label{table:simu.real2}
\end{table}